\documentclass[11pt, a4paper, english, oneside]{book}

\usepackage[intlimits]{amsmath}
\usepackage{amsfonts}
\usepackage{amsmath}
\usepackage{amssymb}
\usepackage{mathtools}
\usepackage{bm}
\usepackage{bbm}

\usepackage[a4paper]{geometry} 
\usepackage{setspace}          
\usepackage{fancyhdr}          
\usepackage[hang, bottom, stable, multiple]{footmisc} 

\usepackage{framed}
\usepackage{xcolor}
\definecolor{lightgray}{gray}{0.5}

\usepackage{rotating}
\usepackage{tabularx}
\usepackage{float}
\usepackage{graphicx}
\usepackage[skip=2pt, font=small]{caption}
\usepackage{subcaption}
\usepackage{booktabs}
\usepackage{enumitem}
\graphicspath{{images/}}
\usepackage{tikz}
\usepackage{multirow}

\usepackage{lmodern}

\newsavebox\CBox
\def\mathBF#1{\sbox\CBox{#1}\resizebox{\wd\CBox}{\ht\CBox}{\ensuremath{\mathbf{#1}}}}
\def\textBF#1{\sbox\CBox{#1}\resizebox{\wd\CBox}{\ht\CBox}{\textbf{#1}}}

\fancypagestyle{firststyle}{%
  \fancyhf{}%
  
  \fancyfoot[C]{\thepage}
}

\usepackage{bookmark}
\usepackage{pdfpages}
\usepackage{rotating}

\usepackage{setspace} 
\counterwithout{footnote}{chapter}

\newcounter{experiment}
\renewcommand{\theexperiment}{\arabic{experiment}}

\usepackage{natbib}
\providecommand{\subfilebibliography}{
\ifSubfilesClassLoaded{
  \bibliography{biblio/thesis, biblio/paperpile}}{}
}

\usepackage{makeidx} 

\usepackage{subfiles} 

\begin{document}

\pagenumbering{roman}


    \thispagestyle{empty}
    
    \centering 
    \vspace*{-1.2cm}
    \includegraphics{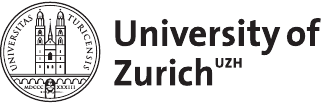}\hfill~\\[0.5\baselineskip]

    \rule{\textwidth}{1.6pt}\vspace*{-\baselineskip}\vspace*{2pt}
    \rule{\textwidth}{0.4pt}\\[0.7\baselineskip]

    {\Large\textbf{neuro2voc: Decoding Vocalizations from Neural Activity}}\\[0.2\baselineskip]

    \rule{\textwidth}{0.4pt}\vspace*{-\baselineskip}\vspace{3.2pt}
    \rule{\textwidth}{1.6pt}\\[1\baselineskip]
    
    \scshape 
    Master's Thesis\\[1\baselineskip]

    Presented to the Faculty of Arts and Social Sciences of\\the University of Zurich for the degree of
    
    Master of Arts \par

\vspace*{2\baselineskip}

Author\\
{\Large ~~Fei Gao
}
\vspace*{1.5\baselineskip}

Principal Investigators\\
{\Large Prof.\ Dr.\ Nathalie Giroud \par}
{\Large Prof.\ Dr.\ Richard Hahnloser \par}
\vspace*{1.5\baselineskip}

Supervisors\\
{\Large Maris Basha \par}
{\Large Dr. Corinna Lorenz \par}
{\Large Dr. Remo Nitschke \par}

\vfill
{\scshape Date of Submission: November 30, 2024}
\newpage



\chapter*{Abstract}
\thispagestyle{empty}
    Accurate decoding of neural spike trains and relating them to motor output is a challenging task due to the inherent sparsity and length in neural spikes and the complexity of brain circuits. This master project investigates experimental methods for decoding zebra finch motor outputs (in both discrete syllables and continuous spectrograms), from invasive neural recordings obtained from Neuropixels. 

    There are three major achievements: (1) XGBoost with SHAP analysis trained on spike rates revealed neuronal interaction patterns crucial for syllable classification. (2) Novel method (tokenizing neural data with GPT2) and architecture (Mamba2) demonstrated potential for decoding of syllables using spikes. (3) A combined contrastive learning-VAE framework successfully generated spectrograms from binned neural data.

    This work establishes a promising foundation for neural decoding of complex motor outputs and offers several novel methodological approaches for processing sparse neural data.


\phantomsection
\addcontentsline{toc}{chapter}{Acknowledgment}
\chapter*{Acknowledgment}
First and foremost, I would like to express my sincere gratitude to Prof.\ Dr.\ Richard Hahnloser for providing me with the opportunity to undertake this project and providing valuable feedback. I am equally grateful to Prof.\ Dr.\ Nathalie Giroud for her continued support throughout my academic journey.

I would like to extend my gratitude to my supervisors: Corinna, Maris, and Remo. I am deeply grateful to Corinna for her excellent data, patient guidance and detailed feedbacks on the intricate details of the project. I thank Maris for his ideas and consistent input on potential directions, as well as valuable feedbacks on presentation slides. His contribution of the NeuralEnCodec repository is the foundation for my subsequent EnCodec implementations. I am grateful to Remo for his thoughtful feedback, help and guidance on the tools, catch-ups, and insights into model interpretability. I would like to thank all three of my supervisors again for their help and feedback. Their collective mentorship have been invaluable to the completion of this thesis.

I would also like to thank Aidan, Jacob, as well as all members in the Birdsong and Natural Language Group for their feedback and patience during my presentations.

My sincere thanks to all amazing administrative team members at the Institute of Computational Linguistics and Institute of Neuroinformatics for their invaluable administrative support for this thesis.

Finally, I want to express my heartfelt gratitude to Zhugu, my friends, and my family for their unwavering support. This thesis would not have been possible without their unconditional support.

\newpage


\tableofcontents
\newpage







\clearpage
\pagenumbering{arabic}



\chapter{Introduction}%
\label{chap:intro}

Recent advances in neural recording technologies and speech analysis methods have created new possibilities to address challenges during analyzing the relationship between neural activity and behaviours. The challenges stem from the complex nature of spike trains -- sparse, long, and high dimensional -- as well as the need to align them with complex behavioural labels or external stimuli. For this project, we worked on bird vocalizations and the corresponding neural signals. We benefited from Neuropixels \citep{Jun2017-uh}, which allows for extremely high temporal and spatial resolution recording of neural spikes in animals, and from methods that enable effective extraction of syllable clusters from the bird vocalizations \citep{Lorenz2022-vm}. With these advances, we attempted to address some challenges in correlating spike train data with vocalization.

\section{Motivation}

This project is motivated by recent technological advances in Machine Learning (ML). First, Natural Language Processing (NLP) techniques have demonstrated remarkable success in text-processing. Transformer models excel at handling sentences, and have been adapted to handle sparse and long sequential data -- a characteristics shared with neural spike data. Second,  recent Mamba architectures \citep{Gu2023-vw, Dao2024-sp} efficiently process long sequences without the quadratic bottleneck which transformers have. These advances offer a perspective to address some key challenges in neural data analysis: the inherent sparsity of spike data and the need for long-range dependencies.

In addition, with emergent representation learning techniques, such as Variational Auto-Encoder (VAE) architecture and contrastive learning method, we would like to investigate correlations between neural data and behavioural data at a finer temporal scale. 

\section{Research Question}
Our project aimed to address three questions:
\begin{enumerate}
    \item Can we validate spike based classification method for syllables of Zebra Finch?
    \item How well can natural language processing methods and different deep learning architectures handle long, sparse, and high-dimensional spike data?
    \item Can we effectively map spike data to latent spaces for classification and reconstruction?
\end{enumerate}

\section{Methodological Framework}
To answer these questions, we developed a three-part methodological framework:

\subsection{Baseline Validation}
We first established performance benchmarks for syllable-based classification through classical ML methods: Support Vector Machine (SVM), Random Forest (RF), and XGBoost. In addition, we used SHAP values to analyze the interaction between neurons.

\subsection{Deep Learning Methods}
Then, we explored various neural data analysis methods. We used a tokenization method to analyze the data with GPT-2. We implemented several models, including Mamba-2 and EEGNet \citep{Lawhern2016-tf}, to deal with long range temporal dependencies over tokenized and time series data.

\subsection{Representational Learning}
We used CEBRA methods and VAE architecture to obtain interpretable latent representations from our neural data. We created a joint VAE with contrastive learning methods to map neural data into vocalizations. We evaluated all obtained representations on syllable classification tasks.

The code is publicly available at \href{https://github.com/askrbayern/neuro2voc_thesis}{https://github.com/askrbayern/neuro2voc\_thesis}.

\section{Thesis Structure}
The remainder of this thesis is divided into the following chapters:

\begin{itemize}
    \item Chapter~\ref{chap:related-work}: Related Work
    \item Chapter~\ref{chap:dataset}: Dataset
    \item Chapter~\ref{chap:methods}: Methodology
    \item Chapter~5--9: Experiments \newline
    Each chapter describes one key experiment, including:
    \begin{itemize}
        \item Experiment~\ref{exp:1} Machine Learning Benchmarks
        \item Experiment~\ref{exp:2} Token-based Deep Learning Methods 
        \item Experiment~\ref{exp:3} Temporal Deep Learning Methods
        \item Experiment~\ref{exp:4} CEBRA Latent Embeddings
        \item Experiment~\ref{exp:5} VAE Latent Representations
    \end{itemize}
   \item Chapter~\ref{chap:conclusion}:Conclusion
\end{itemize}

\subfilebibliography

\chapter{Related Work}%
\label{chap:related-work}
Research has shown that juvenile zebra finch birds learn to sing through a complex sensorimotor process \citep{Doupe1999-nm}. The birds progress from simple and highly variable sub-songs to fixed and distinct songs during learning. A direct motor pathway connects HVC to nucleus RA and controls the production of the song \citep{Gale2010-av}. HVC is also connected to RA through a second pathway, the anterior forebrain pathway (AFP). AFP controls the song variability, song learning and plasticity \citep{Doupe2005-iy}. In this pathway, HVC is connected to RA through area X, DLM, and LMAN in sequence. In this pathway, LMAN projects back to the area X in this circuit, thus forming a recursive loop. Research has shown that lesions from the LMAN area in this sensorimotor process stop adult birds from adapting spectral features of their song by operant conditioning \citep{Andalman2009-qt, Bottjer1984-mr}.

\section{Representation Learning Approaches}
Previous work by \cite{Singh_Alvarado2021-vr} investigated the variability of performance in male zebra finch birds in two conditions: solo practice (undirected singing) and performing (directed singing). They developed finchVAE, a Variational Auto-Encoder (VAE), to extract the latent representations of both neural and vocal data, and analyze the relationship between the latent embeddings. 

While only using undirected song data, we extended their work in two ways. First, with annotated syllables from \citep{Lorenz2022-vm}, we could focus on a finer scale of syllables instead of motifs, which is composed of several different syllables. Second, we adapted the original model on another type of neural recordings. The main challenge comes from the fact that the original paper used Calcium imaging as neural recordings, while our neural data was a time series collected from Neuropixels 1.0 \citep{Jun2017-uh} and then spike-sorted with Kilosort \citep{Pachitariu2016-mi}. VAEs were originally designed to deal with image data \citep{Kingma2013-es, Kingma2019-ng}, which is well-suited for the Calcium imaging data. For our time series data, we were particularly inspired by EnCodec from \cite{Defossez2022-lq}, a VAE structure for high-fidelity neural audio compression that utilizes quantization techniques and codebooks. The concept of codebooks originated from Vector Quantized VAE (VQ-VAE) proposed by \cite{van-den-Oord2017-rz} and was later put into work by \cite{Zeghidour2021-yp}. In VQ-VAE, the codebook is used to map continuous representations into discrete codes. We implemented EnCodec and devised different types of VAEs based on finchVAE to deal with spike data. 

Researchers have also explored contrastive learning methods to decode neural data \citep{Schneider2023-xu, Defossez2022-pq}. \cite{Schneider2023-xu} proposed the CEBRA method which processes time-series neural recordings, and generates latent embeddings for each time point. In the paper, the latent embeddings were used to decode neural responses in the visual cortex of mice that were watching a repeating video, and 95\% of the frames were correctly identified in the test dataset. We aim to use this method, and use different syllables to supervise the training. For human speech, \cite{Defossez2022-pq} proposed brainmagick, a contrastive learning based method to decode the speech heard from non-invasive brain recordings, such as Electroencephalogram (EEG) or Magnetoencephalography (MEG). They successfully aligned the neural responses with acoustic stimuli. For neural data, the best result came from the MEG data due to the high spatial accuracy of MEG. They first generated latent embeddings of non-invasive recordings through a Convolutional Neural Network. The speech embeddings were either from Mel Spectrogram, or taken from pre-trained models like wav2vec $2.0$ \citep{Baevski2020-rj}. The embeddings of two modalities were compared using Contrastive Language-Image Pre-Training (CLIP) loss. During our reproduction of brainmagick on the data of \cite{Gwilliams2022-xx}, the model correctly identified the word heard with 40\,\% accuracy from MEG data and had a Top 10 accuracy of 69\,\%.

In conclusion, the above representation learning methods show great potential for neural decoding. VAE excels at data compression and is generative in nature, while contrastive learning methods like CEBRA offer great decoding capability as well as fine temporal resolution. As a result, we combined VAE with contrastive learning in the last task.

There are several representation learning methods for neural data unexplored in this project: GP-VAE \citep{Gondur2023-eg}, MARBLE \citep{Gosztolai2023-by}, pi-VAE \citep{Zhou2020-om}, TiDeVAE \citep{Huang2024-wj}, etc.

\section{Deep Learning Approaches}
With recent advances in Natural Language Processing (NLP) and Machine Learning (ML) architectures, we aimed to use these new tools to deal with the spike train data. Spike train data is composed of spikes, or action potentials, which are discrete electrical signals generated by neurons as a result of underlying cellular and molecular activities. Spike train data is in nature very sparse. In Natural Language Processing (NLP), sparsity can be used to refer to either data sparsity or model sparsity. Data sparsity refers to large embedding sizes where only a few dimensions have significant values \citep{Farina2024-qh}. The most typical case would be One-Hot-Encodings, where each word is represented by a column vector of the length of the vocabulary, and all the positions are zero, except that the position for the current word has a one. Model sparsity refers to a neural network with a lot of empty weights, which could either increase the performance \citep{Child2019-od} or reduce performance by missing out important information \citep{Farina2024-qh}.

The sparsity in spike train data is similar to data sparsity described above. One drawback brought up by sparse spikes is that the activity patterns are encoded in a long `One-Hot-Encodings' where only a small part is meaningful. Transformer is not the best architecture to deal with long sequences, but there has been some work on this direction \citep{Li2023-ge}. In addition to this, transformers are resource-demanding. The computational complexity is quadratic to the sentence length, coming from the dot product calculation of the attention. 

Nevertheless, there are a few promising work done in this direction. \cite{Azabou2023-os} proposed a transformer-based framework Pre-training On manY neurOns (POYO). POYO treats the neural spikes as discrete tokens, and pretrains the model on large-scale recordings to improve the performance on decoding tasks. In this project, we also explored the tokenization method.

Although transformers can handle the sparse data to some extent, they are not the best models to use. With the quadratic bottleneck of transformers, it consumes a lot of time and computational resources to train and inference. \cite{Gu2023-vw} proposed Mamba, a Structured State-Space Model, that only has linear complexity. Mamba is specialized at dealing with long sequential data with minimum inference time and resource consumption. \cite{Dao2024-sp} proposed Mamba-2, a Structured State-Space Duality, which is more efficient than Mamba in training time. With Mamba models, analyzing sparse spike data by addressing data length seems to be more optimistic.

There are a few other methods. Binning is the most commonly used method to resolve this sparsity \citep{Labach2023-ib}. During binning, finer temporal information is lost. \citep{Tipirneni2021-ag} works on sparse and irregularly sampled data, where they used continuous value embeddings (CVE) to deal with such data. The CVE maps continuous values to high-dimensional spaces using a neural network. 

These works show various approaches to process neural data, and each approach offers unique advantage. In our project, we aimed to address the data sparsity issue. Our neural recordings were collected from Neuropixels 1.0 and spike sorted with Kilosort, which produces precise but sparse data. We explored NLP methods and several time series models including Mamba. We also investigated the potential of supervised representation learning methods, and use different VAEs to reconstruct our data.


\subfilebibliography

\chapter{Dataset}%
\label{chap:dataset}

\section{Data Overview}
The project used neural and vocal data that were previously collected through three recording sessions from the same bird. During each session, neural and vocal data were recorded simultaneously, thus their start points were the same. Table~\ref{tab:recording_durations} shows the length of recorded vocal and neural data. Extra vocal recordings were removed from the end in this project.

\begin{table}[htbp]
   \centering
   \begin{tabular}{lcc}
       \toprule
       Recording Session & Neural Data Length  & Vocal Data Length\\
       \midrule
       Session 0 & 53.29\,min& 53.29\,min\\
       Session 1 & 89.91\,min& 90.06\,min\\
       Session 2 & 70.37\,min& 77.90\,min\\
       \bottomrule
   \end{tabular}
   \caption{Duration of recordings across different sessions}
   \label{tab:recording_durations}
\end{table}

\section{Neural Data}
The preprocessed neural data consists of an array of spike times and an array of spike cluster indices. The data was provided at a sampling rate of $20$\,kHz. The raw data was collected with Neuropixels $1.0$ \citep{Jun2017-uh} and the preprocessed spike data was obtained through spike-sorting with Kilosort \citep{Pachitariu2016-mi}. All recordings contain only non-directed songs. The data includes additional information: (1) depth, (2) channel on Neuropixels, (3) firing rate, (4) spike amplitude, and (5) cluster quality.

\subsection{Characteristics of Neurons}
\begin{figure}[H]
   \centering
   \includegraphics[width=\textwidth]{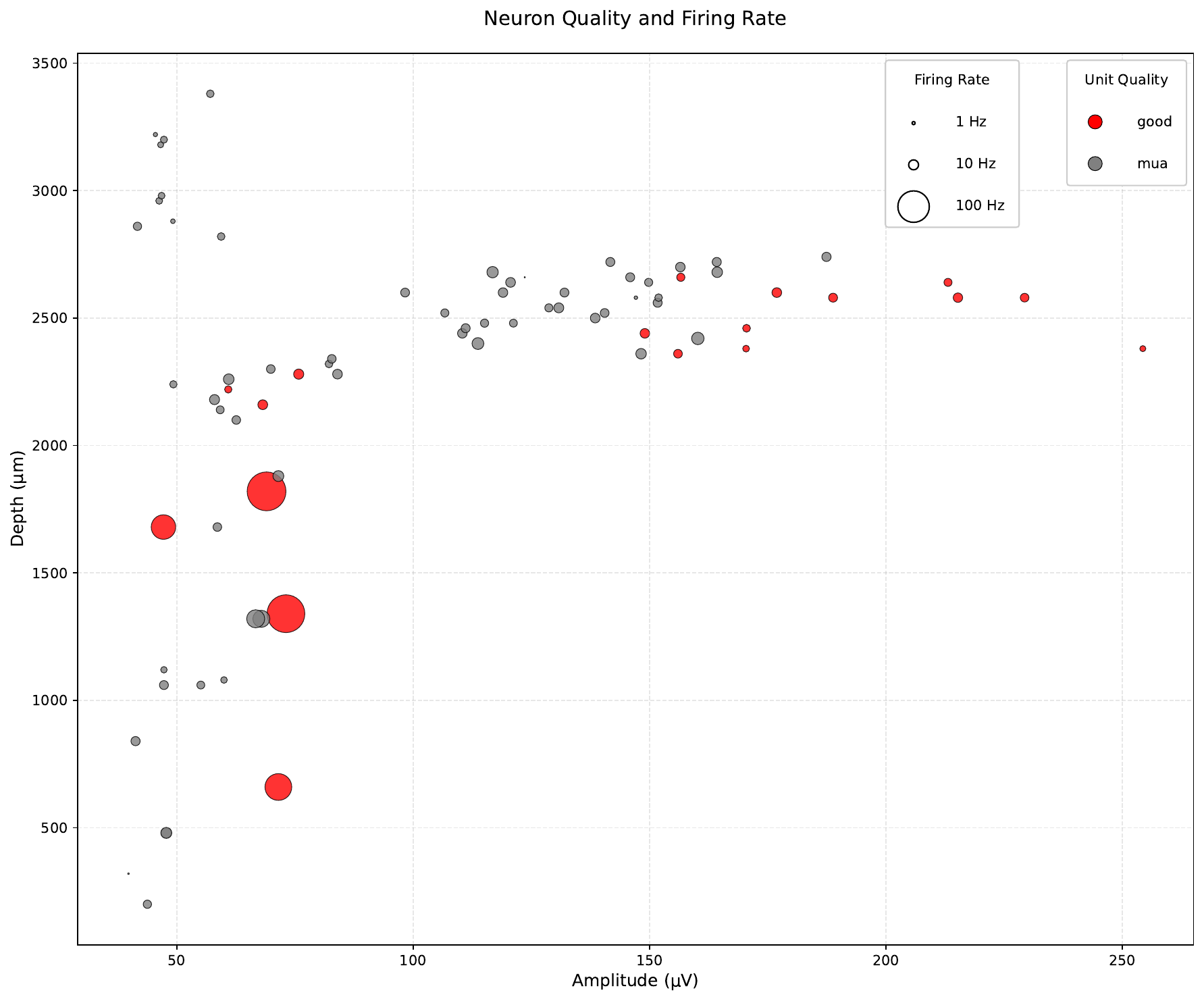}
   \caption{Quality of clusters}
   \label{fig:neuron_quality}
\end{figure}

Spike-sorting produces clusters of different qualities, which are marked with either good, multi-unit acitivities (MUA), or noise. In the data used in this project, all neuronal clusters were labeled as either good or MUA. The dataset consists of reliable neural recordings without noisy clusters, making it suitable for further analysis. Figure~\ref{fig:neuron_quality} shows the quality of clusters in dark grey and red, with circle diameter indicating the firing rate. LMAN neurons roughly distributed between 2300\,$\mu$m and 3000\,$\mu$m.

\section{Vocal Data}
\begin{figure}[htbp]
   \centering
   \includegraphics[width=\textwidth]{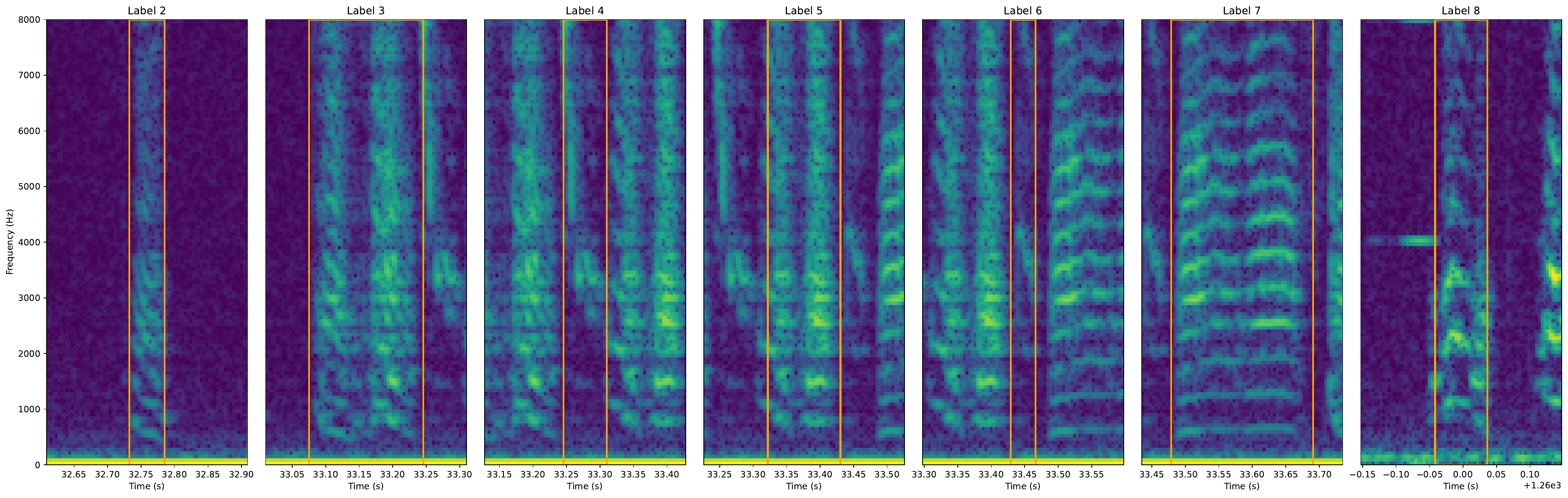}
   \caption{Labeled spectrogram for all 7 syllables}
   \label{fig:labeled_spectrogram}
\end{figure}

The vocal recordings were obtained from recording devices in the lab. The raw data was preprocessed by first downsampling from $20$\,kHz to $250$\,Hz, and then applying a STFT with a $16$\,ms offset ($64$ samples under $250$\,Hz). The data used for this project were prepared in spectrograms in $128$ frequency bins ranging from $[0\,\text{Hz}, -8000\,\text{Hz}]$ in $250$\,Hz. A separate annotation file provides the onset and duration for each vocalization cluster (referred to as syllable in this paper). Our analysis focused on valid vocalization patterns, labeled $2$--$8$. Other syllables were excluded from this investigation. Syllables $2$--$8$ are shown in Figure~\ref{fig:labeled_spectrogram}.

\subsection{Distribution of Syllables}
\begin{figure}[htbp]
   \centering
   \begin{subfigure}[b]{0.85\textwidth}
       \centering
       \includegraphics[width=0.9\textwidth]{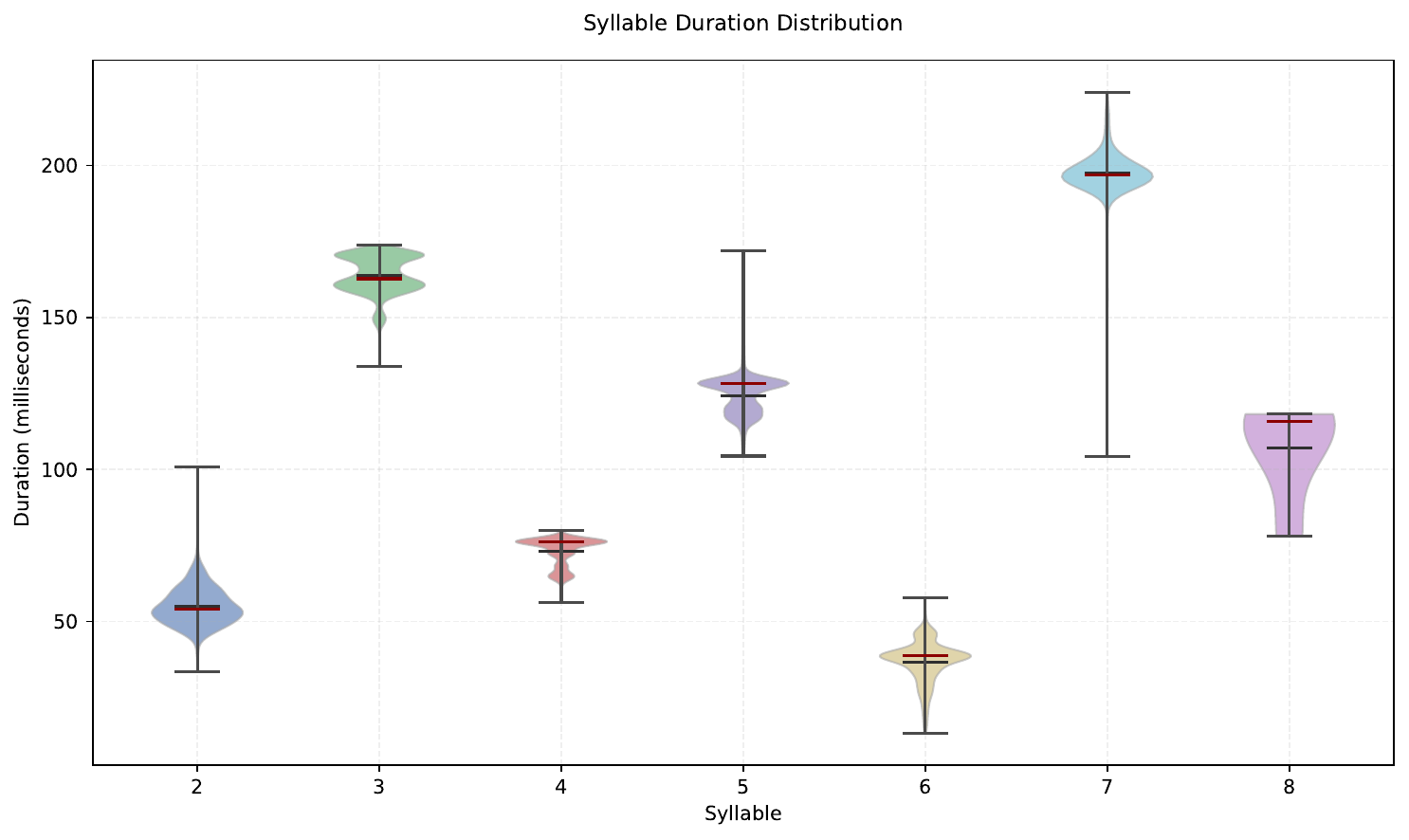}
       \caption{Distribution of syllable lengths}
       \label{fig:length_distribution}
   \end{subfigure}
   
   \begin{subfigure}[b]{0.85\textwidth}
       \centering
       \includegraphics[width=\textwidth]{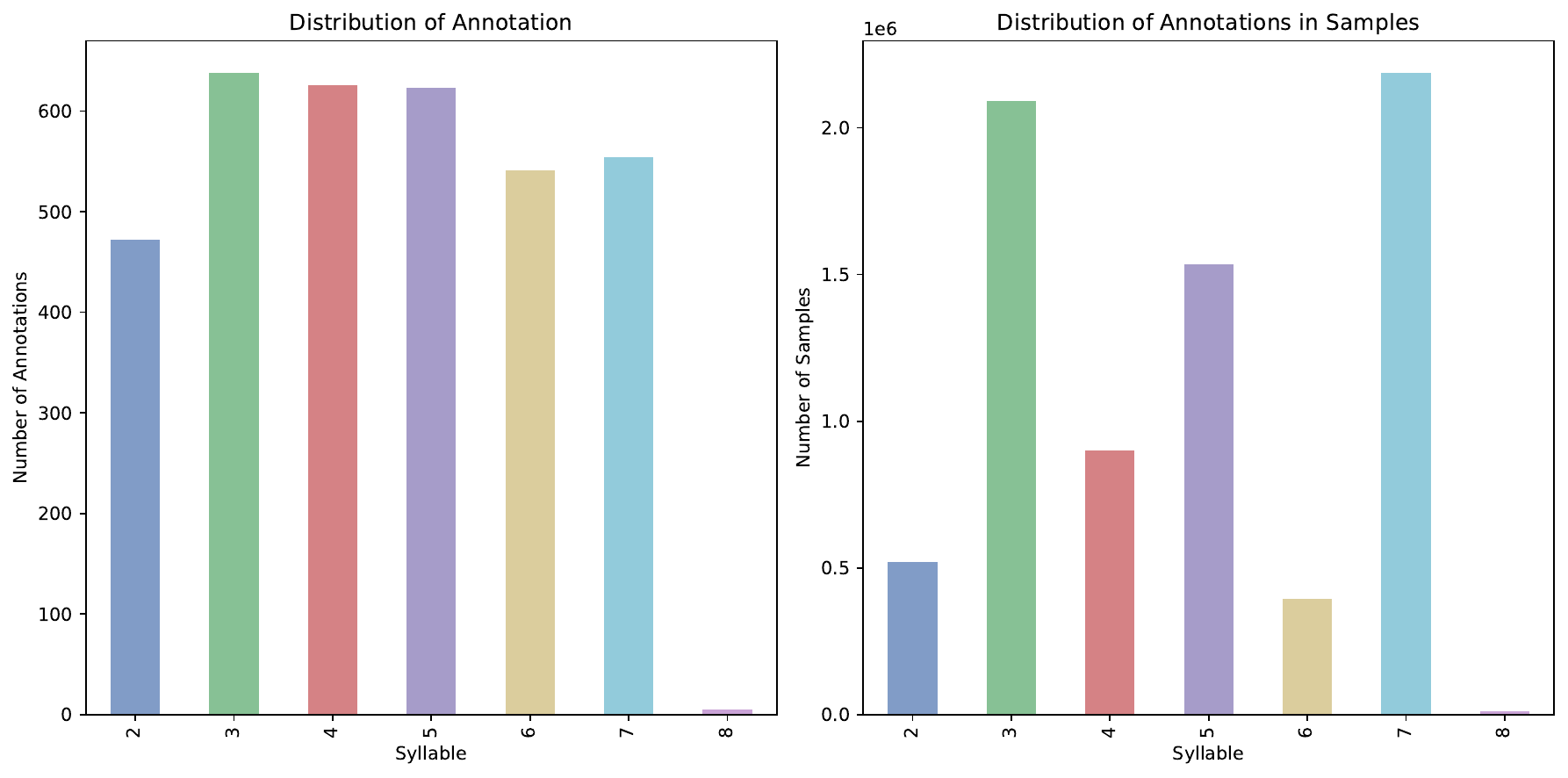}
       \caption{Total distribution of syllables}
       \label{fig:syllable_distribution}
   \end{subfigure}

   \begin{subfigure}[b]{0.85\textwidth}
       \centering
       \includegraphics[width=\textwidth]{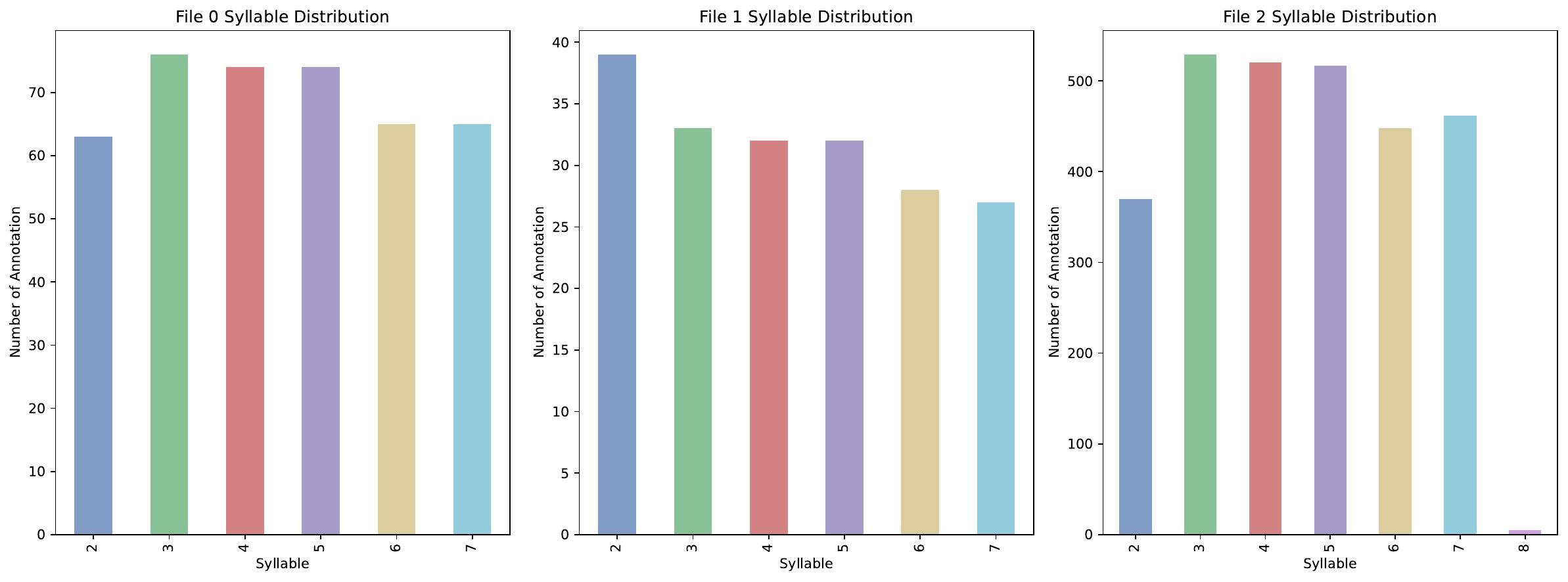}
        \caption{Distribution of syllables per session}
       \label{fig:distributionFile}
   \end{subfigure}
    \caption{Distribution of annotated syllables}
   \label{fig:distribution_syllable}
\end{figure}

Figure~\ref{fig:length_distribution} shows that there is considerable variation in syllable length. Figure~\ref{fig:syllable_distribution} shows syllables $2$--$7$ occur at similar frequencies, while syllable $8$ occurs rarely. On a sampling rate level, however, Figure~\ref{fig:syllable_distribution} shows that the data is imbalanced on a sampling time level, where label $3$, $5$, and $7$ together take up $80$\,\% of the sampling time data. Figure~\ref{fig:distributionFile} shows that the syllable distributions are similar across sessions.

\subfilebibliography

\chapter{Methodology}%
\label{chap:methods}

\section{Mamba}
\subsection{Selective State Space Model}
\begin{figure}[H]
    \centering
    \includegraphics[width=0.7\linewidth]{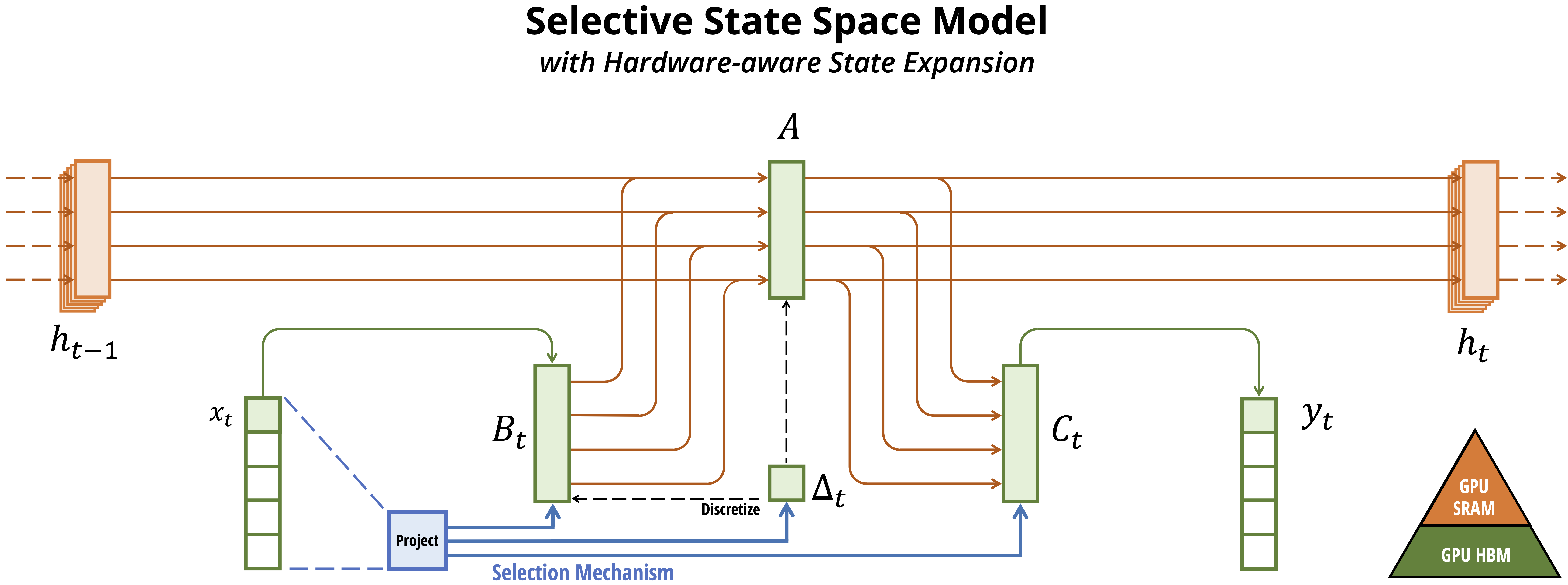}
    \caption{Selective State Space Model}
    \label{fig:ssm}
\end{figure}
\begin{align*}
    h_t &= g(\mathbf{A}_t h_{t-1} + \mathbf{B}_t x_t) \\
    y_t &= \mathbf{C}_t^T h_t
\end{align*}
\begin{align*}
\mathbf{A}_t \in \mathbb{R}^{N,N} &: \text{state to state transition matrices}\\
\mathbf{B}_t \in \mathbb{R}^N &: \text{input to state transition vector}\\
\mathbf{C}_t \in \mathbb{R}^N &: \text{state to output transition vector}\\ 
g &: \text{activation function}
\end{align*}
A selective state space model (SSM) is defined above, where $\mathbf{A}$, $\mathbf{B}$, and $\mathbf{C}$ are the three parameters of the model. Mamba \citep{Gu2023-vw} uses a structured SSM, which is a selective SSM with $\mathbf{A}_t$ being a diagonal matrix. 

\subsection{State Space Duality}
Mamba-2 \cite{Dao2024-sp} uses the state space duality (SSD) structure. SSD is built on SSM with one specification:
$$
    \mathbf{A}_t = a_t\mathbf{I}
$$
$$
a_t \in \mathbb{R}: \text{a learnable scalar} 
$$
This linear mode is also called the SSM mode, which is excellent at inference time due to its computational efficiency. This scalar-identity structured model can be expanded to a quadratic mode, which enables matrix multiplication and dramatically improves the training process. The quadratic mode above can be viewed as:
$$
    y = \mathbf{M} x
$$
$\mathbf{M}$ is a semi-separable matrix similar to causal attention, where $\mathbf{M}_{ij} = 0$ for $i\,<\,j$ and otherwise
$$
    \mathbf{M}_{ij} = \prod_{k=j+1}^{i} a_k \cdot \mathbf{C}_i^\top \mathbf{B}_j
$$

\subsection{Mamba and Mamba-2}
\begin{figure}[H]
    \centering
    \includegraphics[width=0.6\linewidth]{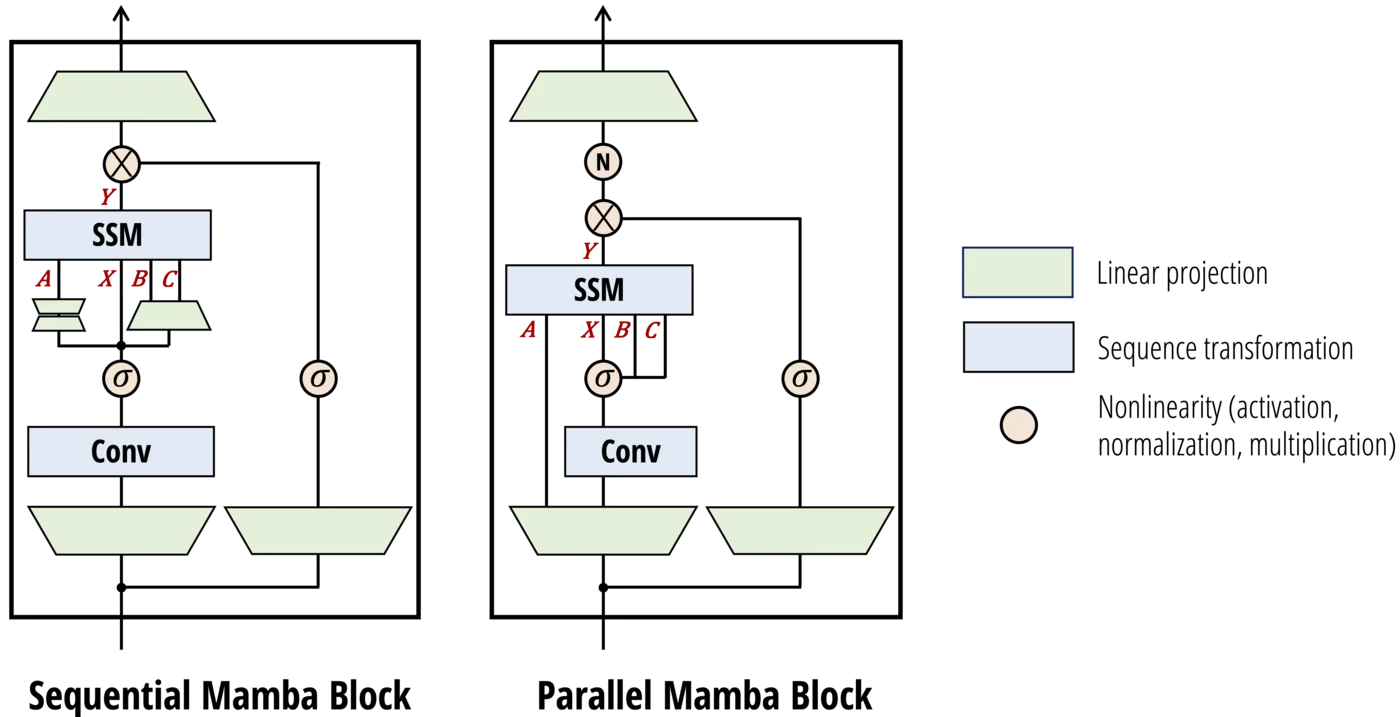}
    \caption{Mamba and Mamba-2 structures}
    \label{fig:mamba_structure}
\end{figure}
Mamba is built upon SSM and uses sequential Mamba block, where $\mathbf{A}$, $\mathbf{B}$, and $\mathbf{C}$ are calculated sequentially. Mamba-2 is built upon SSD and uses parallel Mamba block, where $\mathbf{A}$, $\mathbf{B}$, and $\mathbf{C}$ are calculated in parallel. Inspired by attention-like architectures, such parallelization accommodates tensor parallelism.



\section{SHapley Additive exPlanations Interaction Values}
SHapley Additive exPlanations (SHAP) \citep{Lundberg2017-oo} are used to evaluate the contribution of each feature to model prediction using game theory concepts. SHAP interaction values measure how much the syllable prediction result changes when two neurons' activities are considered together, on top of their individual contributions. A positive pairwise interaction value indicates two neurons work together, and their joint activation provides more information about the syllable than simply adding up their individual contributions. A negative interaction suggests that such joint effect is less than the sum of individual contributions of two neurons, meaning some redundancy in the encoding of information.

\section{Support Vector Machine}
Support Vector Machine (SVM) is a supervised machine learning algorithm \citep{Boser1992-fa} used extensively in this project to establish a benchmark for classification in order to be compared to by all later more complicated methods.

SVM creates an optimal hyperplane by maximizing the margin between classes. The optimal hyperplane is selected by maximizing the margin between classes. A hyperplane can be defined as a homogeneous non-trivial linear equation. In a space of $R^n$, a hyperplane has dimensionality of $R^{n-1}$. The closest data points to the hyperplane are called the support vectors, and these vectors determine the decision boundary for classification.

Among multiple metrics generated from classification tasks (accuracy, F1, precision, sensitivity, etc.) only overall accuracies are used in this project for simplicity. 

$$
\min_{w, b, \xi} \frac{1}{2}\Vert w\Vert^2 + C\sum_{i=1}^n \xi_i
$$
s.t.:
$$
y_i\left(w^\text{T} x_i + b\right) \geq 1 - \xi_i, \quad \xi_i \geq 0, \quad i=1,2,\ldots,n
$$
\begin{align*}
w &: \text{the weight vector} \\
b &: \text{the bias} \\
\xi_i &: \text{the slack variables that allow for misclassification} \\
&\quad \text{and encode Hinge Loss}\\
C &: \text{the regularization parameter}
\end{align*}

SVMs are conducted with a linear kernel and a Radio Basis Function (RBF) kernel.
The linear kernel is defined as 
$$
K(x, x') = \langle x, x'\rangle
$$
The RBF kernel in scikit-learn \citep{scikit-learn} is defined as
$$
K(x, x') = \exp(-\gamma \Vert x - x'\Vert_2^2)
$$
\begin{align*}
\gamma &: \text{the influence range of each training example} \\
\Vert x - x'\Vert_2^2 &: \text{the squared 2-Norm (Euclidean Norm)} 
\end{align*}

For this project, we evaluated both linear and RBF kernels using grid search with scikit-learn. The parameters were: (1) $C$ parameter: [$0.1$, $1$, $10$]
(2) gamma parameter (for RBF kernel): $[0.01, 0.1, \text{`scale'}]$

\section{TF-IDF}
Term Frequency-Inverse Document Frequency (TF-IDF) is a common feature in the domain of NLP. This feature combines two features: Term Frequency (TF) and Inverse Document Frequency (IDF). TF refers to how often a term appears in a document. IDF refers to how unique this term is across all document. When a term frequently appears in a document but rarely in other documents, the weight of this term would increase. The formula is written as:
$$w_{t,d} = tf_{t,d} \times \log(\frac{N}{df_t})$$
\begin{align*}
w_{t,d} &: \text{the weight of term $t$ in document $d$}\\
tf_{t,d} &: \text{the occurrence frequency of term $t$ in document $d$}\\
N &: \text{the total number of documents}\\
df_t &: \text{the number of documents containing term $t$}
\end{align*}
In this project, the same concept has been transferred to neural data. The term refers to the neuron, and the document refers to the data sample with a specific labels. Just like in NLP where the documents can be classified into Law, Medicine, News, etc., the data samples used in this project can be classified into different syllables.

\section{Transformers}
Transformers are introduced by \cite{Vaswani2017-ga} and have since established new performance benchmarks in Artificial Intelligence. 

\begin{figure}[H]
    \centering
    \includegraphics[width=0.4\textwidth]{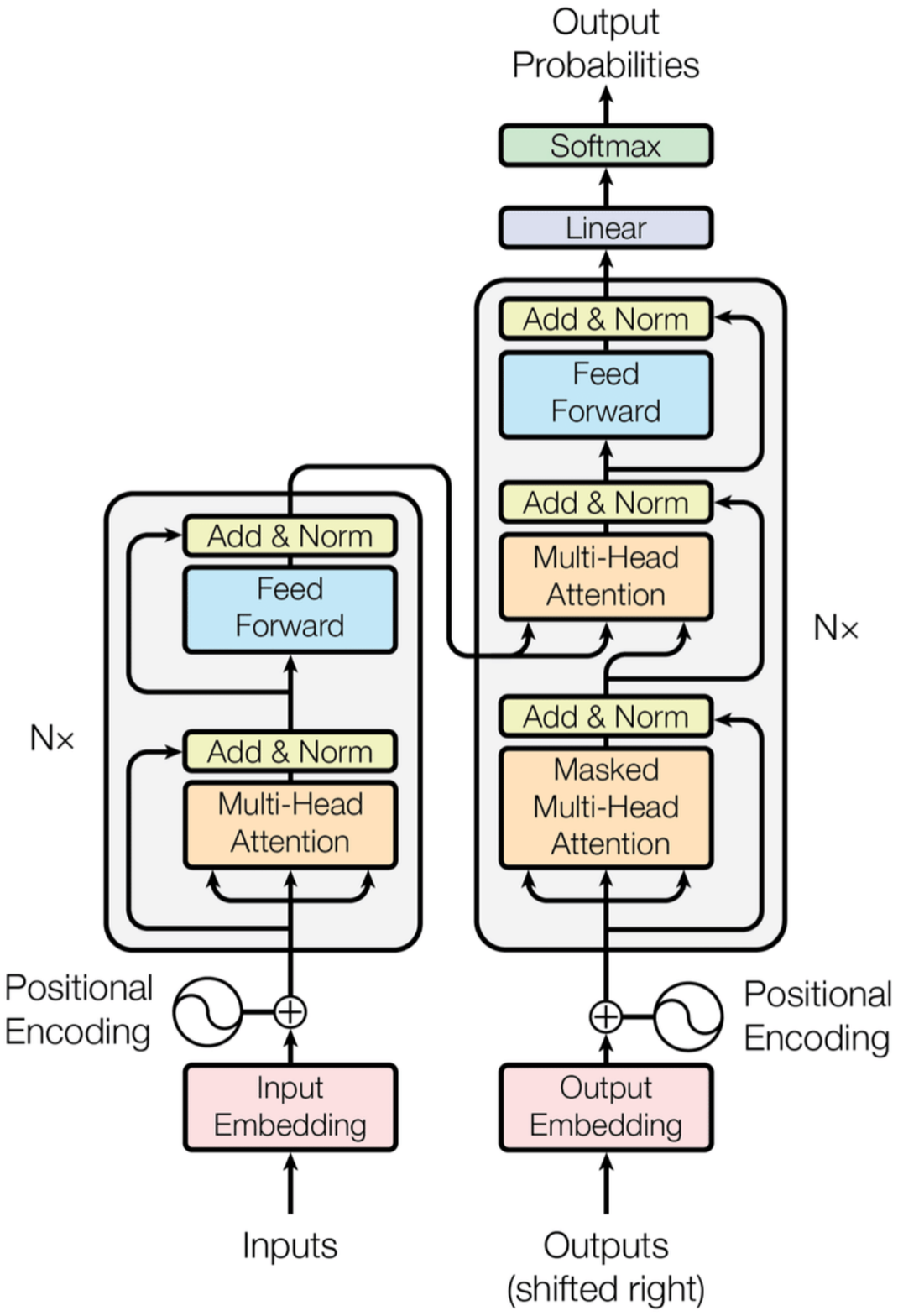}
    \caption{Transformers architecture}
    \label{fig:transformers}
\end{figure}

The Positional Encoding (PE) are added to the input embeddings. Sinusoidal positional encoding is a basic type of PE introduced in the original paper.
\begin{align*}
PE_{(pos,2i)} &= \sin\left(\frac{pos}{10000^{2i/d_{model}}}\right) & pos \in [0, n], i \in [0, d_{model}/2] \\
PE_{(pos,2i+1)} &= \cos\left(\frac{pos}{10000^{2i/d_{model}}}\right) & pos \in [0, n], i \in [0, d_{model}/2-1]
\end{align*}
\begin{align*}
\text{pos} &: \text{the position in the input sequence} \\
d_{\text{model}} &: \text{the dimensionality of the embedding vectors} 
\end{align*}
The word embeddings has a size of $d_{\text{model}}$. The Positional Encoding (PE) matches $d_{\text{model}}$.  In the original transformers \citep{Vaswani2017-ga}, $d_{model} = 512$. In OpenAI Generative Pre-trained Transformer (GPT) 2 \citep{Radford2019-gx} that we used in experiment 2, $d_{model} = 768$.

\begin{align*}
\text{score}(x_i, x_j) &= \frac{q_i \cdot k_j}{\sqrt{d_k}}\\
[\alpha_{ij}] &= \text{softmax}(\text{score}(x_i, x_j)) \quad \forall j \leq i\\
a_i &= \sum_{j\leq i} \alpha_{ij}v_j
\end{align*}
The calculation for masked attention used in GPT is shown above. The attention weights are calculated by doing pair-wise dot product between key and query, and the scores are divided by $\sqrt{d_k}$ to reduce the variance from dot products. Then softmax is applied to the scores, and creates attention weights. The attention weights are multiplied by values and produces the attention output. In Multi-Head Attention (MHA), the outputs of individual heads are stacked.

\begin{align*}
x &= \text{MHA}(x) + \alpha x, \quad \alpha \in [0,1]\\
\text{LayerNorm}(x) &= \gamma \frac{x - \mu}{\sigma} + \beta
\end{align*}
\begin{align*}
\alpha&: \text{scaling factor for the residual connection}\\
\mu&: \text{mean of the input}\\
\sigma&: \text{standard deviation of the input}\\
\gamma \text{ and } \beta&: \text{learnable parameters}
\end{align*}
The add operation is called the residual connection. It ensures that the transformed output is close to the input shape. The normalization is applied after every layers to control the value ranges. The layer norm is the most popular normalization. In layer normalization, the activations are normalized (demeaned and then divided by the standard deviation) over the feature dimensions ($d_\text{model}$) for each example in the batch.

\begin{align*}
x' &= \text{FFN}(\text{LayerNorm}(x)) + \alpha \text{LayerNorm}(x),  \quad \alpha \in [0,1]\\
\text{LayerNorm}(x') &= \gamma \frac{x' - \mu}{\sigma} + \beta
\end{align*}
Then, a feed-forward network (FFN) is applied. The result is passed to another residual connection and norm layer.

\section{Variational Auto-Encoders}
The Variational Auto-Encoder (VAE) is a generative model introduced by \cite{Kingma2013-es}. VAE transforms the input data into a continuous latent space. It includes an encoder and a decoder. The encoder $q_\varphi(z|x)$ learns to approximate the posterior distribution of latent variables $z$ using input data $x$. The decoder $q_\theta(x|z)$ learns to reconstruct the input data $x$ using samples from this latent distribution. Typically, a standard normal distribution is used as the prior distribution $p(z)$.
\begin{figure}[H]
    \centering
    \includegraphics[width=0.75\textwidth]{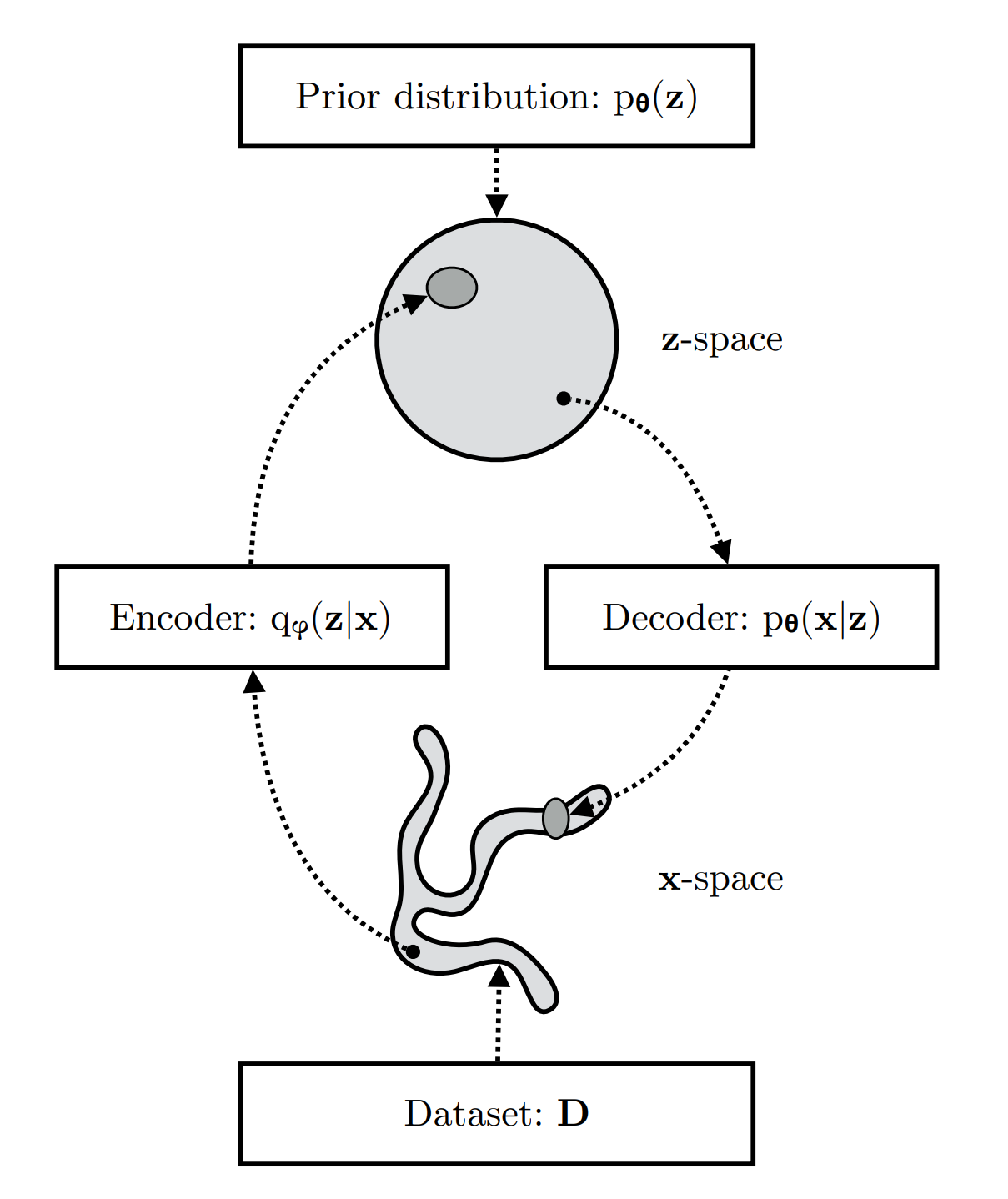}
    \caption{VAE architecture}
    \label{fig:vae}
\end{figure}

\begin{align*}
\mathcal{L}_{\text{VAE}} &= \mathcal{L}_{\text{MSE}}(x, \hat{x}) + \text{KL}(q_\varphi(z|x) \| p(z)) \\[1em] 
q_\varphi(z|x) &= \mathcal{N}(z; \mu_\varphi(x), \sigma^2_\varphi(x)) \\
p(z) &= \mathcal{N}(0, I) \\[0.5em]
\varphi, \theta &: \text{learnable parameters}
\end{align*}

The loss function of VAE combines two terms: (1) the reconstruction loss, and (2) Kullback–Leibler divergence (KL divergence) loss. The first term is the loss function. It uses Mean Squared Error (MSE) to ensure that the output data matches the input data. The second term is a regularization term (KL divergence) that pushes the latent distribution toward a standard normal prior. The encoder produces two learnable parameters for the latent space: mean $\mu_\varphi(x)$ and variance $\sigma_\varphi^2(x)$.

\begin{figure}[H]
    \centering
    \includegraphics[width=\textwidth]{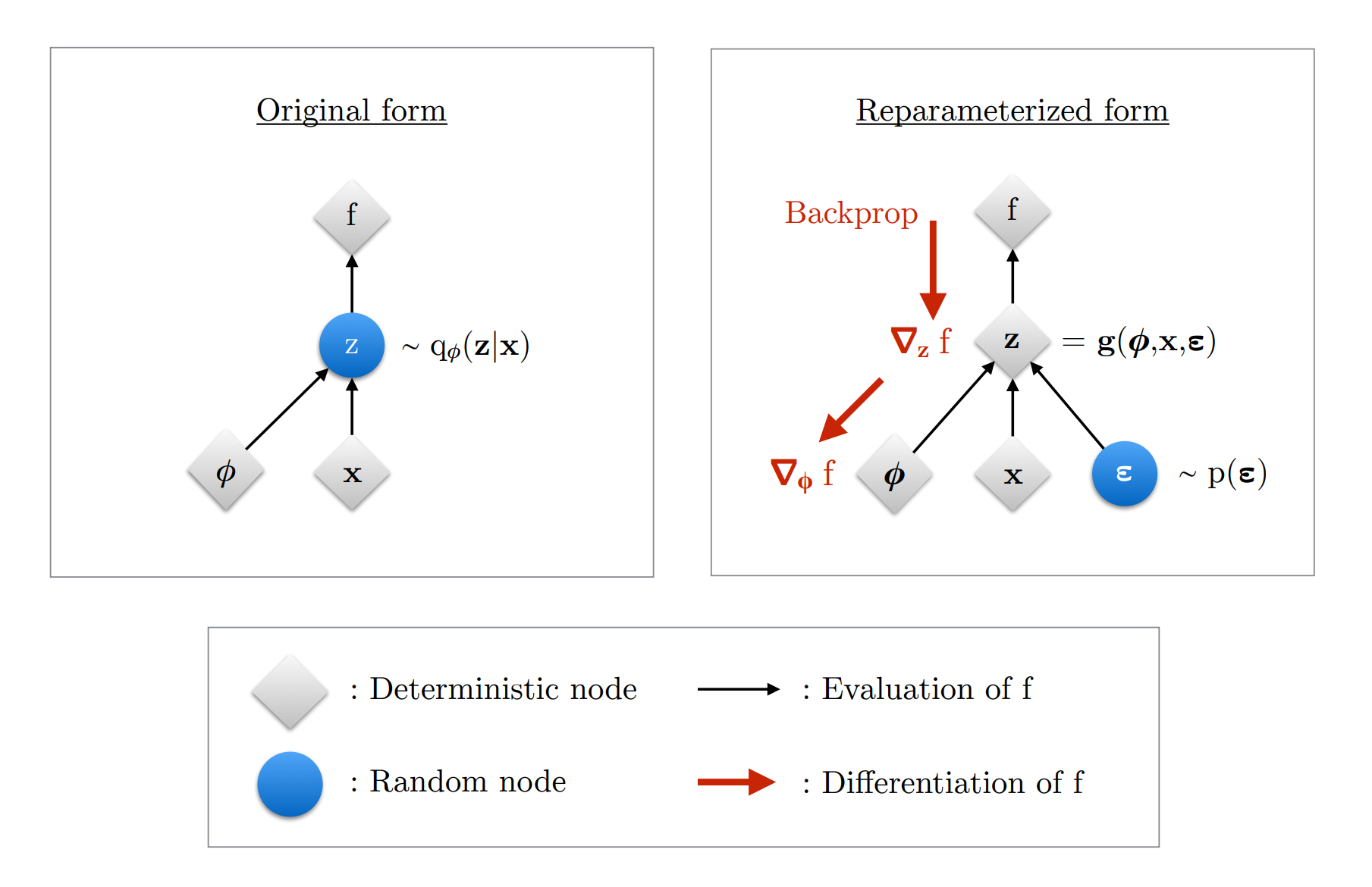}
    \caption{VAE reparameterization}
    \label{fig:vae-bp}
\end{figure}

Back propagation is impossible through the sampling process. To enable back propagation, VAE employs a reparameterization trick. Instead of directly sampling from $q_\phi(z|x)$, VAE samples $\epsilon \sim \mathcal{N}(0, I)$ and then computes $z = \mu_\phi(x) + \sigma_\phi(x) \odot \epsilon$, where $\odot$ refers to element-wise multiplication. By applying this reparameterization trick, VAE successfully converts the previously random sampling into deterministic and differentiable nodes of encoder parameter $\phi$, input data $x$, as well as random noise $\epsilon$.  As a result, the latent embedding z can be expressed as
\begin{align*}
    z = g(\phi, x, \epsilon)
\end{align*}


\subfilebibliography

\chapter{Experiment 1: Machine Learning Benchmarks}%
\refstepcounter{experiment}  
\label{exp:\theexperiment}
Based on annotated recording files, we extracted the onset and duration of each annotated syllable. In this experiment, we engineered different feature representations for each neuron in each annotated syllable, and used these features for syllable classification. This experiment aimed to establish the baseline performance on the classification of different syllables, identifying hub neurons, and investigating how neural information is integrated by the model through time.

\section{Methods}
\subsection{Feature Extraction}
\subsubsection{Single Neuron Level}
We first evaluated the predictive power of each neuron. For each neuron, we calculated firing rates during the entire syllable production. For each neuron, we trained classifiers using its firing rate.

\subsubsection{Neuron Population Level}
We analyzed the predictive ability of neurons on a population level to investigate the correlation and interaction between neurons in classification tasks. We used three types of feature representations. For each representation, we created a feature vector of length $75$, which is equal to the number of neurons. Each element in the feature vector corresponds to one neuron's activity. Each feature representation was independently evaluated using basic classifiers.

The first representation is the total number of spikes of the neuron per syllable. The second representation is the mean firing rate of the neuron per syllable. The third representation is Term Frequency-Inverse Document Frequency (TF-IDF). The TF-IDF indices were calculated by treating each neuron as an individual word and each annotated syllable as a document. Thus, the total number of firing during an annotated syllable sample served as the number of occurrences of the word in the document.

\subsection{Models and Evaluation}
For each of the feature representations above, we trained Support Vector Machine (SVM), Random Forest (RF) and eXtreme Gradient Boosting (XGBoost) classifiers. We evaluated the accuracy of each model for each feature representation. Then, we used three methods to evaluate the feature importance and interaction between neurons. The first is the built-in feature importance visualization method in RF, which identifies key contributing neurons. The second uses SHapley Additive exPlanations (SHAP) \citep{Lundberg2017-oo} values to identify the most important neurons for each syllable. The third is using SHAP interactive values to visualize pairwise interactions between neurons and find potential hub neurons. We first plotted a heatmap for all neurons and then picked top 10 interactive pairs and plotted them on a segment of Neuropixel. We also computed the spike rate correlation for comparison with SHAP interaction patterns. For the second and third methods, SHAP values were calculated using TreeSHAP for XGBoost models only. TreeSHAP is computationally efficient but does not support SVM or RF.

\subsection{Neural Information Integration}
Building upon our previous classification experiments, we conducted two analyses using SVM and population level mean firing rates to further investigate temporal relationships in our data.  

The first analysis quantifies how different temporal segments contribute to classification performance. We performed a parameter sweep to analyze the accuracy of SVM trained on mean firing rates across different time periods. In this parameter sweep:
\begin{itemize}
    \item Window start points vary from $-200.0$\,ms to $-12.5$\,ms
    \item Window end point increments by $2.5$\,ms, until it reaches $100.0$\,ms
\end{itemize}
For instance, the initial window would be $[-200.0$\,ms, $-197.5$\,ms$]$, and the decoding accuracy using this data is stored. The second window would be $[-200.0$\,ms, $-195.0$\,ms$]$, the accuracy is stored again. The window increases until it reaches $[-200.0$\,ms, $100.0$\,ms$]$, where the accuracy is stored, and the second round starts with $[-175.0$\,ms, $-172.5$\,ms$]$, and increases until $[-175.0$\,ms, $100.0$\,ms$]$, etc.

The second analysis aimed to reveal spatially and temporally important patterns of information encoding in the neural population. To achieve this, another window sweep was done to extract fixed $20$\,ms windows between $-150$\,ms and $50$\,ms.

\section{Results}
\subsection{Classification Accuracy}
Table~\ref{tab:classification_results} shows the results of classification accuracy across different feature and models.
\begin{table}[H]
   \centering
   \begin{tabular}{lccc}
       \toprule
       Feature Type & SVM (\%) & RF (\%) & XGBoost (\%)\\
       \midrule
       (Single Neuron) Mean Firing Rate & 39.0 & 66.8 & 64.2 \\
       (Neuron Population) Total Spike Counts & 91.5 & 89.6 & 96.2 \\
       (Neuron Population) Mean Firing Rates & 77.3 & 91.7 & 91.2 \\
       (Neuron Population) TF-IDF & 80.0 & 81.7 & 77.7 \\
       \bottomrule
   \end{tabular}
   \caption{Classification accuracy}
   \label{tab:classification_results}
\end{table}

\newpage
\subsection{Single Neuron Mean Firing Rate}
\begin{figure}[H]
  \centering
  \includegraphics[width=\textwidth]{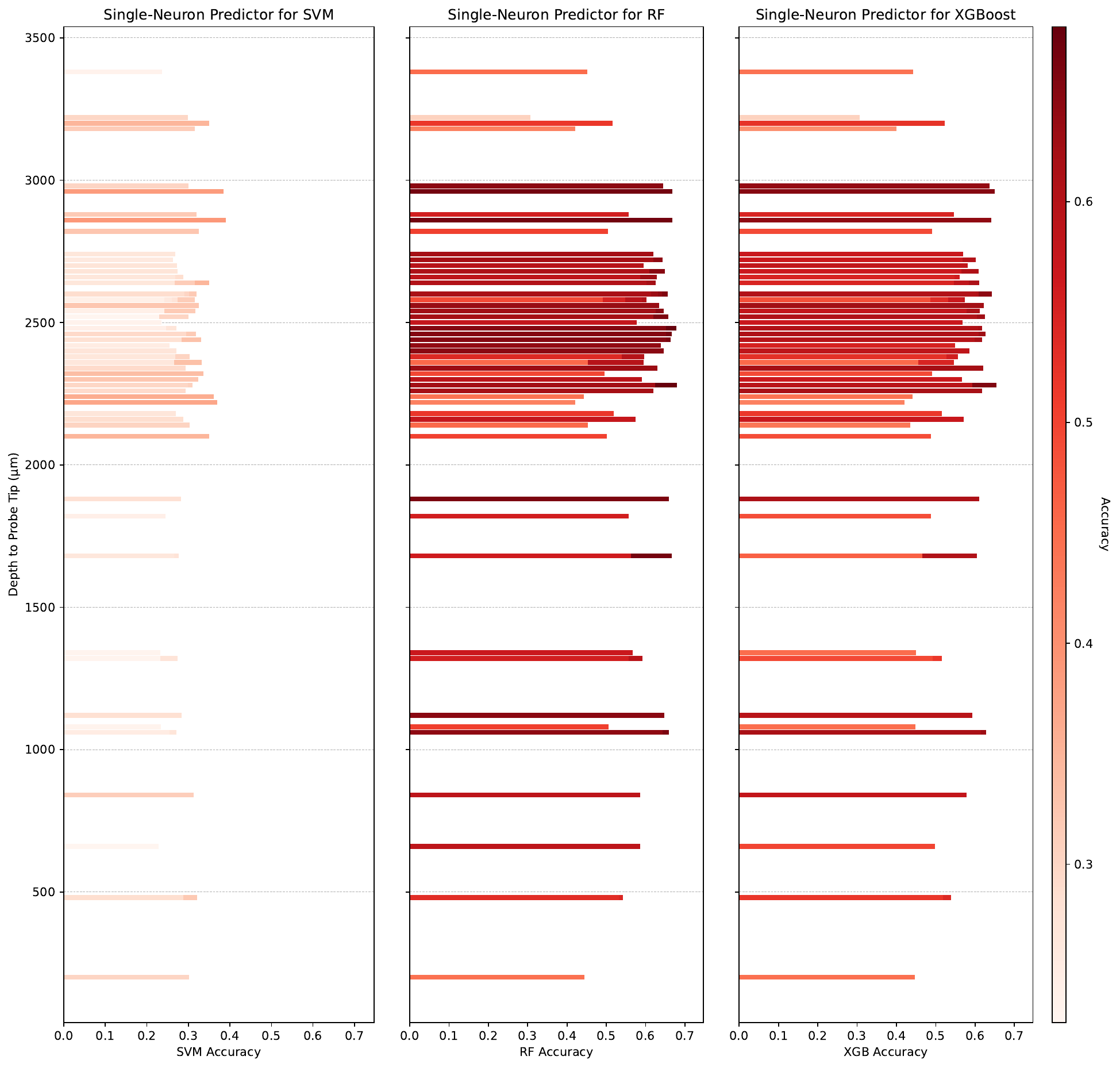}
  \caption{Classification accuracy using single neuron firing rates}
  \label{fig:single_neuron_results}
\end{figure}

From Figure~\ref{fig:single_neuron_results}, we can observe that neurons located at depths of $2000$--$3000\,\mu m$ have higher accuracies than the rest, but the overall variation across neurons of different depth remains small. The result from SVM suggests that the information is not encoded at single neuron level, and the decoding requires population level activity patterns. In the figure, each horizontal bar represents the classification accuracy using only that neuron's firing rate as the predictor. The accuracy is also encoded by the color. Color and length were used together to separate the neurons recorded at the same depth. The x-axis shows the distance of each neuron to the tip of the Neuropixels probe in $\mu m$.

\newpage
\subsection{Neuron Population Total Spike Counts}
\begin{figure}[H]
  \centering
  \begin{subfigure}{0.45\textwidth}
      \includegraphics[width=\textwidth,height=1.5\textwidth]{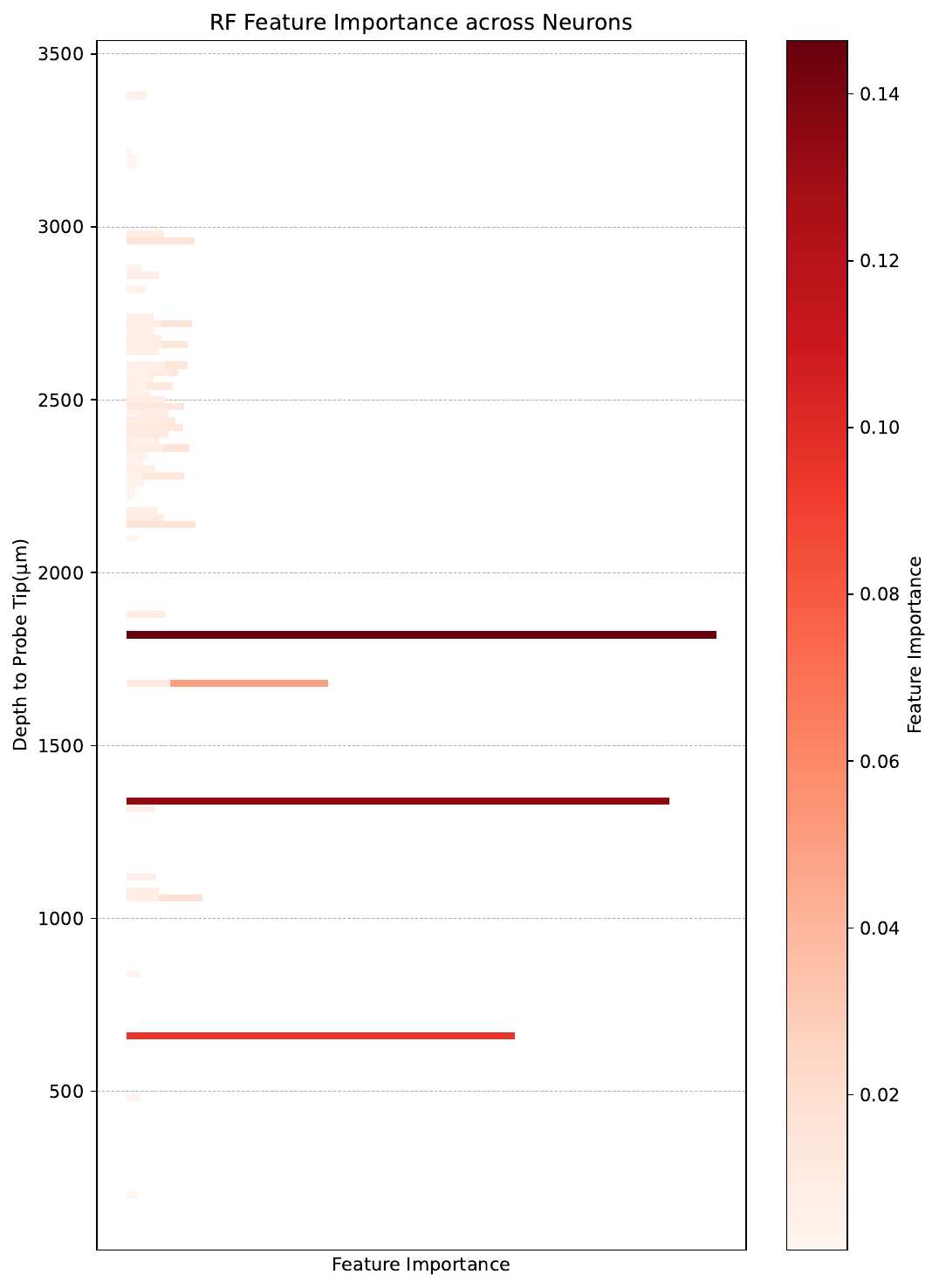}
      \caption{RF Feature Importance}
      \label{fig:rf_total}
  \end{subfigure}
  \begin{subfigure}{0.45\textwidth}
      \includegraphics[width=\textwidth,height=1.5\textwidth]{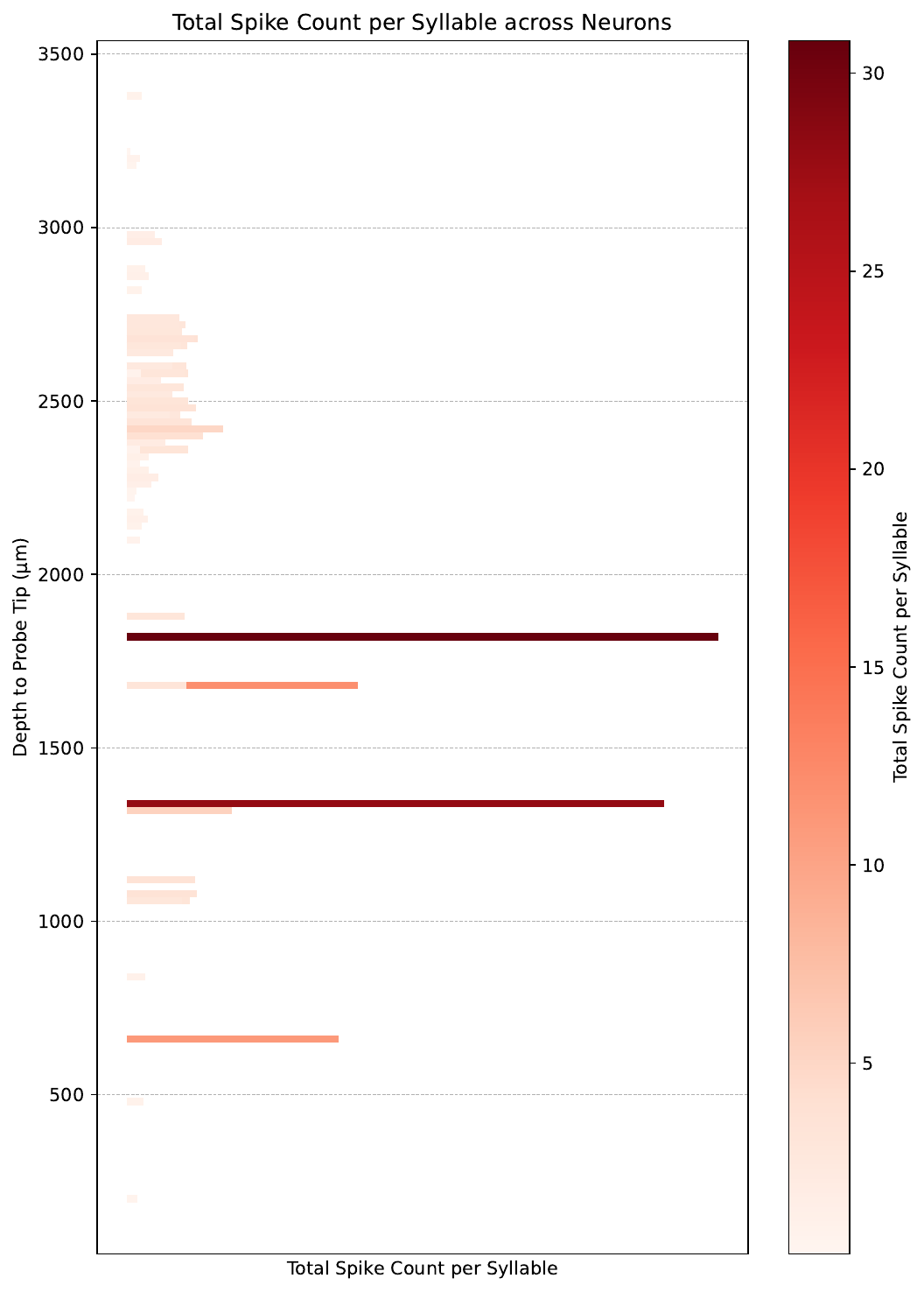}
      \caption{Total Spike Counts}
      \label{fig:spike_count}
  \end{subfigure}

  \caption{Evaluation of total spike counts}
  \label{fig:total_visualization}
\end{figure}

All classifiers achieved high classification accuracy using total spike counts for each syllable. However, this performance should be interpreted with caution. Some neurons exhibit consistent firing patterns, and may encode temporal information more than syllable-specific features. Since each syllable differs in duration, the result of classification using total spike counts may also reflect temporal characteristics across syllables of different lengths instead of syllable-specific neural activities. Figure~\ref{fig:total_visualization} shows the feature importance, and how the most important neurons in Figure~\ref{fig:rf_total} are aligned with the total spike counts shown in \ref{fig:spike_count}. This suggests that the RF simply learns the syllable length instead of focusing on genuine neural patterns.

\subsection{Neuron Population Mean Firing Rates}
\begin{figure}[H]
   \centering
   \begin{minipage}{0.45\textwidth}
      \centering
      \includegraphics[width=\textwidth]{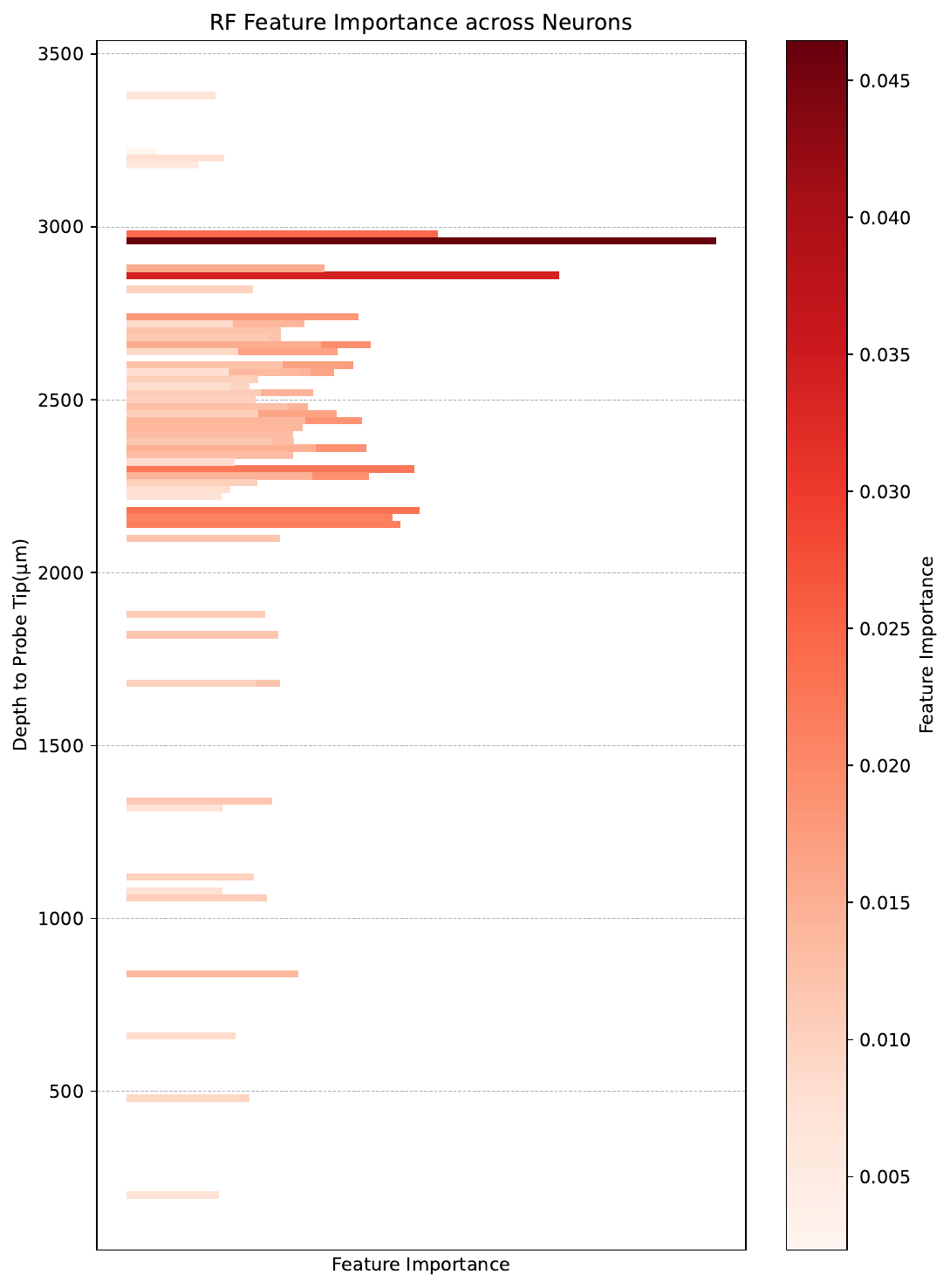}
      \caption{RF feature importance\\\phantom{Figure 1.3: }Mean firing rates}
      \label{fig:rf_feature_importance}
   \end{minipage}
   \hfill
   \begin{minipage}{0.45\textwidth}
      \centering
      \includegraphics[width=\textwidth]{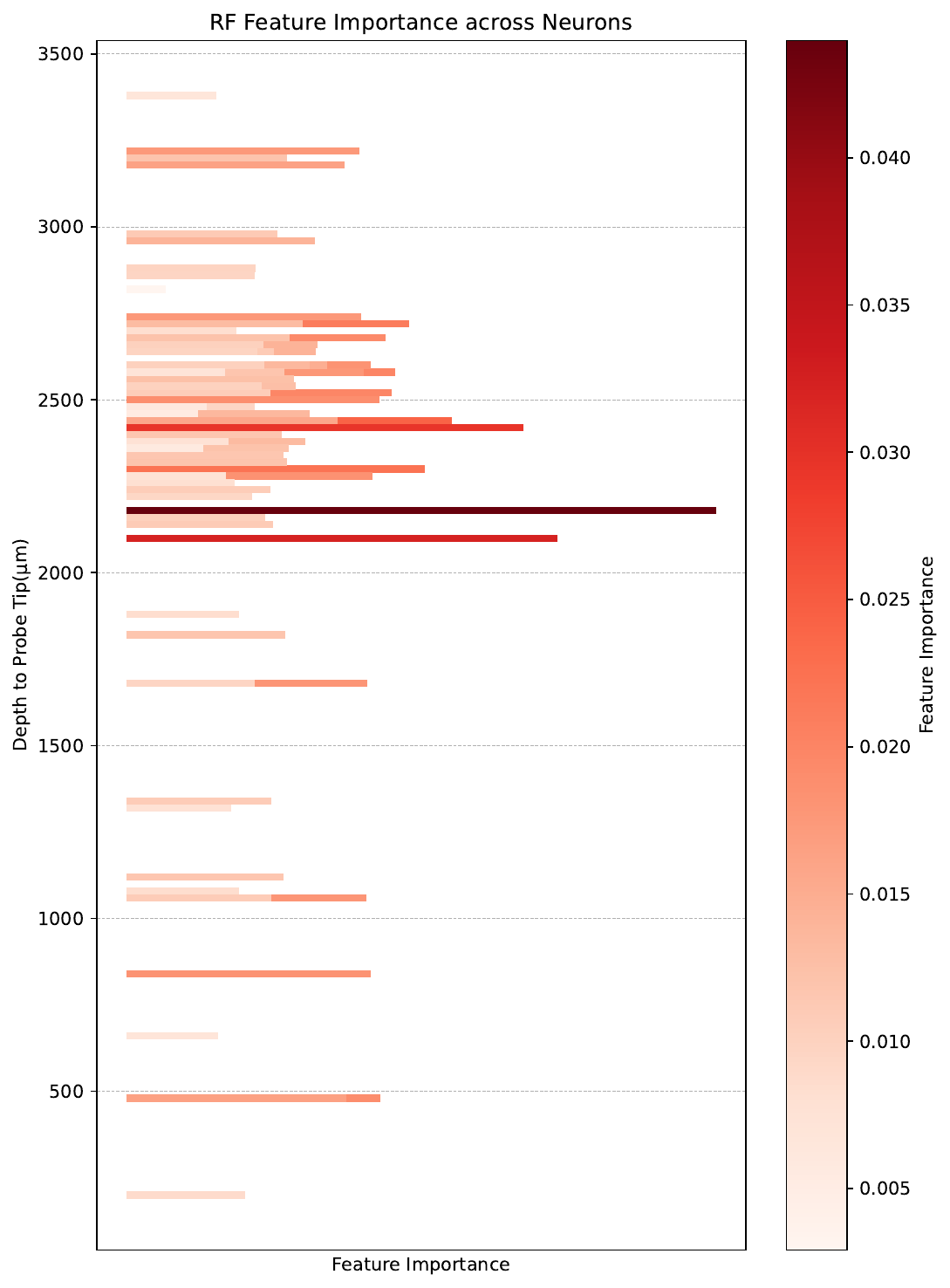}
      \caption{RF feature importance\\\phantom{Figure 1.3: }TF-IDF}
      \label{fig:tfidf_feature_importance}
   \end{minipage}
\end{figure}

SVM achieved 77.3\,\% accuracy while RF reached 91.7\,\% accuracy using mean firing rates. Figure~\ref{fig:rf_feature_importance} shows the importance of each neuron during the classification task in an RF model. Darker red indicates higher contribution to the correct classification.

\subsection{TF-IDF}
SVM achieved 80.0\,\% accuracy and RF achieved 81.7\,\% using TF-IDF as feature representations. Figure~\ref{fig:tfidf_feature_importance} visualizes the feature importance from the RF model. 

\subsection{SHAP Analysis}
\begin{figure}[htbp]
   \centering
   \includegraphics[width=\textwidth]{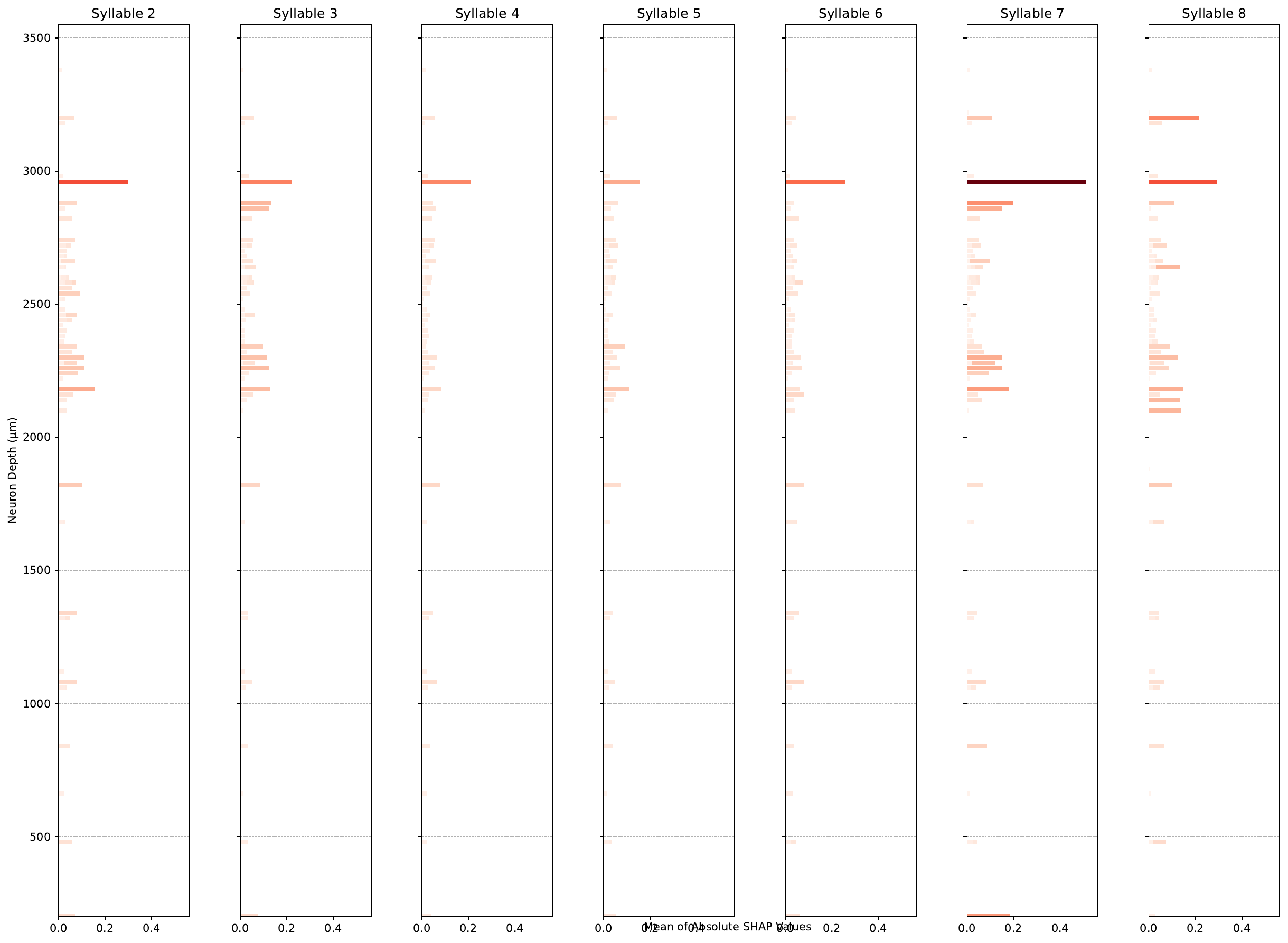}
   \caption{SHAP visualization of neuron importance for classification}
   \label{fig:shap}     
\end{figure}

XGBoost classifier achieved a 91.2\,\% accuracy. The result of XGBoost is visualized using SHAP visualizations to investigate the interaction between neurons. Figure~\ref{fig:shap} visualizes feature importance for each syllable using SHAP values. Higher SHAP values (darker red) indicate stronger positive contribution to the classification of specific syllables. The visualization shows distinct patterns of neuronal contributions to the correct identification of each syllable. Noteworthy the neuron N$6$ at depth $2960\,\mu m$ plays the most critical role among the classification of all syllables. 

\subsection{SHAP Interaction Analysis}
\begin{figure}[H]
   \centering
   \begin{subfigure}{\textwidth}
      \centering
      \includegraphics[height=0.45\textheight]{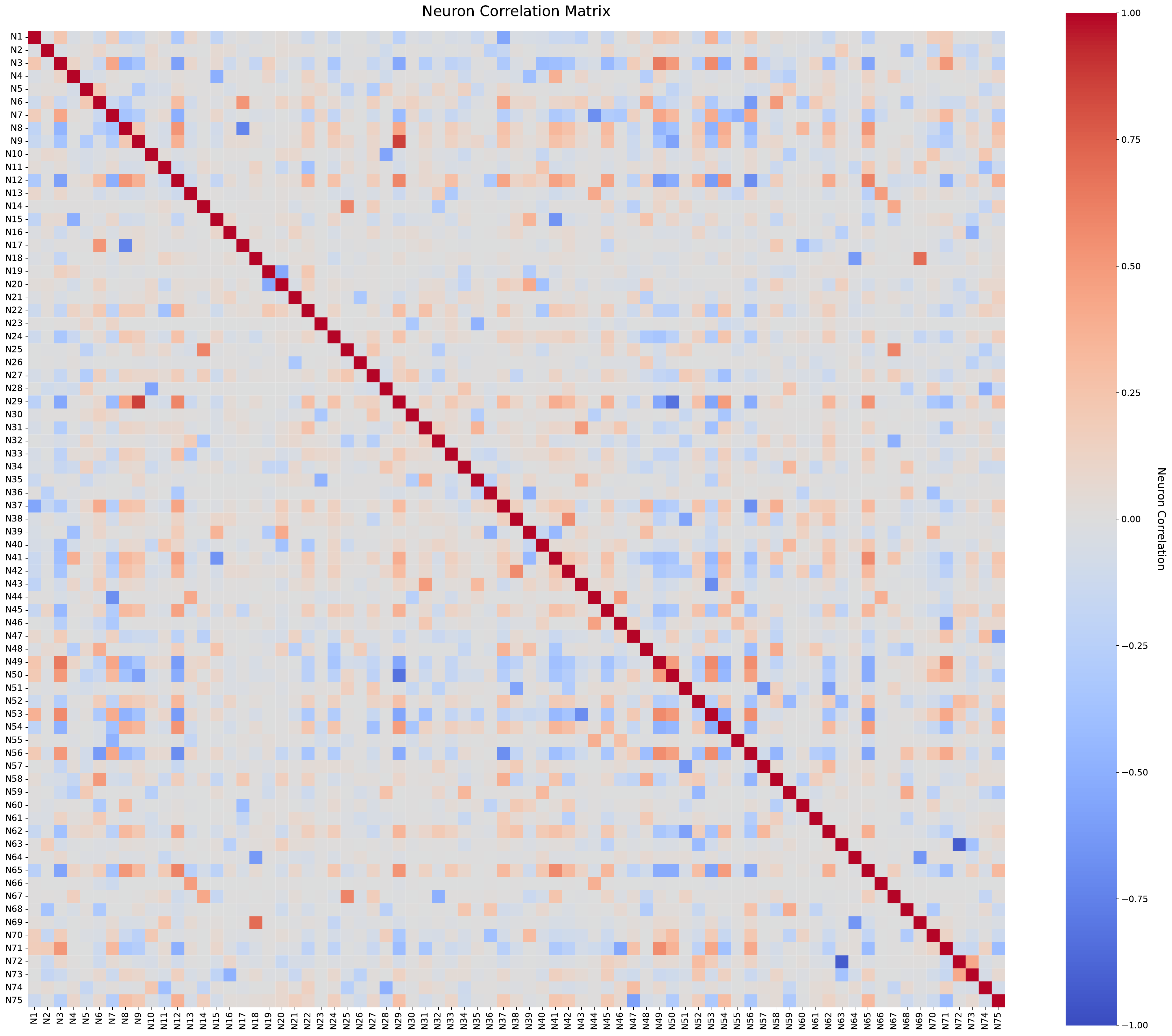}
      \caption{Correlation coefficients between 75 neurons}
      \label{fig:correlation}
   \end{subfigure}
   
   \begin{subfigure}{\textwidth}
      \centering
      \includegraphics[height=0.45\textheight]{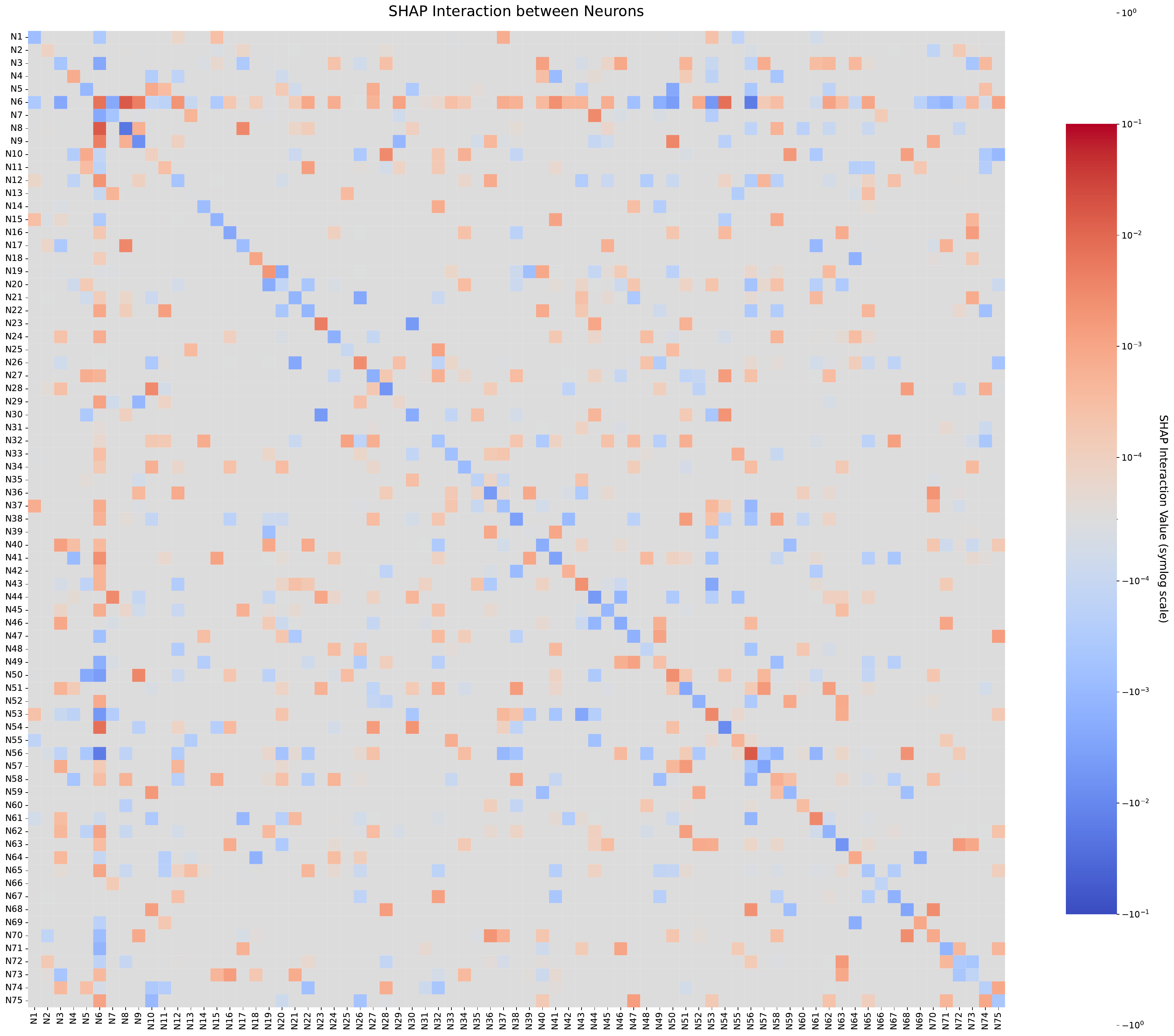}
      \caption{SHAP visualization of interactions between neurons}
      \label{fig:shap_interaction}
   \end{subfigure}
   \caption{Neural correlation and interaction analysis}
   \label{fig:all_neurons}
\end{figure}

Figure~\ref{fig:correlation} shows the correlation coefficients between neuron firing rates. Neurons N$1$($3380$)--N$12$ ($2720\,\mu m$) and N$49$($2320$)--N$56$($2180\,\mu m$) show opposite firing patterns compared to other neurons during syllable production. 

Figure~\ref{fig:shap_interaction} shows a heatmap of the relationship between the neurons calculated from SHAP interaction values. The neurons were inversely ranked by the depth to probe tip. A positive SHAP interaction value in red suggests that when both neurons are active, they contribute more to the syllable classification than the sum of individual contributions. A negative SHAP interaction value in blue suggests that when both neurons are active, their total contribution is less than the sum of their individual contributions. This could result from redundant information encoding, or potential inhibition effect from inhibitory interneurons. We can observe from the figure that Neuron $6$ serves as a hub neuron. When it is active, it can either enhance (red dots) or suppress (blue dots) the contributions of other neurons to syllable classification, depending on the specific neural states. The depth of all neurons can be found in the appendix \ref{tab:neuron_depths}.

\begin{figure}[H]
   \centering
   \includegraphics[width=\textwidth]{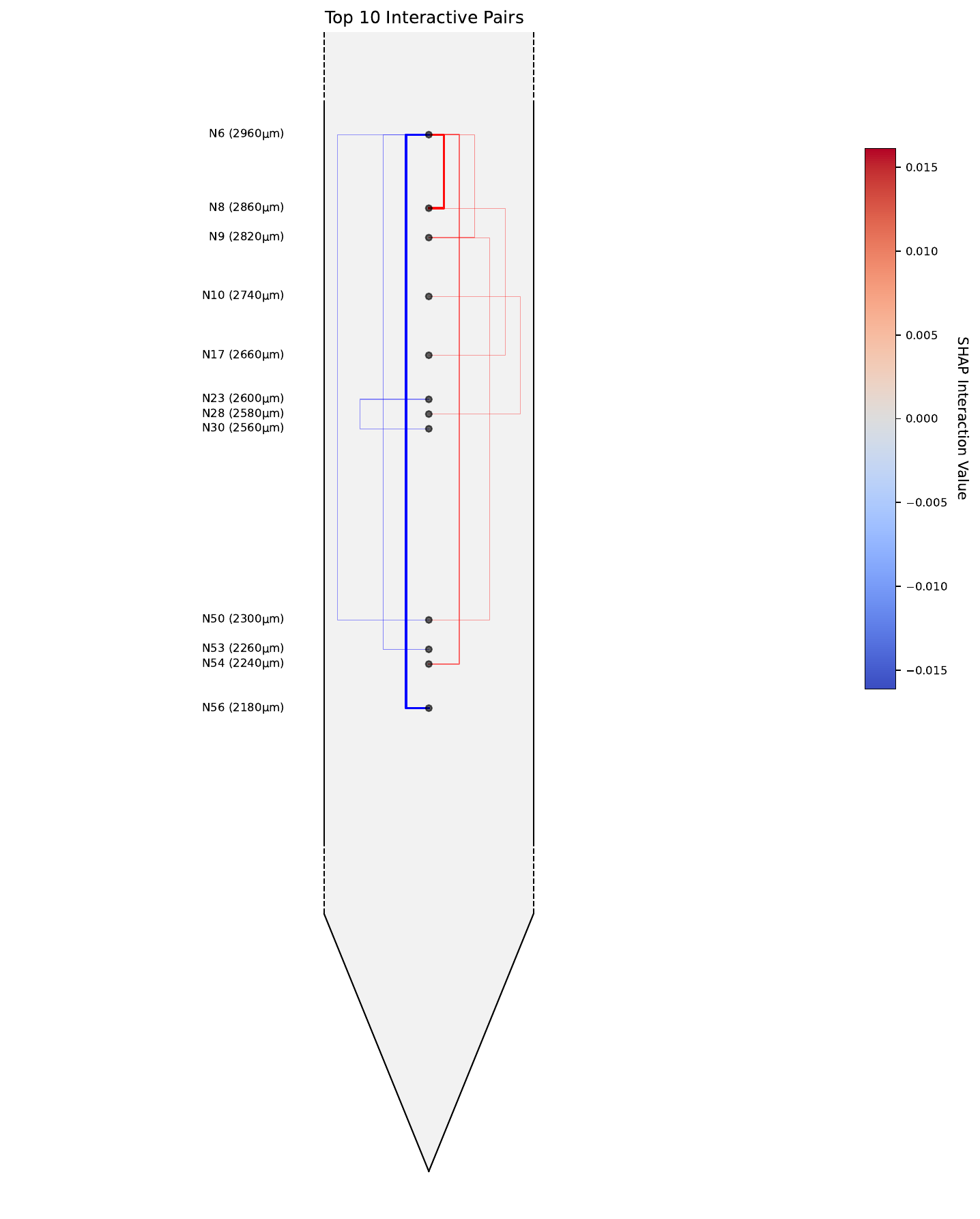}
   \caption{Top 10 interactive pairs plotted on a Neuropixels probe}
   \label{fig:neuropixel}     
\end{figure}

The two figures show certain shared spectral patterns on the distribution of less correlated neurons (in gray). Additionally, Figure~\ref{fig:neuropixel} visualizes the interaction of top 10 pairs of important neurons (called Features in the SHAP package) on a Neuropixels probe. The colored step lines indicate the interaction conditions between neurons. In the figure, N$6$ actively interacts with N$8$ and negatively interacts with N$56$. N$6$ and N$8$ are roughly located on the edge of LMAN, while N$56$ is roughly located at area X. 

\newpage
\subsection{Neural Information Integration} 
\begin{figure}[H]
 \centering
 \includegraphics[width=\textwidth]{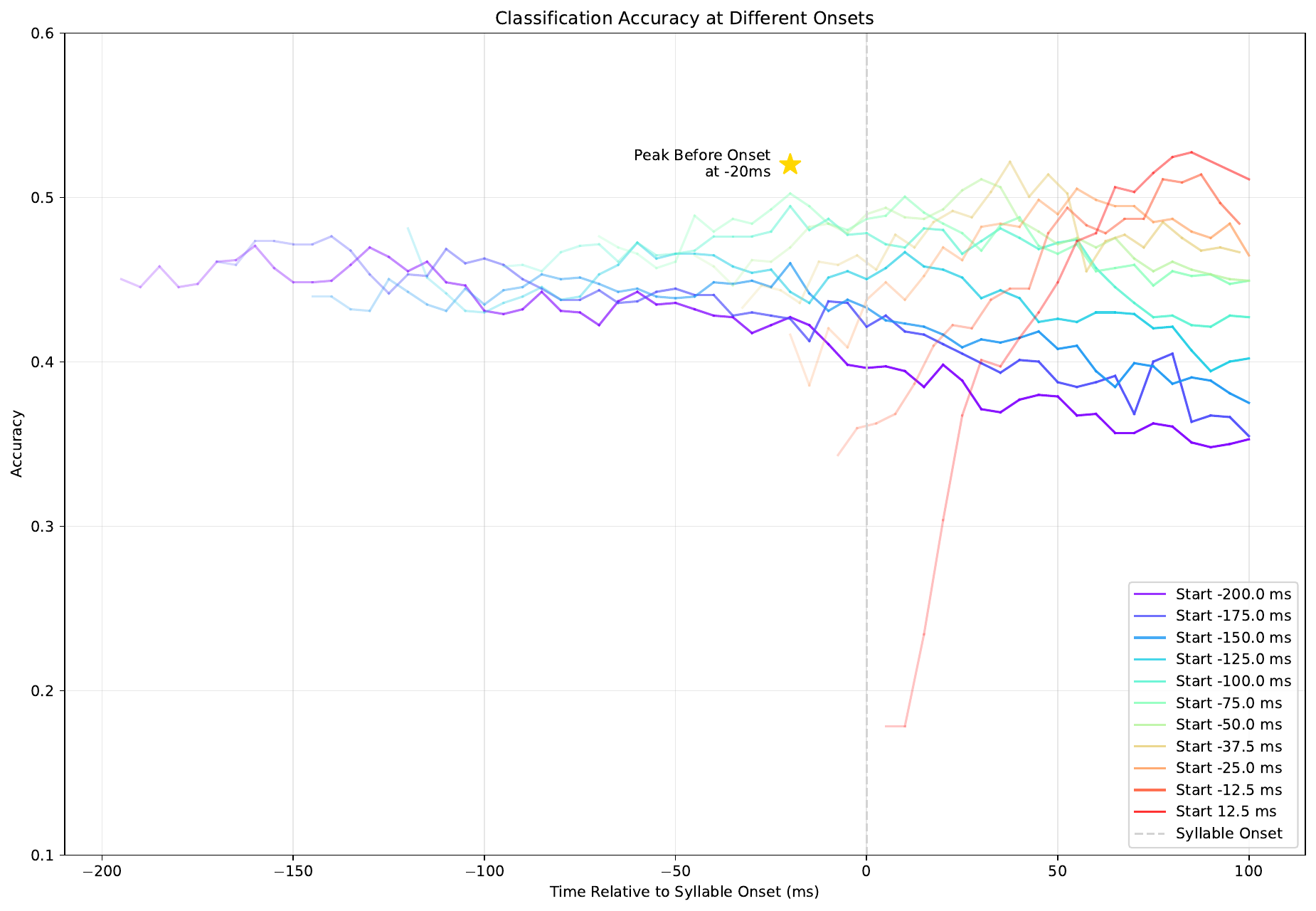}
 \caption{Classification accuracy}
 \label{fig:temporal_accuracy}
\end{figure}

Figure~\ref{fig:temporal_accuracy} shows the decoding accuracy using the mean firing rates of neuron population using a basic SVM.\footnote{Individual plots of each onset is included in Appendix~\ref{appendix:integration}} 

Each colored line indicates how classification accuracy evolves when integrating neural activity from a specific starting point. The transparency indicates the total length of windows used for classification. For instance, the purple line starts at $-200$\,ms. It collects the spikes every $5$\,ms, and uses these spikes to do classification. The first point in the purple line at $-195$\,ms indicates the classification accuracy using spikes collected during $[-200\,\text{ms}, -195\,\text{ms}]$. Then it collects the spikes during $[-200\,\text{ms}, -190\,\text{ms}]$, and the accuracy is reflected by the value of the purple line at $-190$\,ms.

On the other hand, highest classification accuracies occur at $-20$\,ms. Colored lines starting at $-12.5$\,ms show very low accuracy, which indicates that there are homogeneous activities in all syllables around the syllable onset. Based on this observation and inspired by Multivariate Pattern Analysis in EEG, we did the second analysis. 

\begin{figure}[H]
 \centering
 \includegraphics[width=0.6\textwidth]{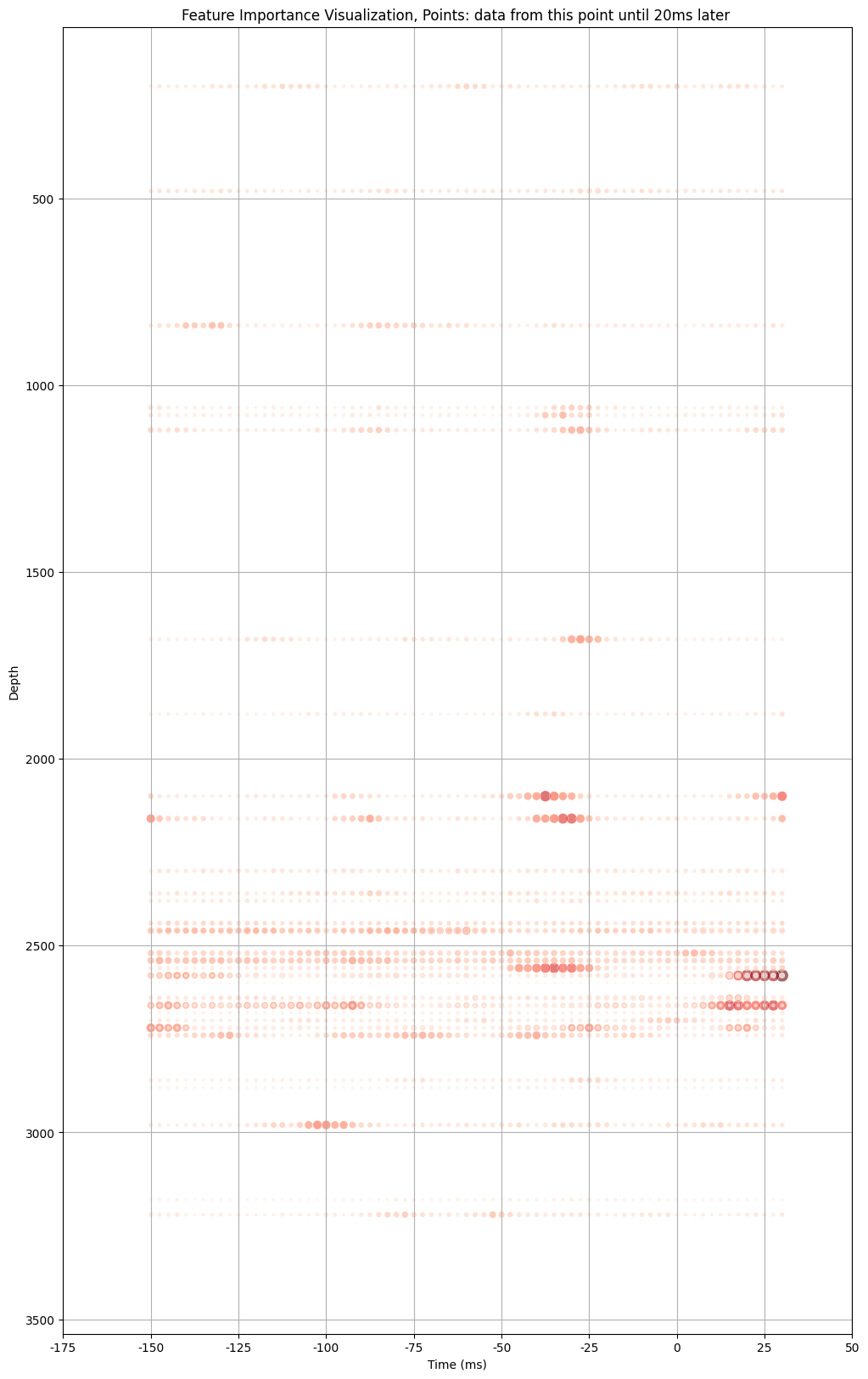}
 \caption{Classification performance in different random time windows}
 \label{fig:random_windows}
\end{figure}

In the second analysis, we binned the firing counts into $2.5$\,ms windows in the range $[-150\,\text{ms}, 30\,\text{ms}]$ and use these as input features to an RF model. Figure~\ref{fig:random_windows} visualizes the feature importance. There are occasionally dark red dots between $[-50\,\text{ms}, -25\,\text{ms}]$ before the onsets, indicating these time windows contain relatively more important information for correct identification of syllables.

\newpage
\section{Discussion}
\subsection{Baseline Methods and Interpretation}
Our benchmark results show that the mean firing rates can be used to classify different syllables of zebra finches. A fair baseline can be established by passing mean firing rates on a neuron population level to SVM, yielding an accuracy of 77.3\,\%. RF and XGBoost have better results, but may come from brute force ensemble averaging rather than understanding true data patterns. Individual neurons show some predictive power, and the population level features suggest that the syllable is mostly encoded at the population level. In addition, TF-IDF is a useful metric and yields a high classification accuracy. SHAP analysis revealed two findings: syllable-specific patterns in neural activity, and neuron interactions. The analysis identified important neuron hubs, and visualized possible relationships between neurons in LMAN and area X. 

\subsection{Information Encoded in LMAN Activity}
The first neural information integration analysis suggests that LMAN activity at $-20$\,ms is the most informative for syllable classification. The second analysis showed that during $[-50\,\text{ms}, -25\,\text{ms}]$, the LMAN activity is the most informative for the vocalization of different syllables. These temporal relationships align well with previous findings of LMAN's pre-motor activity during syllable production. Previous studies have reported LMAN pre-onset bursts occurring at $[-60\,\text{ms}, -10\,\text{ms}]$ \citep{Kojima2018-rw}, $[-42\,\text{ms}, -20\,\text{ms}]$ \citep{Giret2014-kx}, and $-50$\,ms \citep{Kao2005-cl}.

\subsection{Future Direction}
The results indicated LMAN activities contain important timing-specific information around syllable onset. A systematic temporal windows analysis could reveal more precise patterns of information encoding in LMAN. The SHAP interaction analysis and the information integration analysis can be applied to data from HVC and RA, enabling visualization of information flow and integration between AFP and Motor Pathway. The method could also be used on simultaneously recorded data from Area X and DLM to investigate the GABAergic projection from Area X to DLM. These analyses would advance our understanding on how Motor Pathway and AFP coordinate and contribute to motor output from a machine learning perspective.
\subfilebibliography

\chapter{Experiment 2: Token-based Deep Learning Methods}%
\refstepcounter{experiment}  
\label{exp:\theexperiment}
This experiment aimed to process neural data using NLP methods. The spike data were converted into text tokens in the NLP framework.

\section{Methods}
\subsection{Tokenization Method}
We passed in binary data in a text format. For each sampling time, the data were then combined across all 75 neurons from the shallowest to the deepest. For binary data, each value was either 1, which indicates a firing, or 0, which indicates non-existence of firing. Each text token would then be composed of $75$ binary values. Data representing one sampling time ($0.05$\,ms) is shown below:

\begin{equation*}
\underbrace{\texttt{<|START|>}}_{\texttt{bos}}\underbrace{\texttt{0000100000...0000000010}}_{75\text{ Neurons}}\underbrace{\texttt{<|END|>}}_{\texttt{eos}}
\end{equation*}

The first 0 refers to neuron 1 being inactive, and fifth 1 refers to neuron 5 being active. The neurons are re-indexed according to their depth to probe tip from small to large. 

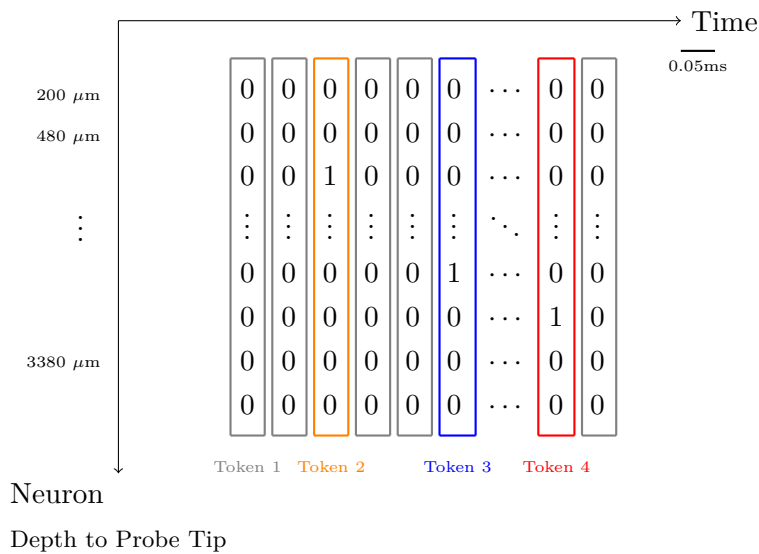
\begin{figure}[H]
\centering
\begin{tikzpicture}
\draw[->] (0,0) -- (7.4,0) node[right, align=left] {Time};

\draw[thick] (7.4,-0.4) -- (7.85,-0.4);
\node[anchor=north] at (7.625,-0.4) {\tiny 0.05ms};

\draw[->] (0,0) -- (0,-6) node[below, align=left] {Neuron\\{\footnotesize Depth to Probe Tip}};

\node[anchor=east] at (-0.1,-1.0) {\tiny 200 $\mu$m};  
\node[anchor=east] at (-0.1,-1.55) {\tiny 480 $\mu$m};  
\node[text=black] at (-0.5,-2.65) {$\vdots$};  
\node[anchor=east] at (-0.1,-4.55) {\tiny 3380 $\mu$m};  

    \node at (4,-3) {$\begin{array}{ccccccccc}
        0 & 0 & 0 & 0 & 0 & 0 & \cdots & 0 & 0\\
        0 & 0 & 0 & 0 & 0 & 0 & \cdots & 0 & 0\\
        0 & 0 & 1 & 0 & 0 & 0 & \cdots & 0 & 0\\
        \vdots & \vdots & \vdots & \vdots & \vdots & \vdots & \ddots & \vdots & \vdots \\
        0 & 0 & 0 & 0 & 0 & 1 & \cdots & 0 & 0\\
        0 & 0 & 0 & 0 & 0 & 0 & \cdots & 1 & 0\\
        0 & 0 & 0 & 0 & 0 & 0 & \cdots & 0 & 0\\
        0 & 0 & 0 & 0 & 0 & 0 & \cdots & 0 & 0\\
    \end{array}$};

\draw[black!50, thick] (1.475,-0.5) -- (1.475,-5.5);
\draw[black!50, thick] (1.925,-0.5) -- (1.925,-5.5);
\draw[black!50, thick] (1.475,-0.5) -- (1.925,-0.5);
\draw[black!50, thick] (1.475,-5.5) -- (1.925,-5.5);

\draw[black!50, thick] (2.025,-0.5) -- (2.025,-5.5);
\draw[black!50, thick] (2.475,-0.5) -- (2.475,-5.5);
\draw[black!50, thick] (2.025,-0.5) -- (2.475,-0.5);
\draw[black!50, thick] (2.025,-5.5) -- (2.475,-5.5);

\draw[orange, thick] (2.575,-0.5) -- (2.575,-5.5);
\draw[orange, thick] (3.025,-0.5) -- (3.025,-5.5);
\draw[orange, thick] (2.575,-0.5) -- (3.025,-0.5);
\draw[orange, thick] (2.575,-5.5) -- (3.025,-5.5);

\draw[black!50, thick] (3.125,-0.5) -- (3.125,-5.5);
\draw[black!50, thick] (3.575,-0.5) -- (3.575,-5.5);
\draw[black!50, thick] (3.125,-0.5) -- (3.575,-0.5);
\draw[black!50, thick] (3.125,-5.5) -- (3.575,-5.5);

\draw[black!50, thick] (3.675,-0.5) -- (3.675,-5.5);
\draw[black!50, thick] (4.125,-0.5) -- (4.125,-5.5);
\draw[black!50, thick] (3.675,-0.5) -- (4.125,-0.5);
\draw[black!50, thick] (3.675,-5.5) -- (4.125,-5.5);

\draw[blue, thick] (4.225,-0.5) -- (4.225,-5.5);
\draw[blue, thick] (4.7,-0.5) -- (4.7,-5.5);
\draw[blue, thick] (4.225,-0.5) -- (4.7,-0.5);
\draw[blue, thick] (4.225,-5.5) -- (4.7,-5.5);

\draw[red, thick] (5.525,-0.5) -- (5.525,-5.5);
\draw[red, thick] (6.0,-0.5) -- (6.0,-5.5);
\draw[red, thick] (5.525,-0.5) -- (6.0,-0.5);
\draw[red, thick] (5.525,-5.5) -- (6.0,-5.5);

\draw[black!50, thick] (6.1,-0.5) -- (6.1,-5.5);
\draw[black!50, thick] (6.55,-0.5) -- (6.55,-5.5);
\draw[black!50, thick] (6.1,-0.5) -- (6.55,-0.5);
\draw[black!50, thick] (6.1,-5.5) -- (6.55 ,-5.5);

\node[anchor=north] at (1.7,-5.7) {\textcolor{black!50}{\tiny Token 1}};
\node[anchor=north] at (2.8,-5.7) {\textcolor{orange}{\tiny Token 2}};
\node[anchor=north] at (4.4625,-5.7) {\textcolor{blue}{\tiny Token 3}};
\node[anchor=north] at (5.7625,-5.7) {\textcolor{red}{\tiny Token 4}};
    
\end{tikzpicture}
\caption{Transforming neural data in sampling time into tokens}
\label{fig:tokens}
\end{figure}

We collected all such combinations, and used Byte-Pair Encoding (BPE) to train a tokenizer. The goal was to represent each sampling time by one input token ID. Since BPE creates both complete units and subunits with varying lengths, we aimed to keep only complete tokens with a length of $75$ (which is equal to the number of neurons). Since the tokenizer was trained on all possibilities, it automatically prioritized single complete tokens over multiple subunit combinations. Thus in practice, the incomplete subtokens had no impact on performance and were kept in the vocabulary. Figure~\ref{fig:tokens} shows how the data were tokenized.

We then used this tokenizer to tokenize the whole data, and we added three special tokens: (1) begin of sequence (bos) token \texttt{<|START|>}, (2) end of sequence (eos) token \texttt{<|END|>}, and (3) divider \texttt{<|>}. The divider \texttt{<|>} was used to combine multiple sampling times together. A data for two sampling times (0.1\,ms) combined together looks like:
\begin{equation*}
\texttt{<|START|>}\underbrace{\texttt{0000100000...0000000010}}_{\text{ Data 1}}\underbrace{\texttt{<|>}}_{\text{divider}}\underbrace{\texttt{0000000000...0010000000}}_{\text{Data 2}}\texttt{<|END|>}
\end{equation*}

This tokenization method allowed us to represent binary spike train data into limited numbers of tokens for subsequent analysis. The obtained vocabulary size is $4087$ ($7107$ including subunits), and $\approx$ 90\,\% of the data is empty data consisting of 75 zeros.

\subsection{Data preparation} 
We used the following data lengths: 0.05\,ms (with and without empty tokens), 0.5\,ms, 1.5\,ms, 2.5\,ms, 3\,ms, 5\,ms, 6\,ms, 15\,ms, 30\,ms, 45\,ms, and 60\,ms. Since 60\,ms exceeded the lengths of some annotated syllables, padding was introduced to maintain consistent input dimensionalities. 
\subsection{Models and Evaluation} 
We randomly initialized the weights of two Hugging Face models to borrow their mature structures. The first model is OpenAI GPT-2 model \citep{Radford2019-gx} with 142M parameters, and the second model is a Mamba \citep{Gu2023-vw} model with 130M parameters with a sequence classification head. The models were evaluated using classification accuracies across different data lengths and model architectures. We then compared the best performing model against our previously established SVM benchmark.
\section{Results}
\subsection{GPT-2 performance}
\begin{figure}[htbp]
    \centering
    \includegraphics[width=0.8\textwidth]{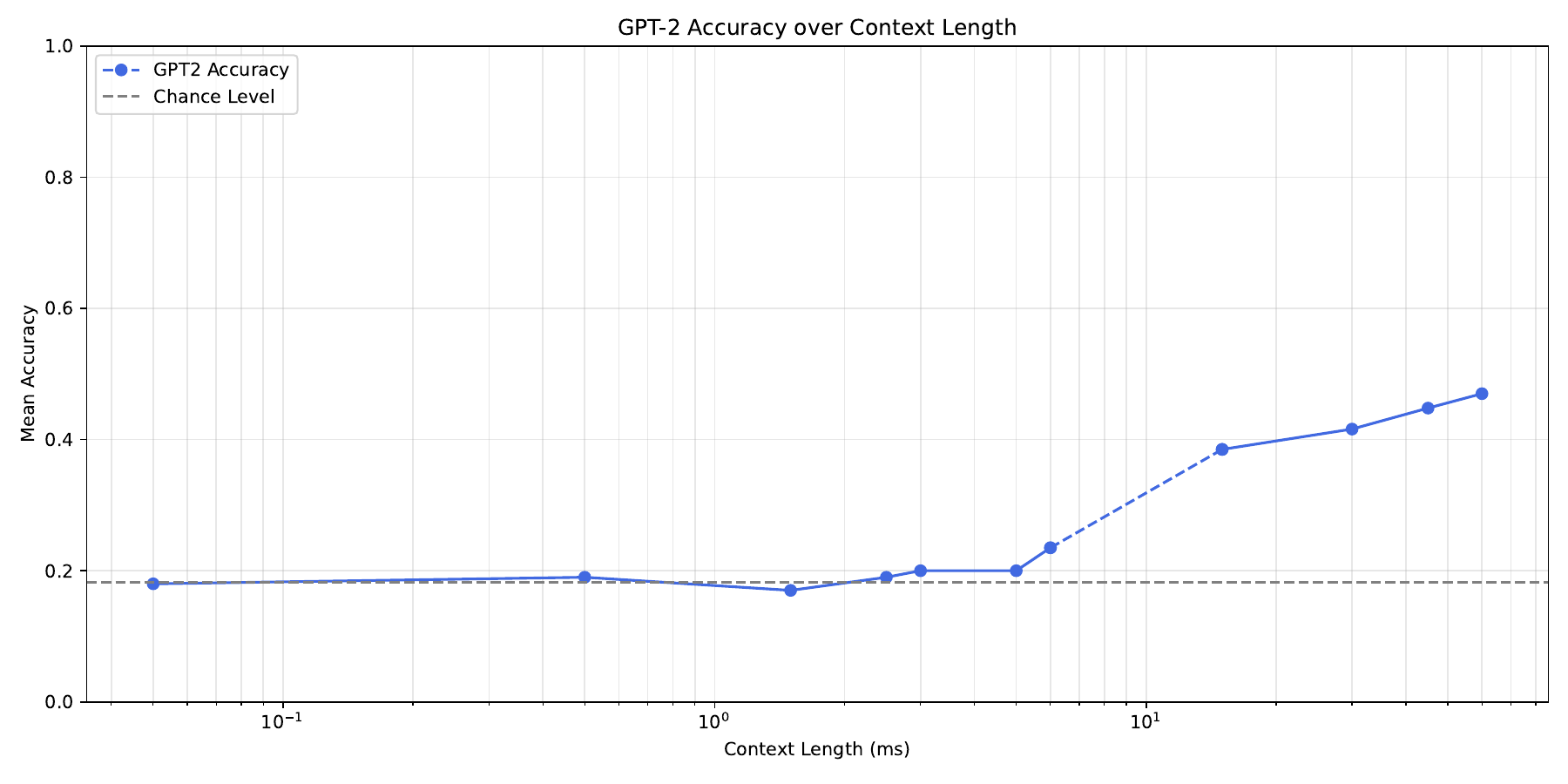}
    \caption{Performance of GPT-2}
    \label{fig:GPT}
\end{figure}

Table~\ref{tab:gpt2_performance} and Figure~\ref{fig:GPT} show the classification accuracy of GPT-2 increases with input data length from 15\,ms, and reaches 47\,\% with information from 60\,ms. On the other hand, data shorter than 15\,ms failed for the task, even when the sparsity was removed.
\begin{table}[htbp]
\centering
\begin{tabular}{cc}
\toprule
Time (ms) & Accuracy (\%) \\
\midrule
15.0 & 38.5 \\
30.0 & 41.6 \\
45.0 & 44.8 \\
\mathBF{60.0} & \mathBF{47.0} \\
\bottomrule
\end{tabular}
\caption{GPT-2 classification accuracy with sequences from 15\,ms to 60\,ms}
\label{tab:gpt2_performance}
\end{table}

\subsection{Mamba performance}
\begin{figure}[H]
    \centering
    \includegraphics[width=0.8\textwidth]{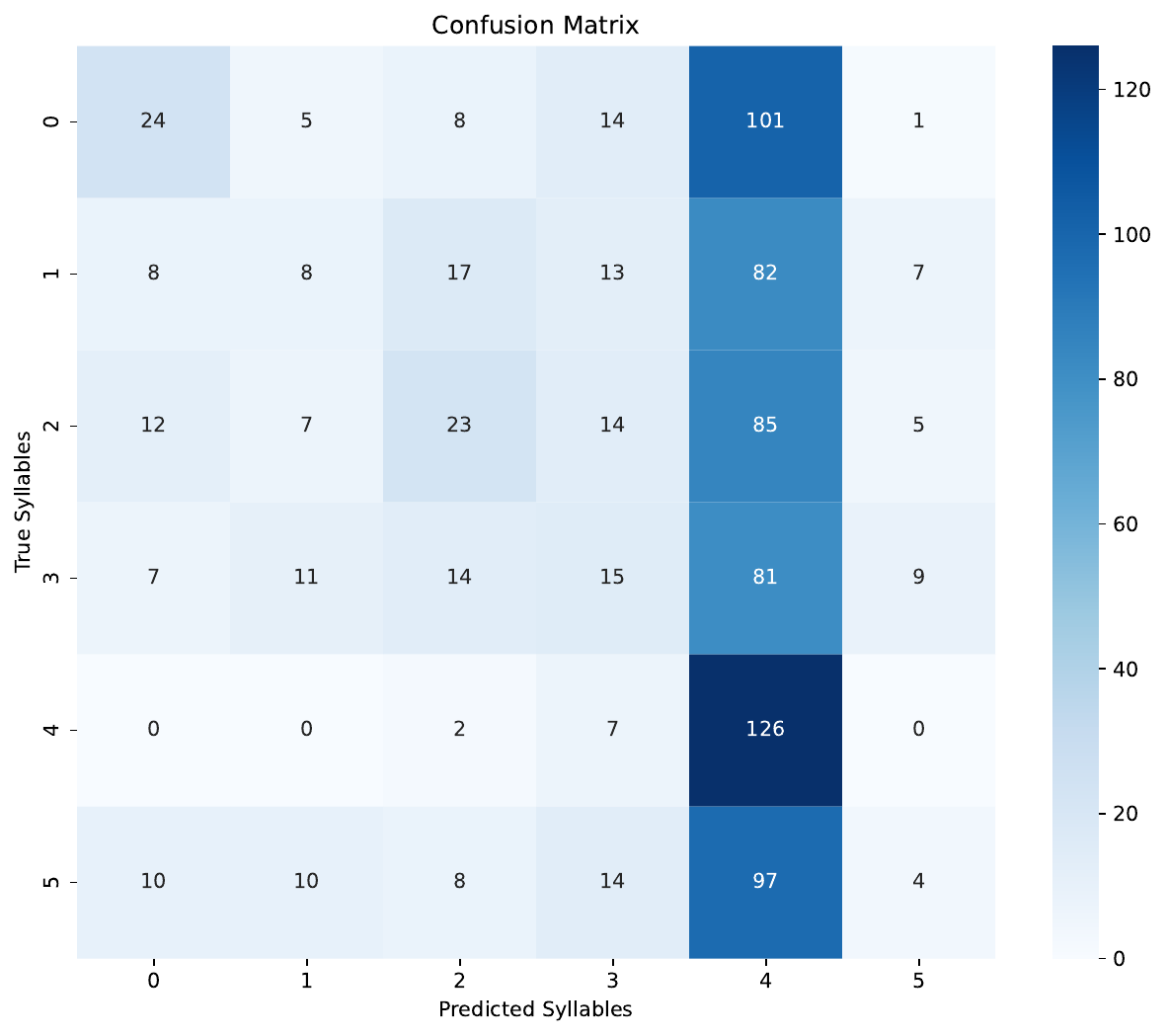}
    \caption{Mamba confusion matrix with 45\,ms sequences}
    \label{fig:MambaText}
\end{figure}

Figure~\ref{fig:MambaText} shows that Mamba 130M failed the task with sequence length 45\,ms.

\section{Discussion}
\subsection{Achievements on GPT-2 performance}
Figure~\ref{fig:GPT} shows the potential of GPT-2 in neural decoding tasks. The performance of the model increases to 47.0\,\% with longer input sequence length. Our findings demonstrate the validity of the approach for transformers to deal with neural data as text tokens. 
\subsection{Limitations}
The performance of GPT-2 model is not optimal. The final accuracy of 47\,\% is below our established SVM benchmark 77.3\,\% using mean firing rates. Additionally, the small Mamba model failed to decode $45$\,ms length data. The performance of large language models indicate that more data is needed.

\subsection{Future Directions}
The current decoding accuracy can be improved by training with data from the motor pathway instead of AFP. In addition, a pre-trained model could be developed using synthesized \citep{Zhou2020-om} or open \citep{de-Vries2020-ld} data from other animals and experiment paradigms, and then fine-tuned using real vocalization data. Introducing mixed transformer-mamba layers \citep{Lieber2024-hk} is also a promising approach to enhance the decoding accuracy.

\subfilebibliography

\chapter{Experiment 3: Temporal Deep Learning Methods}%
\refstepcounter{experiment} 
\label{exp:\theexperiment}
This experiment explored deep learning architectures for syllable classification using neural data. We compared the performance of different neural network architectures (CNN-LSTM, EEGNet, LSTM, and Mamba-2) against the benchmark established from previous experiments. Both sampling time binary data and binned data were evaluated. We adopted the newly released Mamba-2 architecture \citep{Dao2024-sp} rather than the original Mamba \citep{Gu2023-vw}, as Mamba-2 is computationally more efficient during training.

\section{Methods}
\subsection{Data} 
Data was selected around the onset using a moving window. Raw data in its original format $[\text{neuron}, \text{sampling time}]$ were used as binary data. We also binned the binary values and tested various bin sizes, ranging from $0.1$\,ms to $20$\,ms. We picked corresponding windows using a parameter sweep, such that different windows around the onset could be picked and compared. The step size was fixed at 1.

For the parameter sweep, we tested bin sizes from $0.05$\,ms to $20$\,ms. For each bin size, we tested different temporal windows relative to syllable onset:
\begin{itemize}
    \item pre-onset: from $-125$\,ms to $0$\,ms, incrementing by $12.5$\,ms
    \item post-onset: three values $37.5$\,ms, $50.0$\,ms, and $62.5$\,ms
\end{itemize}
The sequence length is determined by total time window length (ranging from $37.5$\,ms to $187.5$\,ms, computed as the sum of absolute values of pre-onset and post-onset times) and the bin size (ranging from no bin, which is $0.05$\,ms, to $20$\,ms). As a result, sequence lengths varied by up to three orders of magnitude, primarily driven by the bin size. When using original binary data or small bin sizes, the resulting data was considerably sparser when compared to those data processed with larger bin sizes.

\subsection{Models and Evaluation}
We used a two-layer LSTM model, a parallel CNN-LSTM model, a two-layer Mamba-$2$ model, and CNN-based EEGNet\citep{Lawhern2016-tf}. The models were evaluated by classification accuracies across different pre-onset and post-onset parameters, bin sizes, and model architectures. The results were then compared against results in previous experiments.

\section{Results}
All models but EEGNet successfully completed the parameter sweep analysis. EEGNet encountered memory constraints when processing smaller bin sizes due to long sequence lengths. For the $10$\,ms bin size, we were only able to test pre-onset windows of $-125$\,ms and $-112.5$\,ms due to technical limitations. We therefore focused our bin size analysis on a single parameter combination (pre-onset: $-125$\,ms, post-onset: $62.5$\,ms). The overall trends remained consistent across other parameter combinations, indicating that this partial testing did not affect the reported observations. Results from the parameter sweep are presented below.

\subsection{Top-3 Results for Each Model}

The Top-3 results for each model are included in Table~\ref{tab:model_comparison}. EEGNet achieved a peak decoding accuracy of $89.0\,\%$\,±\,$1.7\,\%$, with a bin size of $15$\,ms, a step size of $0.05$\,ms, and data extracted from $-125\,\text{ms to } 62.5\,\text{ms}$. 

\begin{table}[H]
   \centering
   \begin{tabular}{lrccc@{ $\pm$ }l}
       \toprule
       Model & Bin (ms) & Pre-onset (ms) & Post-onset (ms) & \multicolumn{2}{c}{Accuracy (\%)} \\
       \midrule
       \multirow{3}{*}{LSTM} 
           & 2.5 & $-112.5$ & 67.5 & 50.4 & 1.9 \\
           & 3.8 & $-125.0$ & 67.5 & 50.3 & 1.1 \\
           & 5.0 & $-125.0$ & 67.5 & 50.3 & 1.2 \\
       \midrule
       \multirow{3}{*}{Mamba-$2$}
           & 5.0 & $-125.0$ & 67.5 & 57.7 & 1.4 \\
           & 10.0 & $-125.0$ & 67.5 & 57.7 & 1.2 \\
           & 10.0 & $-112.5$ & 67.5 & 57.1 & 1.5 \\
       \midrule
       \multirow{3}{*}{\textBF{EEGNet}}
           & \mathBF{15.0} & $\mathbf{-}$\mathBF{125.0} & \mathBF{67.5} & \mathBF{89.0} & \mathBF{1.7} \\
           & 10.0 & $-125.0$ & 67.5 & 88.6 & 0.5 \\
           & 7.5 & $-125.0$ & 67.5 & 88.3 & 0.8 \\
       \midrule
       \multirow{3}{*}{CNN--LSTM}
           & 7.5 & $-125.0$ & 67.5 & 58.0 & 0.8 \\
           & 5.0 & $-125.0$ & 67.5 & 58.0 & 1.7 \\
           & 5.0 & $-112.5$ & 67.5 & 57.9 & 1.0 \\
       \bottomrule
   \end{tabular}
    \caption{Top-3 parameter combinations for each model on accuracy}
    \label{tab:model_comparison}
\end{table}

\subsection{Model Performance Across Different Windows Periods}
\begin{figure}[h]
    \centering
    \includegraphics[width=\textwidth]{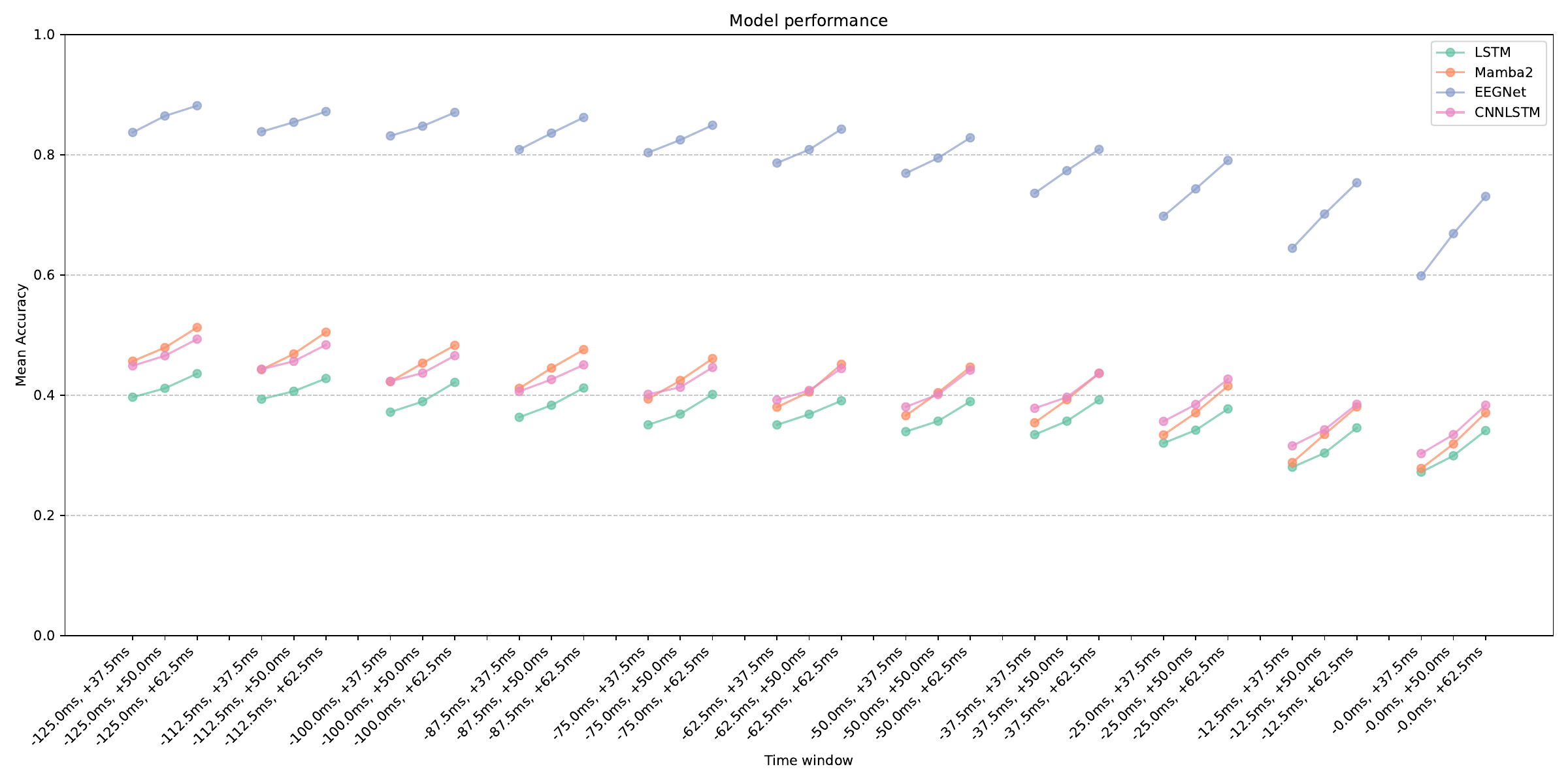}
    \caption{Model performance across different window periods}
    \label{fig:window_sweep}
\end{figure}

Figure~\ref{fig:window_sweep} shows the performance of models with different pre-onset and post-onset combinations. Each grouped column represents one pre-onset time. Within each group, there are three connected data points showing the classification accuracies using different post-onset times.

The grouped columns of pre-onset times show a pattern from left to right. As the pre-onset time moves closer to the syllable onset, the classification accuracy decreases. Within each group, three connected dots of post-onset times form an ascending line. Larger post-onset time on the right consistently outperforms smaller post-onset time on the left. 

In conclusion, longer data windows in both pre-onset and post-onset directions improve the decoding accuracy. This improvement appears stronger in shorter data, as shown on the right side of the figure. EEGNet has the best performance in all windows, and demonstrates the strongest sensitivity to data duration.

\subsection{Model Performance Across Different Bin Sizes} 
\begin{figure}[H]
    \centering
    \includegraphics[width=\textwidth]{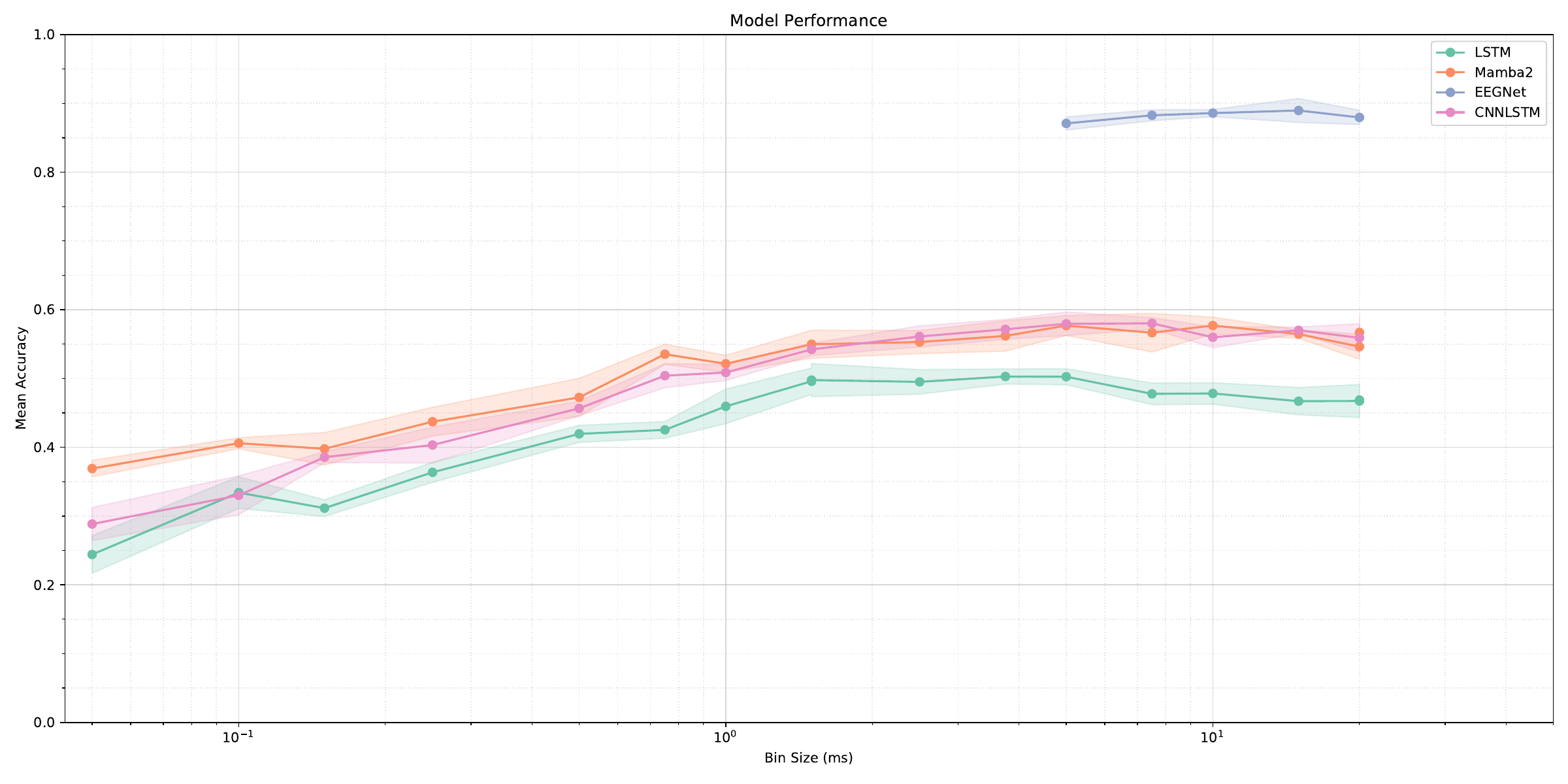} 
    \caption{Model performance across different bin sizes}
    \label{fig:binning_sweep}
\end{figure}

Figure~\ref{fig:binning_sweep} shows the performance across different bin sizes. The time window used is [$-125$\,ms, $62.5$\,ms]. The performance improves as bin size increases until the bin size hits 1.5\,ms. Mamba-2 shows a better performance in small bin sizes. 

\section{Discussion}

\subsection{EEGNet Performance on Binned Spike Data}
Our results show that EEGNet's architecture, originally designed to deal with the temporal and spatial characteristics of EEG recordings, successfully transfers its effectiveness to spike data analysis. It achieves a decoding accuracy of $89.0$\,\%, surpassing the $77.3$\,\% benchmark previously established in Experiment~\ref{exp:1}. The results suggest that deep learning models can effectively capture the complex patterns in neural spike data.

\subsection{Effect of Binning}
The performance increase until $1.5\,\text{ms}$ bin size can be attributed to reduction in data sparsity and decrease in input sequence length. The performance plateux for all models beyond the 1.5\,ms threshold indicate that this bin size achieves a balance between data sparsity, input sequence length, and fine-grained neural patterns. This plateau point at $1.5\,\text{ms}$ aligns with the typical duration of an action potential ($1$--$2$\,ms), suggesting that bin sizes beyond the timescale of individual spikes provide no significant benefit for performance. Different bin sizes have been used in literature, such as $1$\,ms, $8.33$\,ms, and $25$\,ms \citep{Schneider2023-xu, Zhou2020-om}, reflecting data-specific choices that balance trade-offs between fine-grained temporal patterns and computational efficiency.

\subsection{Limitations on Model Performance}
Other models we implemented were less effective at capturing information in the neural data. Their best decoding accuracies are lower than the benchmark from Experiment~\ref{exp:1}. With sampling time data, their performances are also lower than the NLP methods, as shown in Table~\ref{tab:compare_time_GPT}.
\begin{table}[H]
   \centering
   \begin{tabular}{lr}
       \toprule
       Model & Accuracy (\%) \\
       \midrule
       \textBF{GPT-2} & \mathBF{47.0} \\
       Mamba-$2$ & $26.2$ \\
       CNN--LSTM & $24.8$ \\
       LSTM & 21.1 \\
       \bottomrule
   \end{tabular}
   \caption{Model performance on sampling time}
   \label{tab:compare_time_GPT}
\end{table}

\subsection{Future Directions}
Mamba-$2$ showed some potential for dealing with the sparsity and long-range dependencies inherent in unbinned spike train data while maintaining computational efficiency. Future work could explore more sophisticated Mamba-$2$ architectures optimized for unbinned spike train data, and Conv2D layers could be incorporated to enhance the spatial pattern recognition capabilities.

\subfilebibliography

\chapter{Experiment 4: CEBRA Latent Embeddings}%
\refstepcounter{experiment}  
\label{exp:\theexperiment}
Consistent EmBeddings of high-dimensional Recordings using Auxiliary variables (CEBRA) is a contrastive learning method to compress time series data and reveal its hidden structures \citep{Schneider2023-xu}. The model can be self-supervised by temporal adjacency of neural activity, or supervised by behavioral labels, or both. CEBRA consistently generates interpretable latent embeddings in (1) synthesized spikes \citep{Zhou2020-om}, (2) hippocampal place encoding \citep{Grosmark2016-qy}, (3) somatosensory cortical activity during reaching \citep{Chowdhury2019-ev}, and (4) visual cortical responses to videos \citep{de-Vries2020-ld}. We expect CEBRA to reveal the underlying structure of LMAN activity during vocalization as well. With annotated syllables, the neural data can be supervised by a simple categorical system of $7$ syllable types. Thus we investigated the potential of latent embeddings produced by CEBRA for syllable classification.

\section{Methods}
\subsection{Data} 
We extracted the neural activity during the vocalization periods, and binned the data using a $10$\,ms window. Although this method interrupted continuous neural signals, it allowed us to focus on the neural activities of interest. Prior to concatenating, we shuffled the data at the syllable level to break the natural sequence of song. 

\subsection{Models}
CEBRA learns patterns in five ways: 
\begin{enumerate}
    \item over time
    \item over label
    \item over label, shuffled on a label level so that the labels are not in original order
    \item over label, shuffled on a sample level
    \item hybrid, jointly over time and label
\end{enumerate}
We focused on exploring the hyperparameter `offset' which defines the distance (in time) between positive pairs \citep{Schneider2023-xu}. A positive pair is composed of two timepoints separated by the offset value. Time points outside this interval become negative pairs. For example, an offset of $10$ refers to the fact that time point $t\pm10$ form positive pairs, and time point outside of this period will be used as negative pairs. Lower offset values lead to more stable learning, since temporally adjacent neural activities are more correlated. Higher offset values might lead to less stable learning, as the temporal correlation between neural activities decreases with longer time intervals. 

In the experiment, the dimension of latent embedding is set to 3, the learning rate is set to $3e-4$, the number of epochs is dynamically adjusted between 5000 to 50000 by observing the training loss. Train-test split was 8--2. The model was trained on the training set and we obtained the final embeddings by passing all data (including the test set) through the trained CEBRA model.

\subsection{Evaluation}
Our evaluation process consisted of three steps. First, we visualized trained embeddings\footnote{Only the third model is visualized in the experiment. Appendix~\ref{appendix:CEBRA} shows the visualization for four other models, their training loss, and data distribution.} to evaluate how data points from different labels distributed in the embedding space. 

Second, we performed single time point classification. We passed individual CEBRA embeddings to SVM, RF, and XGBoost, and compared the best result to benchmarks from Experiment~\ref{exp:1} using total count of spikes over $10$\,ms. By doing this, we ensured that we were comparing the same temporal resolution of 10\,ms. The performance of all the models, including the ones in the Appendix, were included in the classification accuracy table.

Finally, the latent embeddings were evaluated using sequence classification. For this, we combined the embeddings into sequences, and applied CNN-LSTM to these sequences. Identical window and parameters of data processing from Experiment~\ref{exp:3} were used in the comparison. The CNN-LSTM was picked due to its ease of implementation and its potential for further improvement with enhanced input features.

\section{Results}
\subsection{Visualization of CEBRA latent embeddings}
\begin{figure}[H]
    \centering
    \includegraphics[height=\textheight]{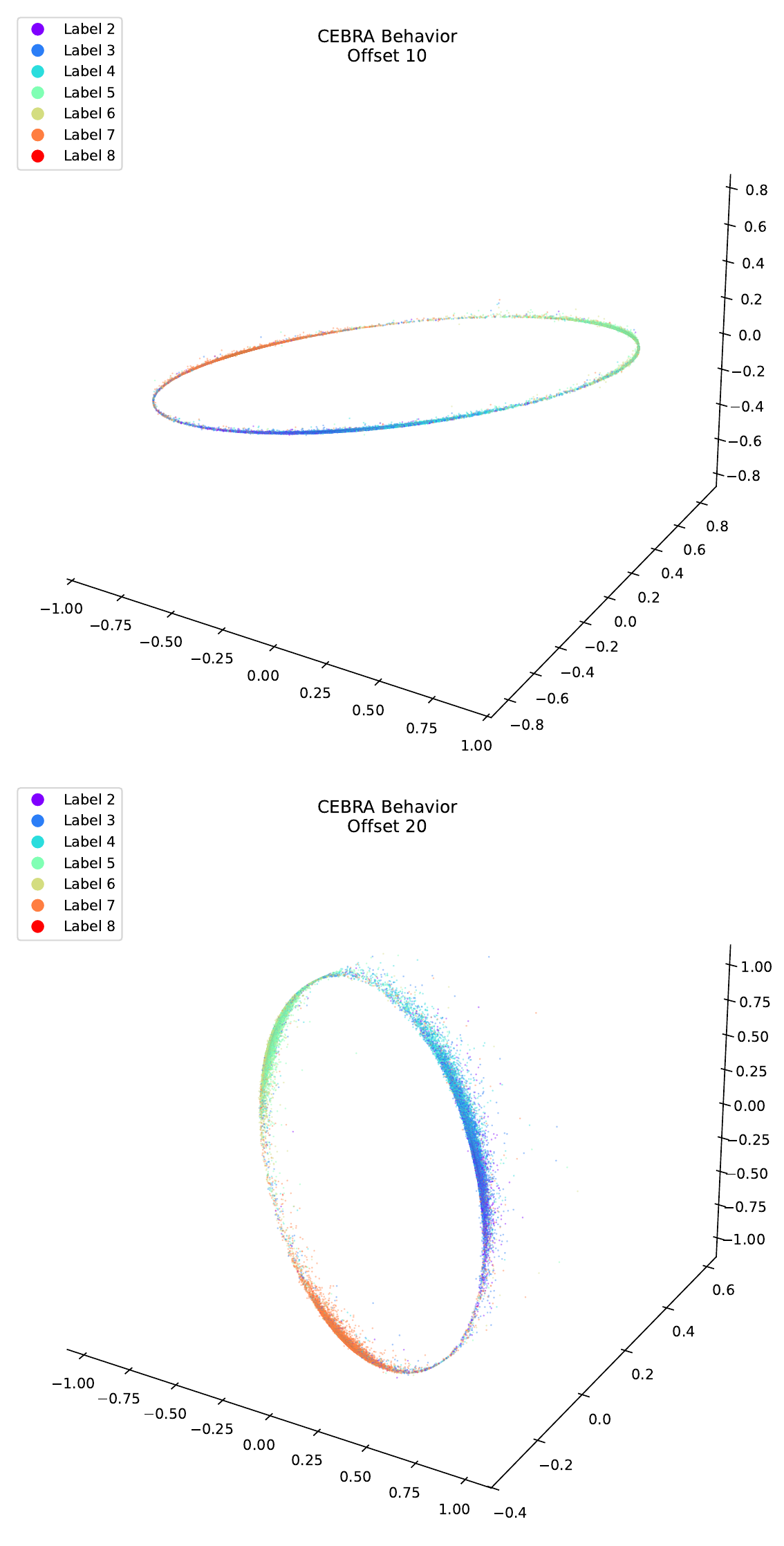}
    \caption{CEBRA embeddings trained with shuffled behavior labels}
    \label{fig:cebra_3d}
\end{figure}

Figure~\ref{fig:cebra_3d} visualizes the latent embeddings of all binned data, generated by CEBRA model trained on shuffled syllables.

The visualization shows excellent clustering with parameter offset 10 and offset 20, where the latent embeddings form a circular pattern. The spatial arrangement of neural data in the embedding space preserves their original sequence order in songs, despite the fact that the syllables are shuffled. The boundaries between different syllables appear to be thin and discontinuous. In comparison, unshuffled data in Appendix~\ref{appendix:CEBRA} shows more distinct boundaries between syllable regions.

\begin{sidewaystable}[htbp]
\centering
\caption{Model performance across different CEBRA parameters}
\begin{tabular}{lccccccccc}
\toprule
\multirow{2}{*}{\textbf{Accuracy (\%)}} & \multicolumn{3}{c}{\textbf{Offset 1}} & \multicolumn{3}{c}{\textbf{Offset 10}} & \multicolumn{3}{c}{\textbf{Offset 20}} \\
\cmidrule(l){2-4}
\cmidrule(l){5-7}
\cmidrule(l){8-10}
& SVM & RF & XGBoost & SVM & RF & XGBoost & SVM & RF & XGBoost \\
\midrule
Time & 27.4 & 27.2 & 26.2 & 26.5 & 25.8 & 25.2 & 25.3 & 25.5 & 24.3 \\
Behavior & 36.2 & 35.7 & 34.6 & 69.2 & 70.4 & 69.8 & 75.3 & 75.7 & 75.2 \\
\textBF{Behavior shuffled over labels} & 36.3 & 35.6 & 34.5 &\mathBF{ 64.0} & \mathBF{63.7} & \mathBF{64.0} & \mathBF{64.0} & \mathBF{64.2} & \mathBF{64.2} \\
Behavior shuffled over samples & 25.3 & 25.6 & 24.1 & 24.7 & 25.2 & 23.8 & 26.9 & 27.2 & 26.0 \\
Hybrid & 28.0 & 27.7 & 26.2 & 28.9 & 29.3 & 28.2 & 27.0 & 26.9 & 24.6 \\
\bottomrule
\end{tabular}
\label{tab:offset_comparison}
\end{sidewaystable}

\subsection{Single Time Point Classification}
CEBRA assigns each labeled neural data point (obtained by binning the spikes over $10$\,ms) a 3-dimensional latent embedding. We passed these latent embeddings to three classifiers: SVM, RF, and XGBoost. Table~\ref{tab:offset_comparison} shows the results for all models\footnote{The result visualizations were included in Appendix~\ref{appendix:CEBRA}.}. The behavior models achieved the best classification performance, and larger offset parameter increases the model performance. The result visualized previously is the third model: Behavior shuffled over labels. The best result is 64.2\,\% with parameter offset 20.

The result from the behavior models outperformed the baseline method. Using the same $10$\,ms window size, we tested the classification accuracy by counting the total number of spikes during fixed $10$\,ms windows as features to the same SVM. The best accuracy was around $36.7\,\%$ from time window $[-20\,\text{ms}, -10\,\text{ms}]$, which is well below the result obtained using CEBRA latent embeddings.

\subsection{Sequence Classification}
We applied the CNN--LSTM model from Experiment~\ref{exp:3} to sequences of CEBRA embeddings, and obtained an accuracy of $81.8$\,\%. The embeddings were from behavior models with shuffled labels and an offset of $10$.

Table~\ref{tab:cebra_lstm_compare} shows CNN--LSTM with CEBRA embeddings outperforms the same CNN--LSTM using binned spikes. The data had a similar window size and the models have the same structure. The CNN--LSTM with CEBRA embeddings also outperform EGGNet with binned spikes.

\begin{table}[H]
    \centering
    \caption{Model performance comparison on sequences}
    \label{tab:cebra_lstm_compare}
    \begin{tabular}{lrlcc}
        \toprule
        Input & Dimensions & Window (ms) & Model & Accuracy (\%) \\
        \midrule
        \textBF{CEBRA embeddings} & \mathBF{3} & \mathBF{60} & \textBF{CNN--LSTM} & \mathBF{81.8} \\
        Binned spikes & 75 & 62.5 & CNN--LSTM & 24.8 \\
        Binned spikes & 75 & 62.5 & EEGNet & 69.8 \\
        \bottomrule
    \end{tabular}
\end{table}

\section{Discussion}
\subsection{Achievements of CEBRA Embeddings}
Classification results from CEBRA embeddings achieve accuracies of 64.0\,\% and 81.8\,\%, outperforming all previous benchmarks. CEBRA effectively squeezes the original $75$ dimensions to $3$, which greatly reduces the sparsity in the data. Offset $20$ models consistently yield the best decoding results, possibly because the timescale of offset $20$ (200\,ms) captures the lengths of syllables.

The circular shape suggests that CEBRA is able to capture the syllable transitions even when trained on shuffled data. Unshuffled data contains information about syllable transitions. Training on unshuffled data achieves higher decoding accuracy of $75.7$\,\%. 

\subsection{Limitations}
CEBRA has been validated in paradigms with clearly defined boundaries in time. Vocal productions are spontaneous and are not triggered by temporally bounded stimuli. Thus, data pre-processing is required to remove the nonactive parts in the dataset, as simply allowing the model to self-supervise over the continuous recording session leads to low efficiency and low decoding accuracy. 

In addition, syllable transitions are continuous during vocalization, and pre-motor neural activity might overlap between adjacent syllables in one motif. Using data from the motor pathway would provide better results.

\subsection{Future Directions}
Future work could focus on contrastive learning frameworks due to their high potential to decode neural data. However, categorical labels like syllables, or redundant and high-dimensional labels like spectrograms might not be the most natural or efficient behavioural labels to use. MFCCs or latent embeddings from pre-trained models (for either human or bird vocalization) should be considered, as they are denser and have shown better performance in decoding tasks \citep{Defossez2022-lq}. 

\subfilebibliography

\chapter{Experiment 5: VAE Latent Representations}%
\refstepcounter{experiment}  
\label{exp:\theexperiment}
Our analysis focused on reconstructing the data using latent representations obtained from Variational Auto-Encoder (VAE). We attempted to capture patterns using both small segments as well as the whole syllable as units. We evaluated how well these generated neural latent embeddings preserve the original information through reconstruction and classification experiments. 

\section{Methods}
The following section describes three experiments conducted. We aimed to reconstruct both neural and vocal data, investigating the correlation between paired neural and vocal data, and investigate the possibility to generate vocalization from neural data.
\subsection{Generation Experiment}
We would like to investigate whether the patterns in the neural data could be captured by VAEs. In this experiment, we examined different methods to extract neural and vocal data separately to avoid encoding temporal information in the data. With the data obtained, we evaluated by visualizing the reconstructed data, investigating the latent embeddings, and training basic ML models on syllable classification tasks using the latent embeddings of either neural of vocal data obtained from the trained models. 

\subsection{Correlation Experiment}
For the second experiment, we applied linear warping to both neural data and spectrograms and mapped them to the maximum sequence length. With warped data, we aimed to investigate the correlation between latent embeddings generated, so that we can investigate the variance between neural and vocal data, as indicated in \cite{Singh_Alvarado2021-vr}. The results of warped data were also included for reconstruction visualization, latent embedding visualization, as well as syllable classifications for completeness.

\subsection{Cross-Modal Transfer: neuro2voc}
We investigated the possibility of generating speech data with neural data. We first created a joint VAE with a speech encoder, a neural encoder, as well as a speech decoder. We trained the VAE first using only speech data to obtain a satisfactory latent representation of speech data. Then, we froze the parameters of the speech encoder, and passed in both speech and neural data together. We used contrastive learning to generate the latent embeddings of neural data to make it align with the neural data as close as possible. The result is evaluated using reconstruction visualization, latent embedding visualization, as well as syllable classification. 

The quality of cross-modal transfer using a mapper function depends on not just the mapper function, but also the quality of reconstructed data in both modalities. The above method was chosen due to the unsatisfactory reconstruction quality of neural data.

\section{Data Preparation}
This following section describes the details of data preparation for binary neural data, binned neural data, and vocal data.
\subsection{Generation Experiment: Binary Neural Data}
For binary data, we selected neural data during active vocalization. We divided the activity into $30$\,ms segments. A moving window of step size $10$\,ms was used. The choice of segment length was made due to computation constraints.

\subsection{Generation Experiment: Binned Neural Data}
For binned data, we used three approaches to extract data:
The first approach was similar to what we did to the binary neural data. We divided the data into $40$\,ms segments, and we used a moving window of $4$\,ms step size. Since some data were less than $40$\,ms, we padded all the data using real neighboring neural activities to increase the amount of the training data. The neural data extraction began $20$\,ms before each annotated syllable onset and ended earlier by the same amount.

The second approach is cutting the all the sequences according to the shortest data sequence. The smallest data sequence is $60$\,ms from syllable 6. However, $40$\,ms was first chosen to compare to the results from sliding windows above, and $80$\,ms were then chosen after visualizing the distribution of data, as $97.5$\,\% data are longer than or equal to $80$\,ms. We would like to investigate the effect of the extra $40$\,ms on the classification accuracy. The neural data extraction began $50$\,ms before each annotated syllable onset. 

The third approach is padding all sequences according to the longest data with zeros. The longest sequence is $270$\,ms. The neural data extraction also began $50$\,ms before each annotated syllable onset. The padded data is excluded from discussion as it encodes temporal information. Nevertheless, the result is presented together with other types of extracting methods.

\subsection{Generation Experiment: Vocal Data}
For vocal data, spectrograms were provided in $128$ frequency bins. The approaches were the same as the approaches for Binned Neural Data above.

The first approach was the sliding window extraction approach for binned neural data, and we made sure that neural data and vocal data are correctly aligned in content and in length through visualization.

The second approach was similar to trimming binned neural data, $40$\,ms was chosen in order to align with the neural data as well as the length of segments from sliding windows.

At the end, we padded all vocal data to the the maximum sequence, which was $224$\,ms. The neural data had a maximum sequence length of $260$\,ms. It was longer because (1) binning was employed on the original sample time data, (2) $50$\,ms were added to the original neural data before the onset. This had no effect for the generation experiment as only single modalities were used. Similar to padded neural data, the result for padded vocal data is excluded from textual discussion, but the visualizations were presented in the result section as well for completeness. 

\subsection{Correlation Experiment: Neural and Vocal Data}
We first normalized the time to $0$--$1$, and we created interpolation function for individual frequency bin. Then we sample at new time points. The vocal data were warped to a maximum sequence of length $224$\,ms. Neural data were warped similarly to a maximum sequence of length $260$\,ms.

\subsection{Cross-Modal Transfer: Neural and Vocal Data}
For cross-modal transfer, we used segmented neural and vocal data on three levels: $4$\,ms, $40$\,ms, and $80$\,ms. Both data were extracted such that they are aligned in sequence length. The neural data were binned with a window size of $4$\,ms to match the sampling frequency of vocal data ($250$\,Hz).

\section{Models and Evaluation}
The data were passed to the following models:
\begin{enumerate}
    \item 1D VAE: a VAE composed of Conv1D layers for neural data
    \item 2D VAE: a VAE composed Conv2D layers for vocal data, based on \cite{Singh_Alvarado2021-vr}
    \item NeuralEnCodec: a VAE aiming at lossless reconstruction of neural data, based on EnCodec of \cite{Defossez2022-lq}
    \item Joint VAE: a VAE composed of Conv1D for neural data and Conv2D for vocal data
\end{enumerate}

We first briefly describe the training progress through the observation of the reconstruction loss, Mean Squared Error (MSE). Then we included in the results section a simple visualization of the final reconstructed data using random data on the test set (an $8/2$ split was used).

After the models were trained, all data were combined and passed again into the encoder to obtain the latent representations. The latent representations (which were 32-dimensional) of all data were visualized in 2D using Uniform Manifold Approximation and Projection (UMAP) and color-coded with syllable types. The latent presentations were also passed to SVM, RF, and XGBoost for syllable classification tasks. The classification accuracies of neural data were compared to CEBRA under the same conditions.

The warped neural and vocal latent representations were used to perform a Canonical Correlation Analysis. The first few Canonical Components were visualized and we reported the correlation coefficient. 

\section{Visualization Results}

\subsection{Generation Experiment: Visualization\footnote{All the neurons used in the visualization in this experiment were ranked by the depth to the tip, where N0 is the closest to the probe tip.}}
In all experiments, all the neural models showed convergence early and the loss were high, suggesting more training data would be needed. Since 2D VAE and NeuralEnCodec repeatedly showed insatisfactory reconstruction results on binary and warped neural data, they were not used for further analysis in generation experiments for neural data. Some reconstruction visualizations were included in the results section for completeness, but only 1D VAE was used for neural data for further processing as well as detailed evaluation.


\subsubsection{Binary Neural Data - Segments of 30\,ms}
\begin{figure}[H]
    \centering
    \includegraphics[width=\textwidth]{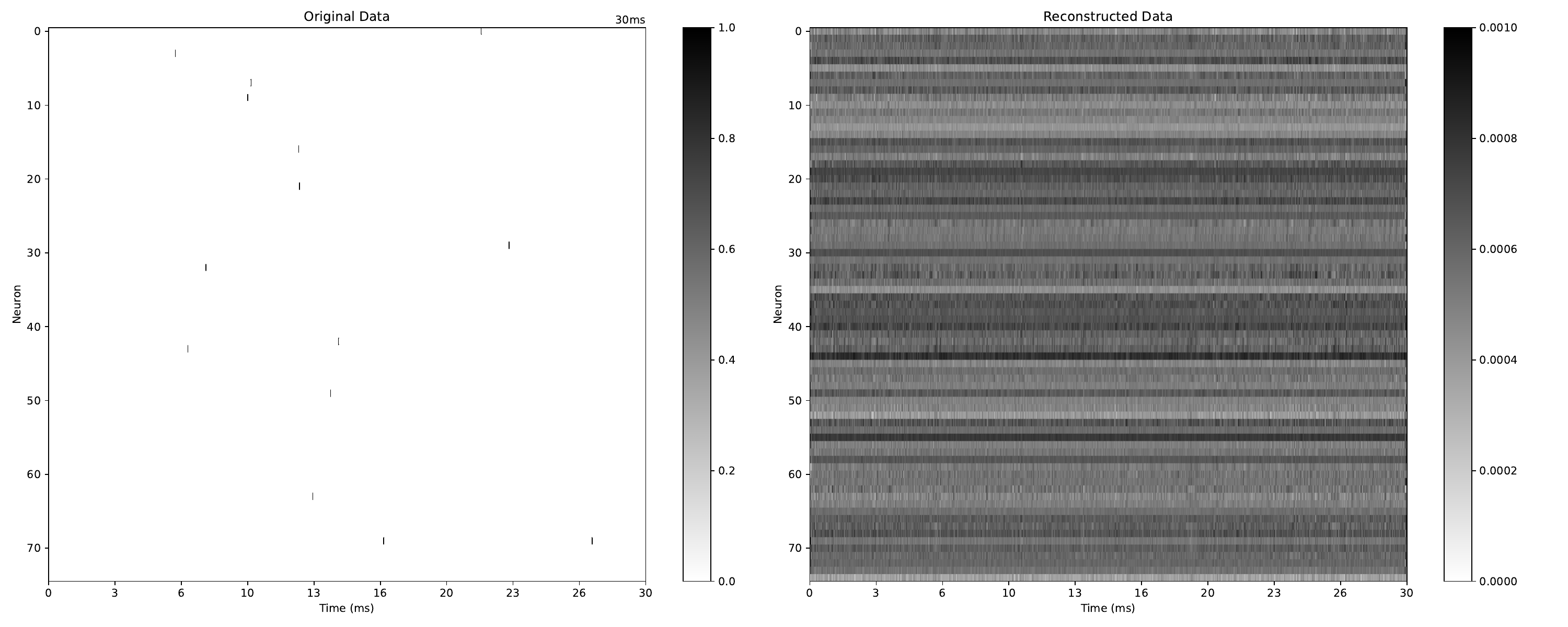}
    \caption{Reconstruction of binary neural data -- 1D VAE}
    \label{fig:binary_1d}
\end{figure}

\begin{figure}[H]
    \centering
    \includegraphics[width=\textwidth]{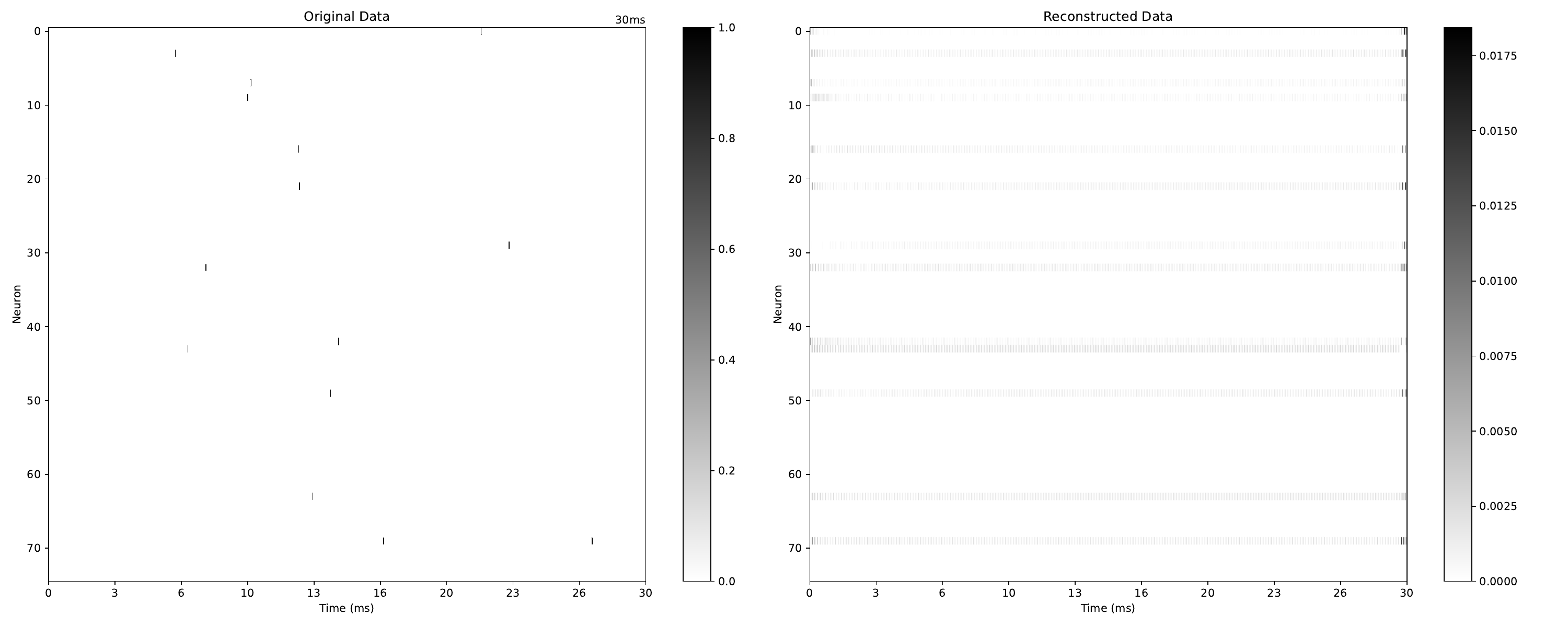}
    \caption{Reconstruction of binary neural data -- NeuralEnCodec}
    \label{fig:binary_encodec}
\end{figure}

1D VAE showed little amount of learning and produced poor reconstruction quality. Figure~\ref{fig:binary_1d} shows the original and the reconstructed binary neural data. Different scales are used for the reconstructed neural data, because the results are visually imperceptible in the reconstructed data using the original scale.
NeuralEnCodec reached the best result early, and showed limited learning progress. The reconstruction in Figure~\ref{fig:binary_encodec} shows temporal smoothing patterns. Instead of reproducing the precise spikes, the model generates an averaged spiking pattern across time. Similarly, different scaled were used in the figure, because the results were visually imperceptible in the reconstructed data using identical scales.

\subsubsection{Binned Neural Data - Segments of 40\,ms}
\begin{figure}[H]
    \centering
    \includegraphics[width=\textwidth]{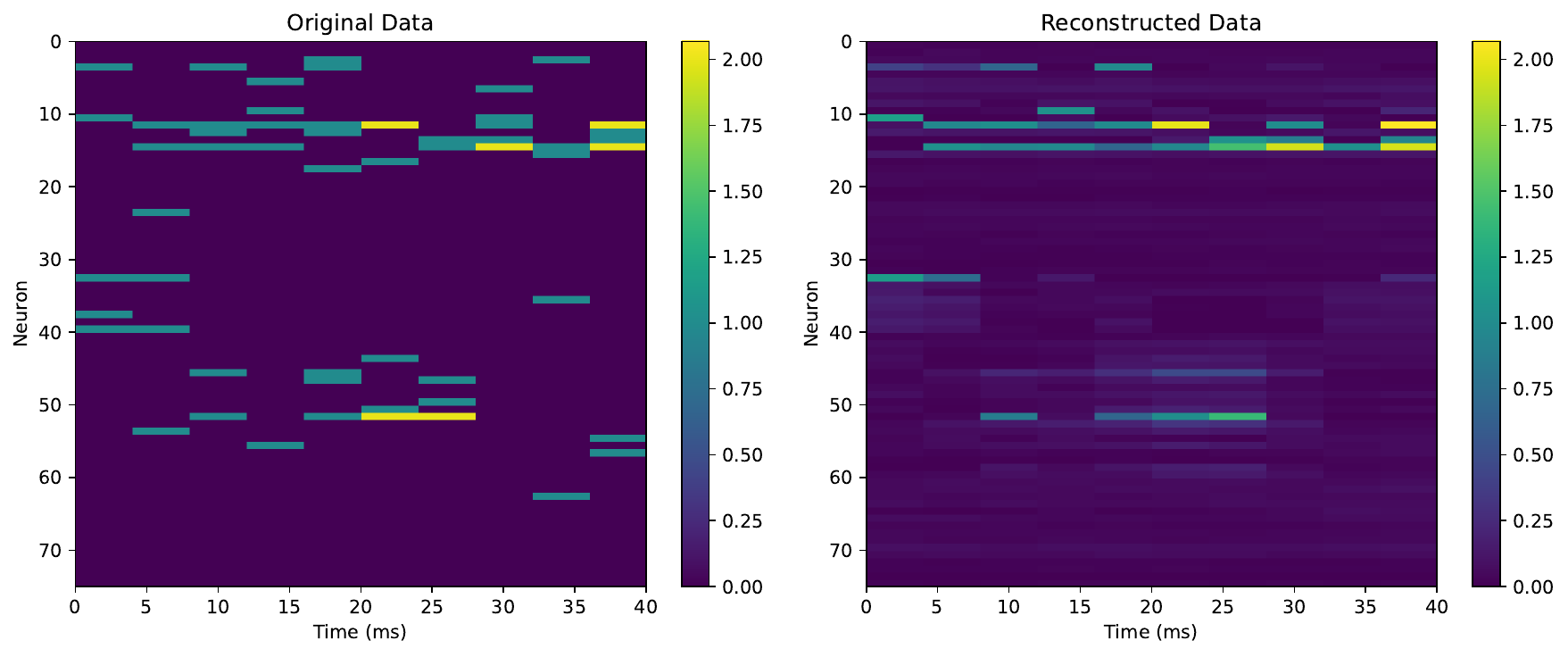}
    \caption{Reconstruction of binned neural data -- segments of 40\,ms}
    \label{fig:neural_segmented}
\end{figure}

Figure~\ref{fig:neural_segmented} shows the reconstruction results of 1D VAE trained on binned neural data. The results were not good enough, and some basic patterns in the original data were not captured. A binning window of $4$\,ms was used.

\subsubsection{Binned Neural Data - Trimmed to 40\,ms and 80\,ms}
\begin{figure}[H]
    \centering
    \includegraphics[width=\textwidth]{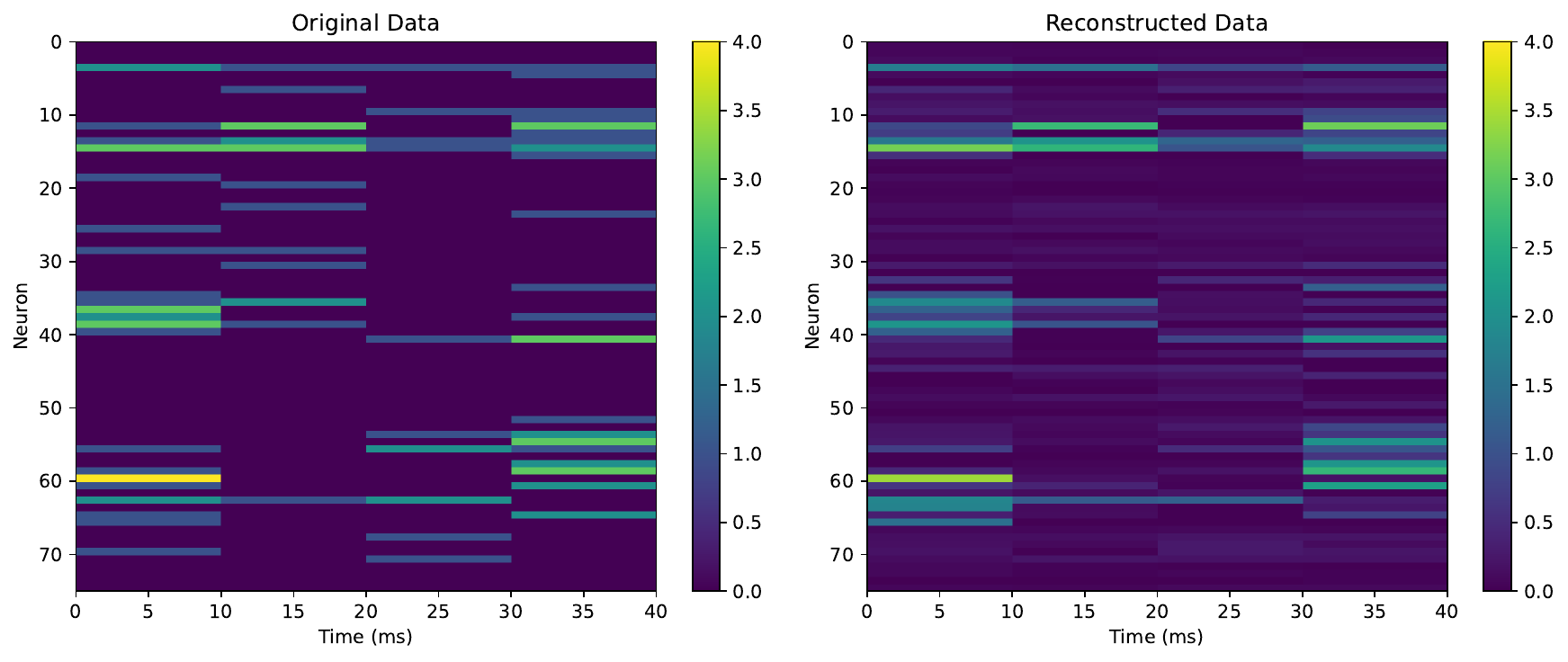}
    \caption{Reconstruction of binned neural data -- trimmed to 40\,ms}
    \label{fig:neural_trimmed_40}
\end{figure}

\begin{figure}[H]
    \centering
    \includegraphics[width=\textwidth]{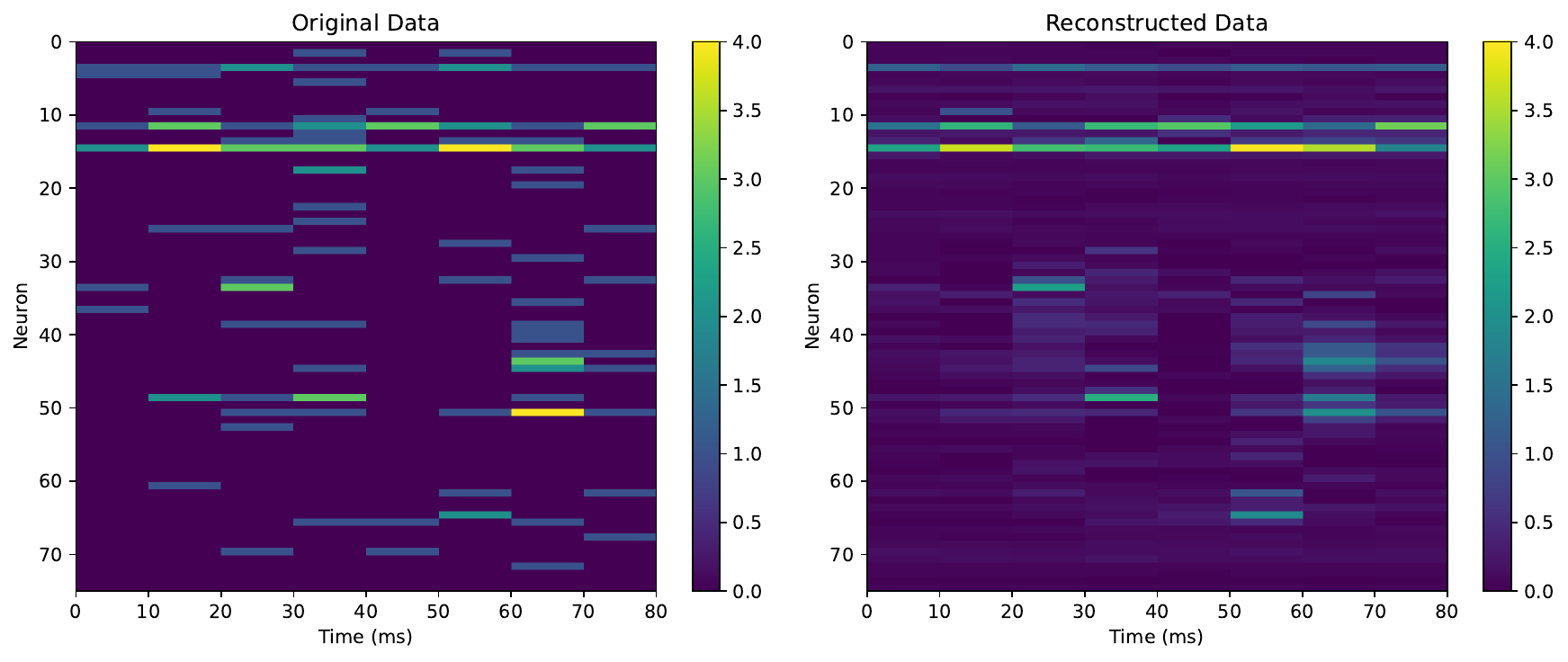}
    \caption{Reconstruction of binned neural data -- trimmed to 80\,ms}
    \label{fig:neural_trimmed}
\end{figure}

Figure~\ref{fig:neural_trimmed_40} and Figure~\ref{fig:neural_trimmed} show the reconstruction results using trimmed neural activities at syllable onsets using 1D. The binning window is $10$\,ms.

\subsubsection{Binned Neural Data - Padded to 270\,ms}
\begin{figure}[H]
    \centering
    \includegraphics[width=\textwidth]{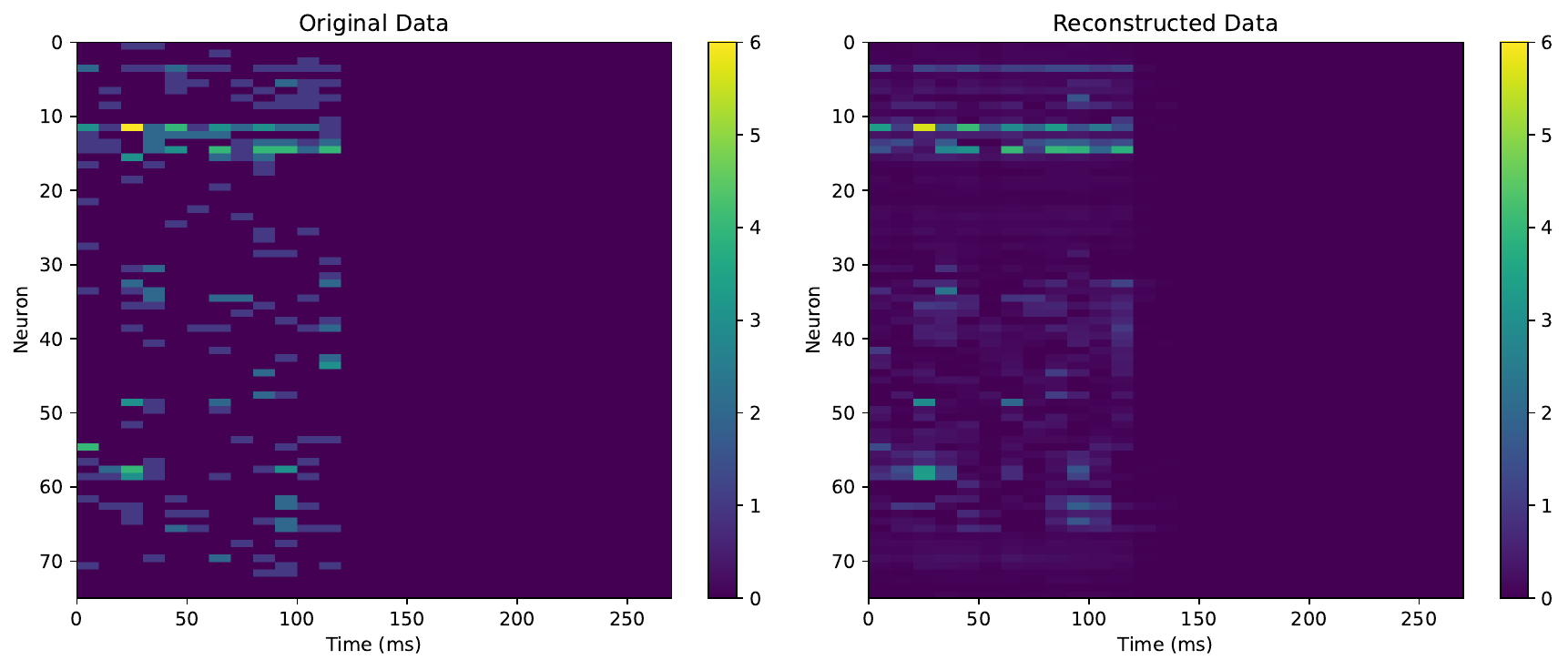}
    \caption{Reconstruction of binned neural data -- padded to 270\,ms}
    \label{fig:neural_padded}
\end{figure}

Figure~\ref{fig:neural_padded} shows the reconstruction results using 1D VAE. Each neural data is padded to the maximum sequence length. Data were binned using a $10$\,ms window.

\newpage
\subsubsection{Vocal Data - Segments of 40\,ms}
\begin{figure}[H]
    \centering
    \includegraphics[width=\textwidth]{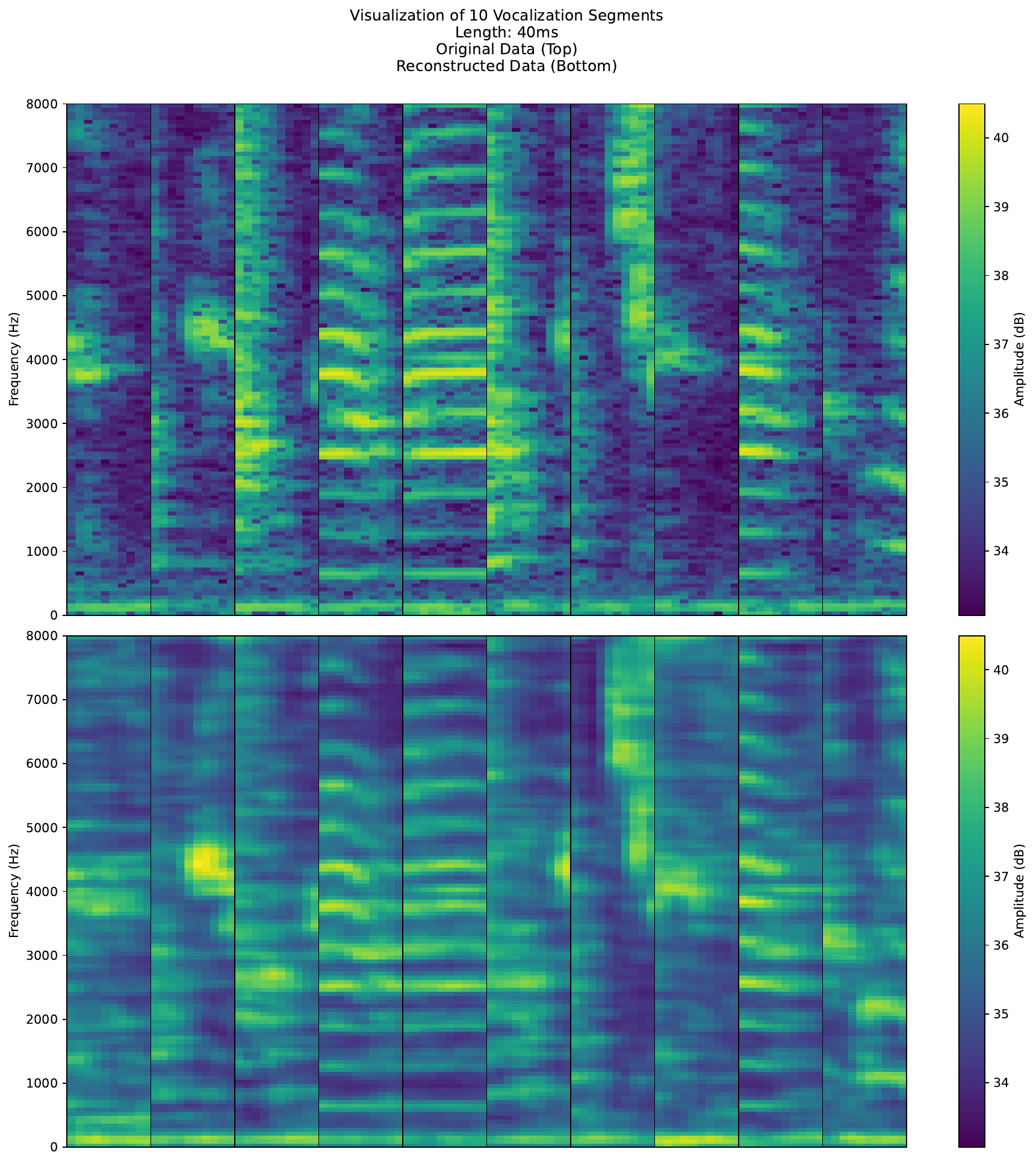}
    \caption{Reconstruction of segmented vocal data -- segments of 40\,ms}
    \label{fig:vocal_segmented}
\end{figure}

Figure~\ref{fig:vocal_segmented} shows the reconstruction results for 10 random vocalization segments using 2D VAE. Each segment is extracted from sliding windows and lasts for $40\,\text{ms}$.

\newpage
\subsubsection{Vocal Data - Trimmed to 40\,ms}
\begin{figure}[H]
    \centering
    \includegraphics[width=\linewidth]{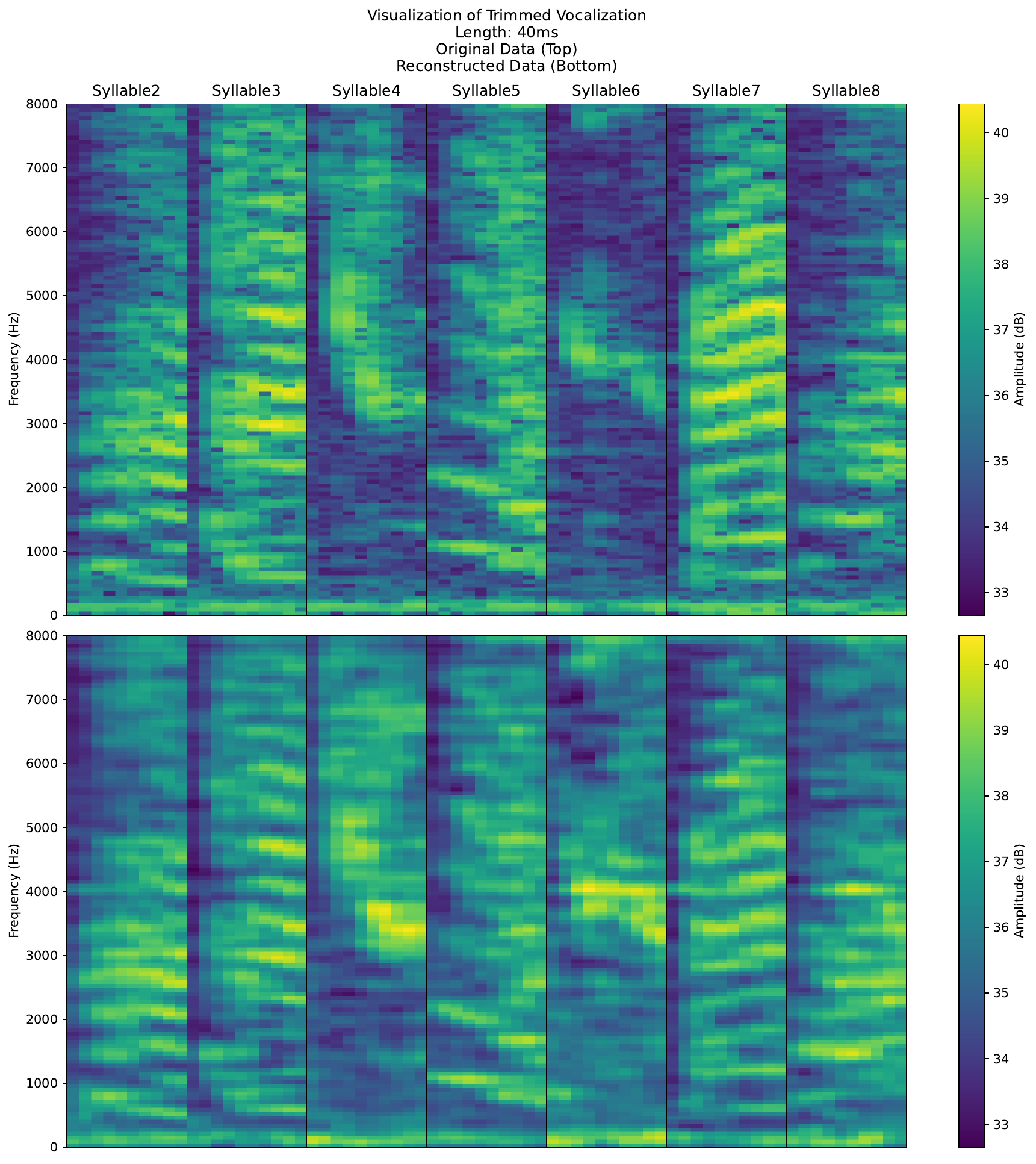}
    \caption{Reconstruction of trimmed vocal data -- trimmed to 40\,ms}
    \label{fig:vocal_trimmed}
\end{figure}

Figure~\ref{fig:vocal_trimmed} shows the reconstruction results from 2D VAE of random vocalization syllables using 2D VAE, from 2 to 8. Only the first $40\,\text{ms}$ was kept for each annotated syllable.

\newpage
\subsubsection{Vocal Data - Padded to 224\,ms}
\begin{figure}[H]
    \centering
    \includegraphics[width=\linewidth]{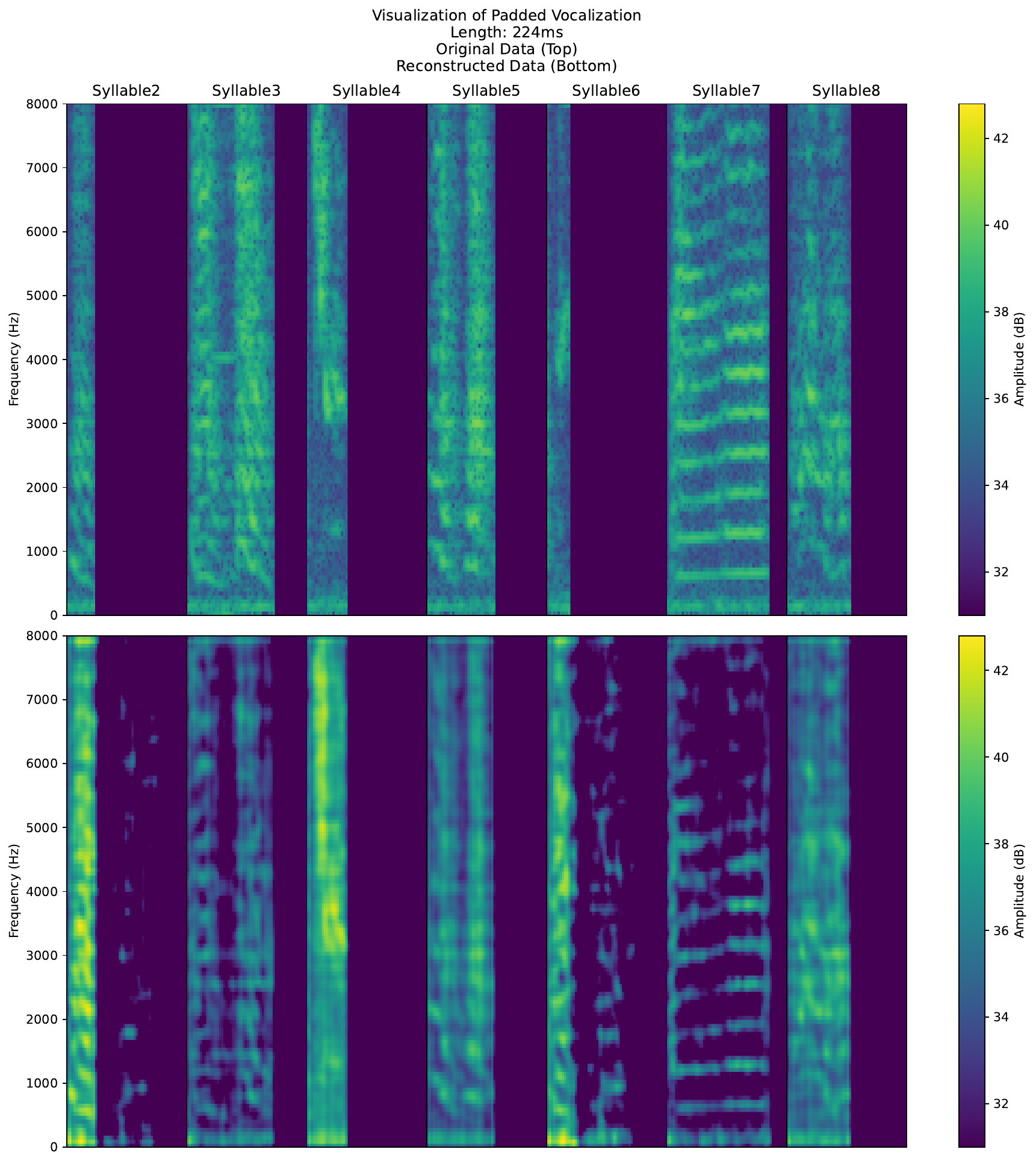}
    \caption{Reconstruction of padded vocal data -- padded to 224\,ms}
    \label{fig:vocal_padded}
\end{figure}

Figure~\ref{fig:vocal_padded} shows the reconstruction results from 2D VAE of random vocalization syllables. Each syllable was padding to $224\,\text{ms}$.

\newpage
\subsection{Correlation Experiment: Visualization}
\subsubsection{Binned Neural Data - Warped}
\begin{figure}[H]
    \centering
    \includegraphics[width=\textwidth]{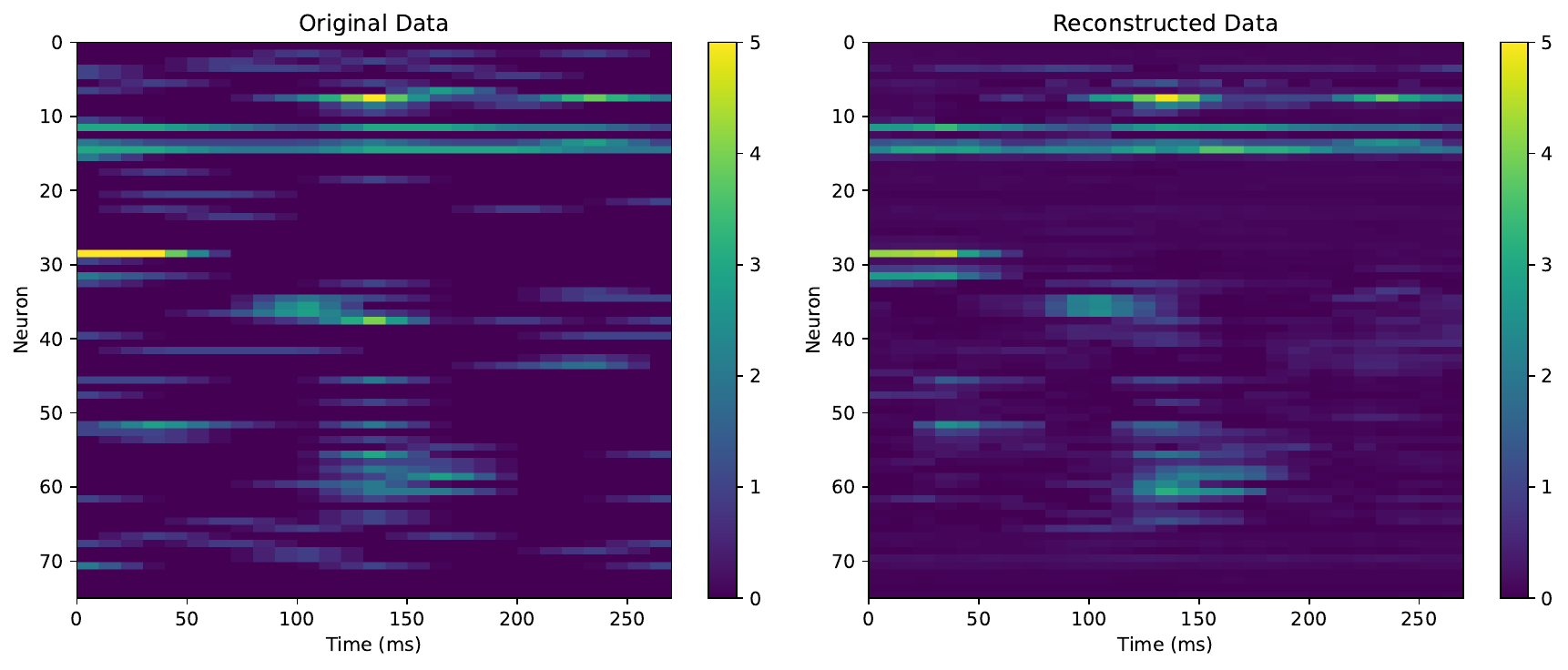}
    \caption{Reconstruction of warped and binned neural data with 1D VAE}
    \label{fig:binned_1d}
\end{figure}

The reconstruction loss from 1D VAE was consistently decreasing. Figure~\ref{fig:binned_1d} showed the original neural data on the left, and the reconstructed neural data on the right from 1D VAE. Neural data showed satisfactory reconstruction results upon visual inspection.
NeuralEnCodec reached the best result as early as Epoch 12. The result is shown below. Similarly to its performance on binary data shown in Figure~\ref{fig:binary_encodec}, NeuralEnCodec captured some characteristics of data, but smoothed out the data over time.

\begin{figure}[H]
    \centering
    \includegraphics[width=\textwidth]{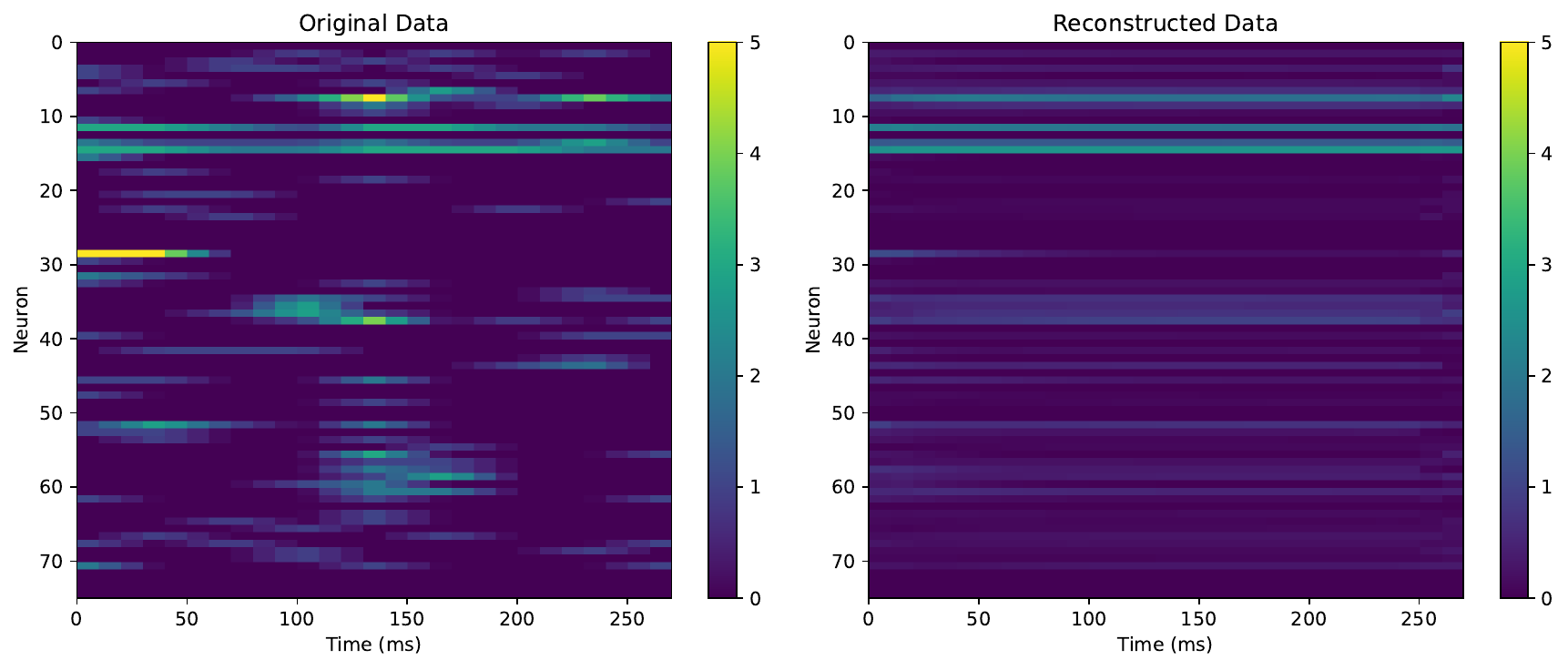}
    \caption{Reconstruction of binary neural data with NeuralEnCodec}
    \label{fig:binned_encodec}
\end{figure}

\newpage
\subsubsection{Vocal Data - Warped}
\begin{figure}[H]
    \centering
    \includegraphics[width=\linewidth]{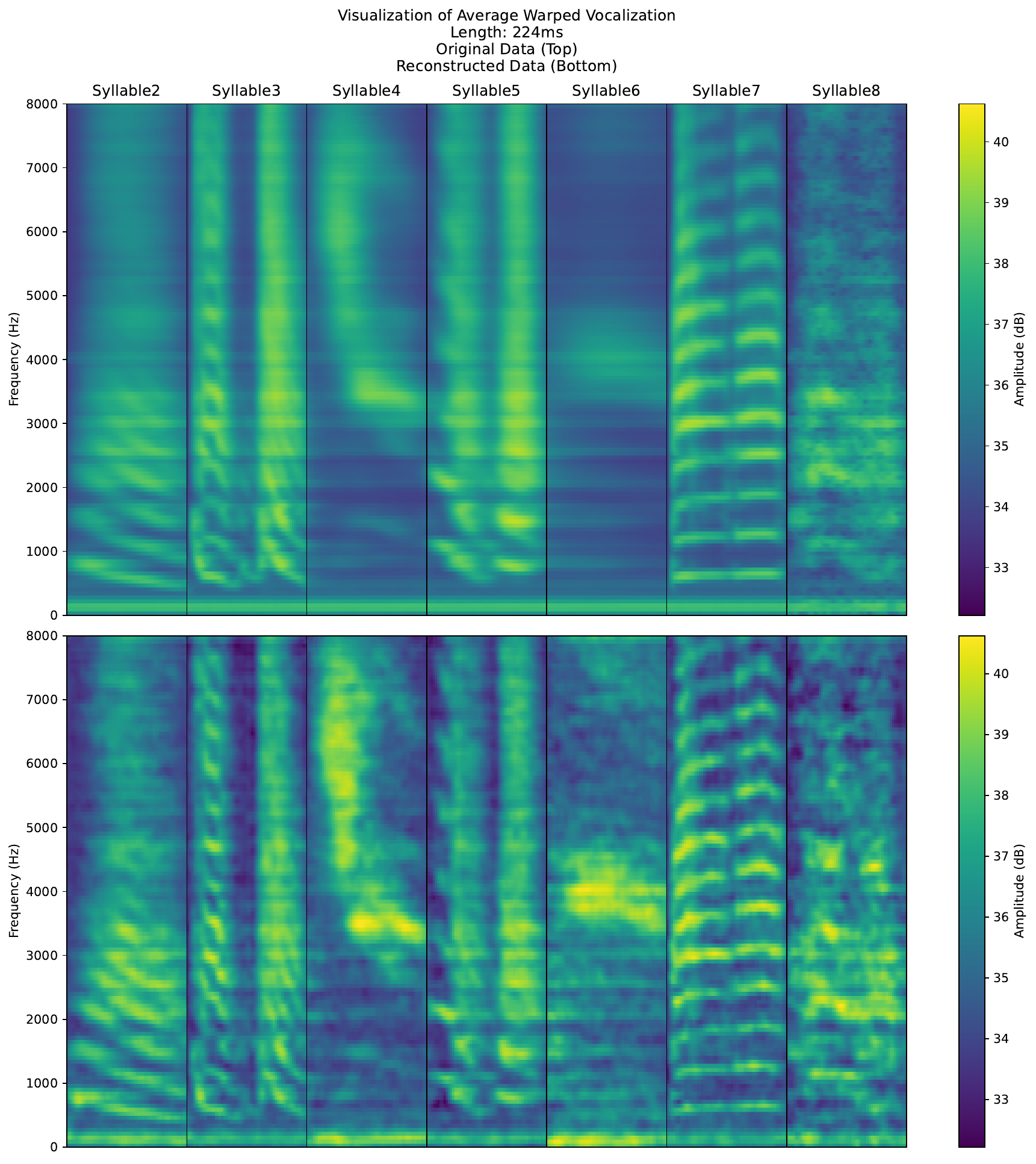}
    \caption{Reconstruction of warped vocal data}
    \label{fig:whole_motif_average}
\end{figure}

Figure~\ref{fig:whole_motif_average} shows original and reconstructed data for all different syllables using 2D VAE. The visualization was averaged over all warped samples. \footnote{Individual visualizations of averaged data and random samples are included in Appendix~\ref{appendix:VAE}}

\newpage
\subsection{Cross-Modal Transfer: neuro2voc}
\begin{figure}[H]
    \centering
    \includegraphics[width=\linewidth]{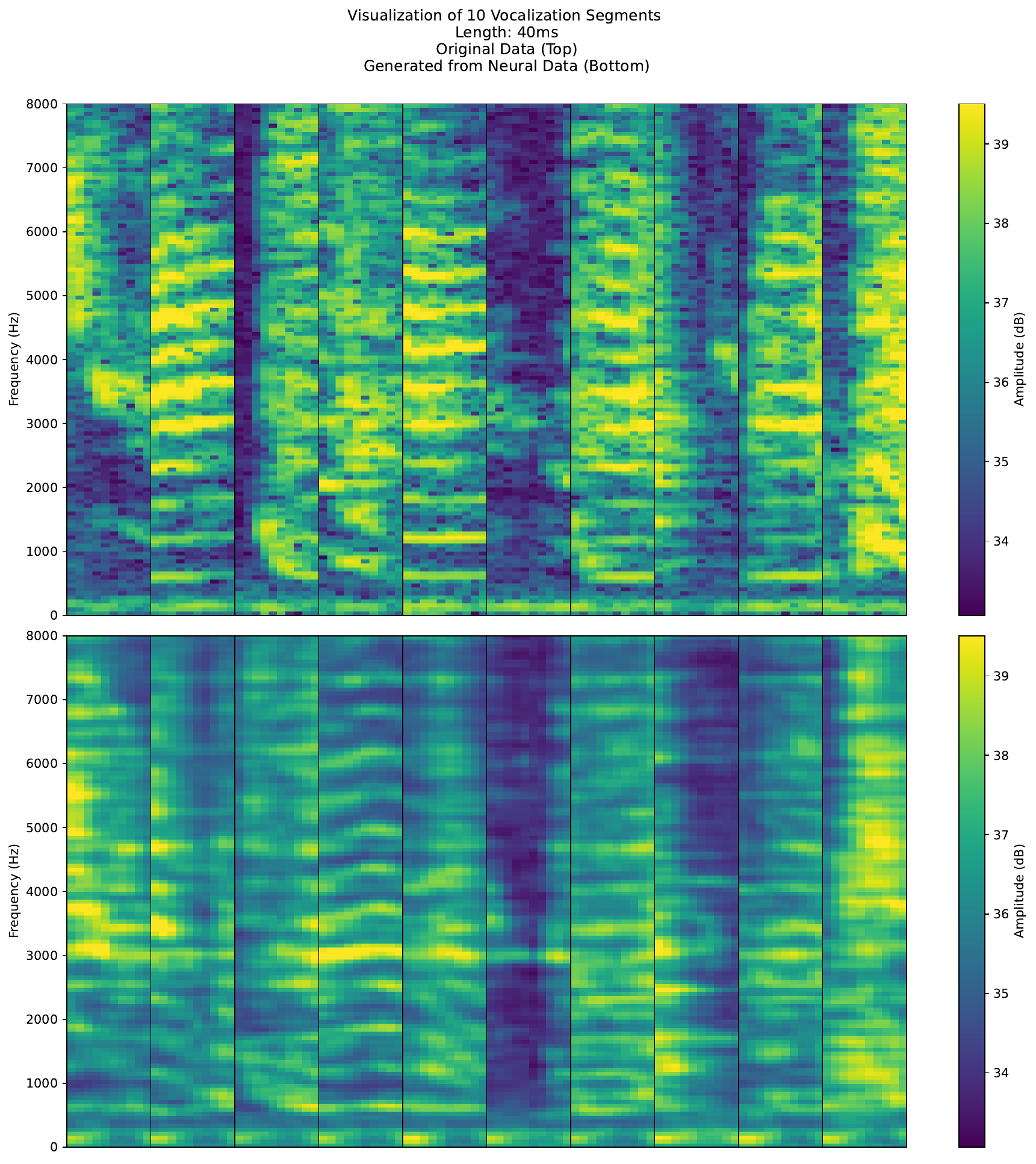}
    \caption{Generate 40\,ms vocalization from 40\,ms neural data}
    \label{fig:neuro2voc_40ms}
\end{figure}

Figure~\ref{fig:neuro2voc_40ms} shows the generated vocal data from neural data inputs. The neural data were obtained with sliding windows with 40\,ms\footnote{Visualization for 80\,ms is included in Appendix~\ref{appendix:VAE}} sequence length.

\section{Latent Representation Evaluations}
\subsection{Visualization of Latent Representations}
\begin{figure}[H]
    \centering
    \begin{subfigure}{0.32\linewidth}
        \includegraphics[width=\linewidth]{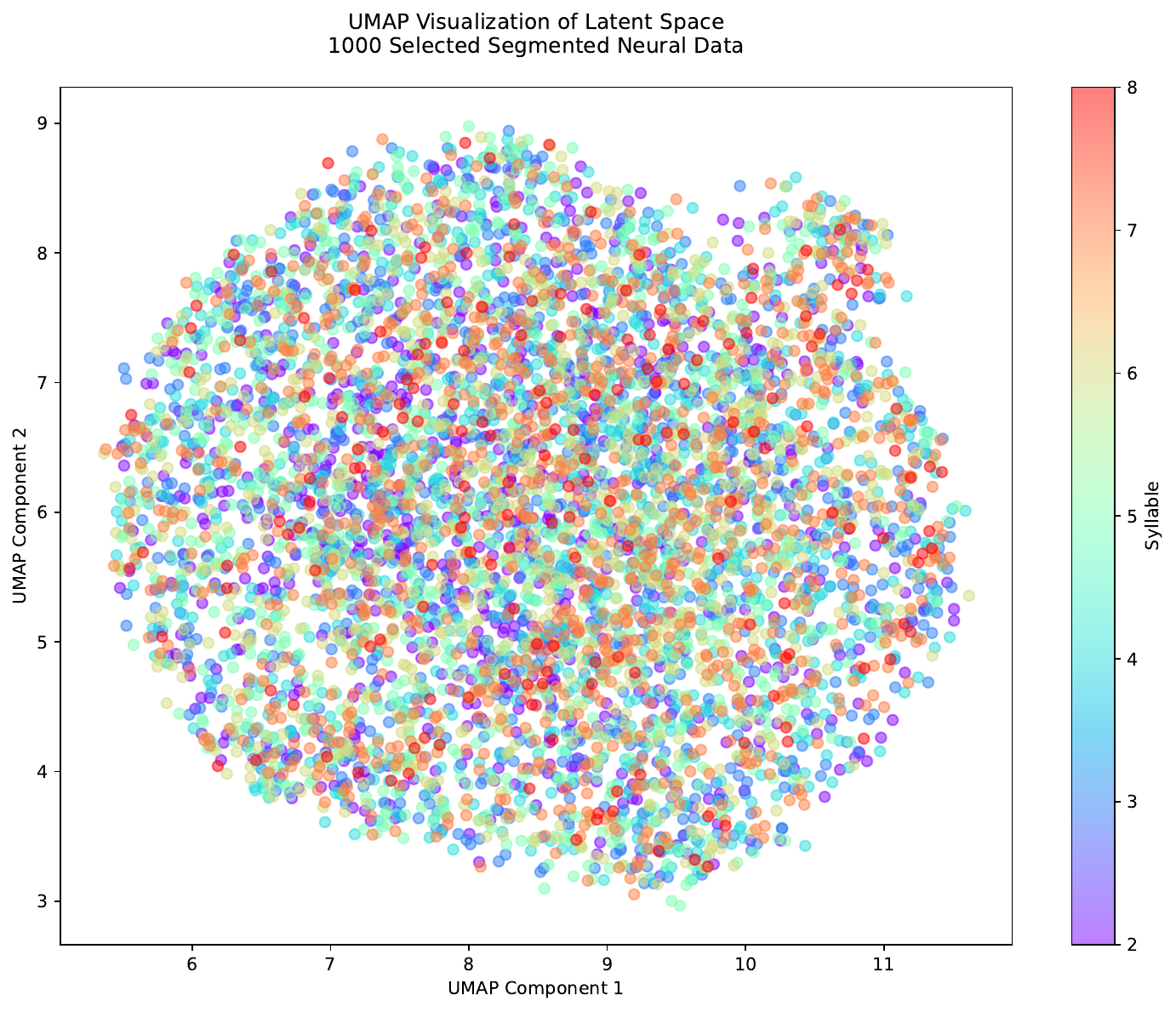}
        \caption{Neural 40\,ms segments}
        \label{fig:umap_segmented_neural}
    \end{subfigure}
    \hfill
    \begin{subfigure}{0.32\linewidth}
        \includegraphics[width=\linewidth]{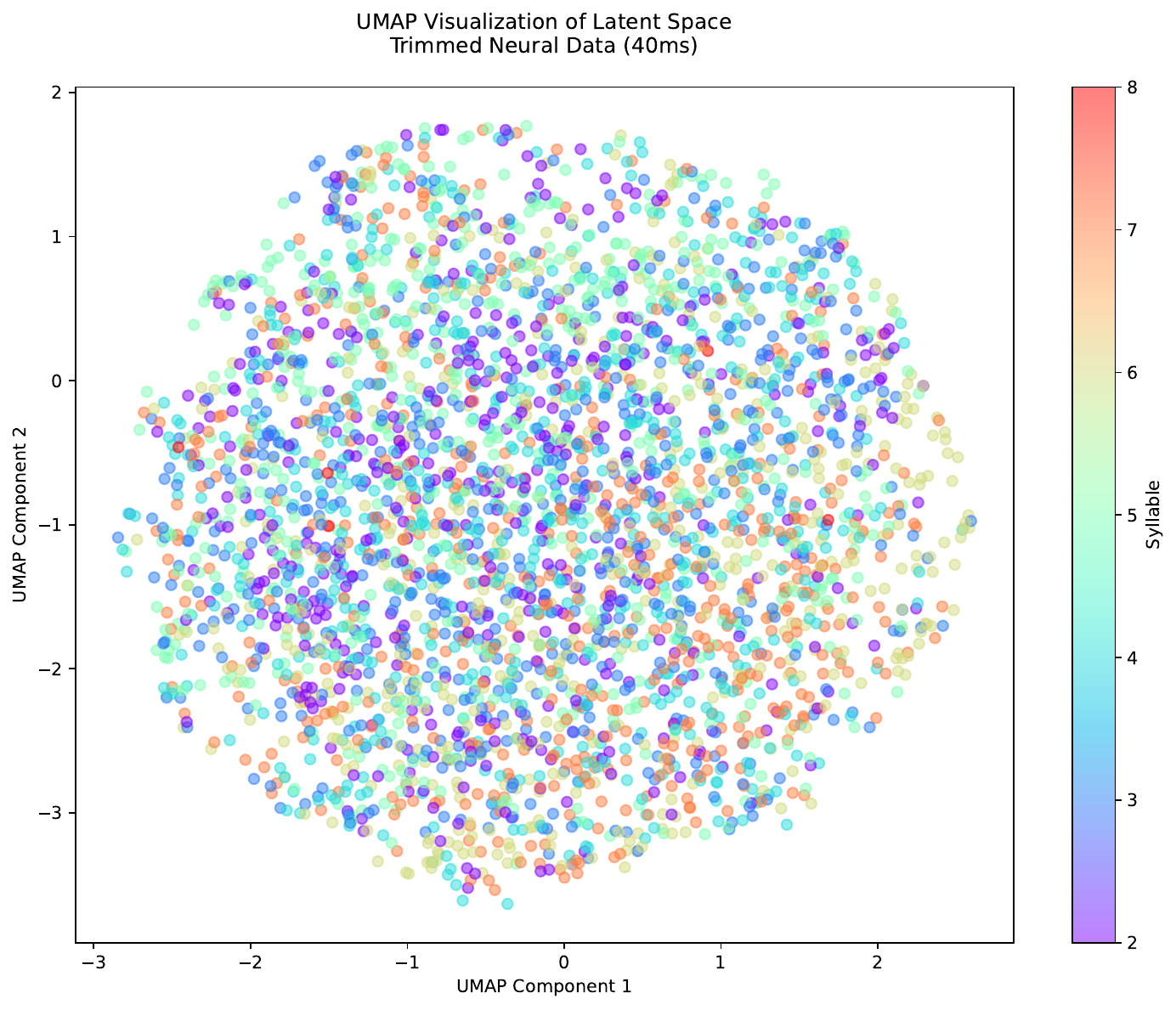}
        \caption{Neural - trimmed to 40\,ms}
        \label{fig:umap_trimmed_neural_40}
    \end{subfigure}
    \hfill
    \begin{subfigure}{0.32\linewidth}   
        \includegraphics[width=\linewidth]{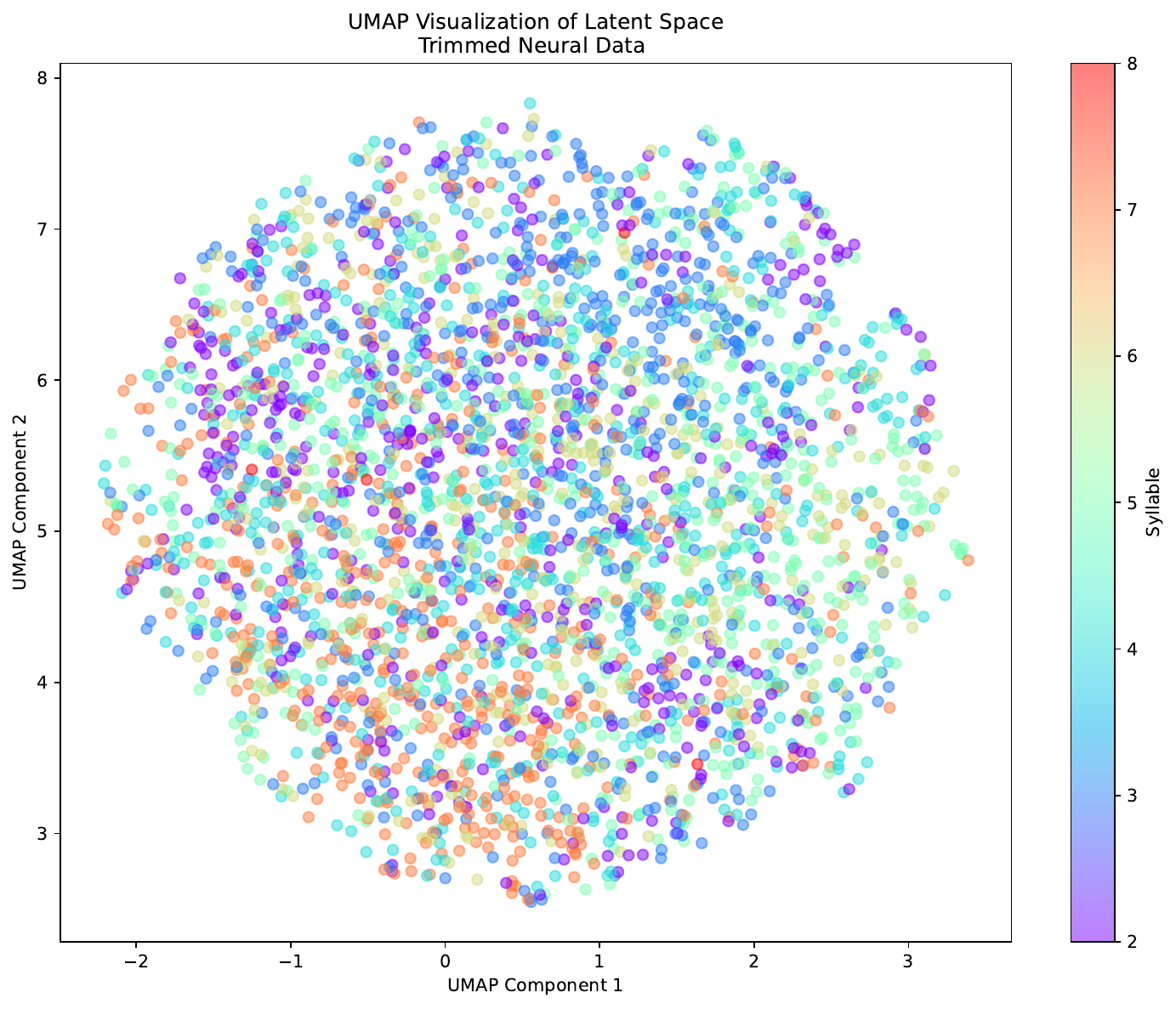}
        \caption{Neural - trimmed to 80\,ms}
        \label{fig:umap_trimmed_neural_80}
    \end{subfigure}

    \begin{subfigure}{0.32\linewidth}
        \includegraphics[width=\linewidth]{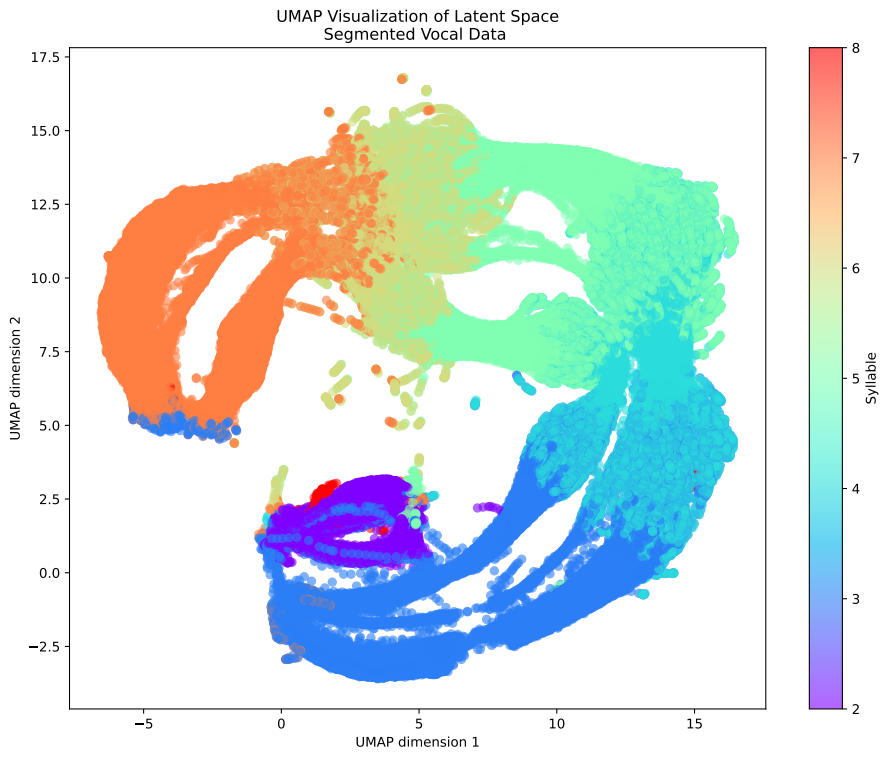}
        \caption{Vocal 40\,ms segments}
        \label{fig:umap_segmented_vocal}
    \end{subfigure}
    \hfill
    \begin{subfigure}{0.32\linewidth}
        \includegraphics[width=\linewidth]{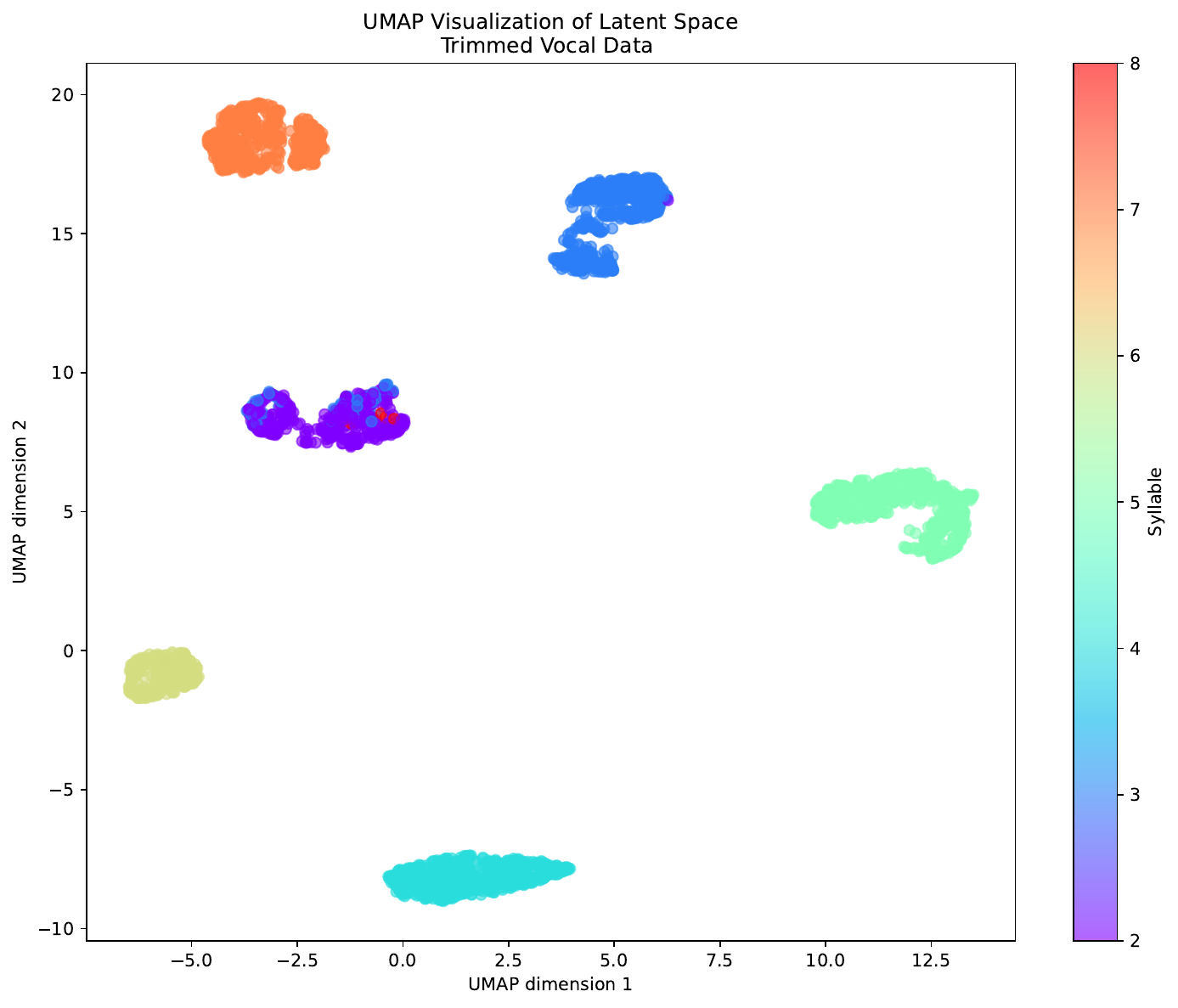}
        \caption{Vocal - trimmed to 40\,ms}
        \label{fig:umap_trimmed_vocal}
    \end{subfigure}
    \hfill
    \begin{subfigure}{0.32\linewidth}
        \phantom{\includegraphics[width=\linewidth]{images/UMAP_segmented_vocal_1dvae.jpg}}
    \end{subfigure}

    \caption{UMAP visualization of latent representations from generation experiment}
    \label{fig:umap_all_1}
\end{figure}

\begin{figure}[H]
   \centering
   \small
   
   \begin{subfigure}{0.32\linewidth}
       \includegraphics[width=\linewidth]{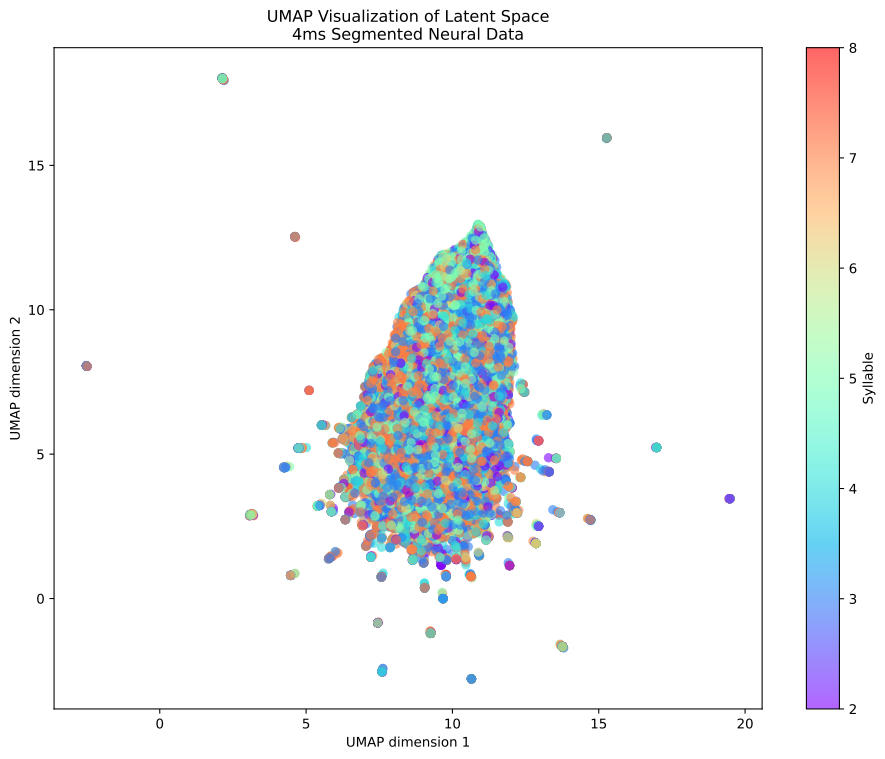}
       \caption{Neural - 4\,ms segments}
       \label{fig:umap_4ms_neural}
   \end{subfigure}
   \hfill
   \begin{subfigure}{0.32\linewidth}
       \includegraphics[width=\linewidth]{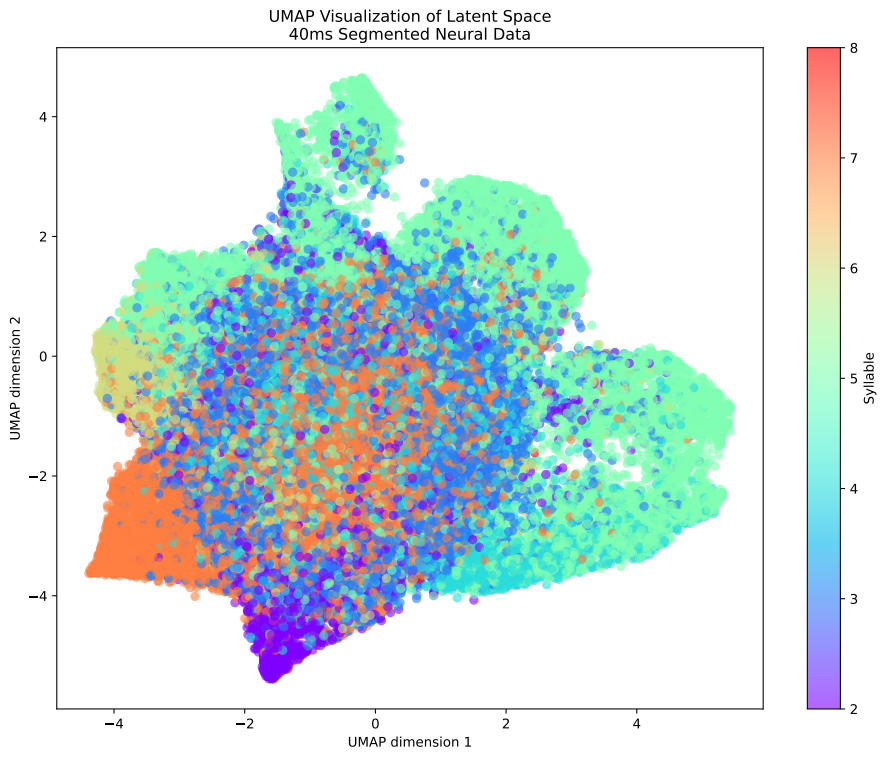}
       \caption{Neural - 40\,ms segments}
       \label{fig:umap_40ms_neural}
   \end{subfigure}
   \hfill
   \begin{subfigure}{0.32\linewidth}
       \includegraphics[width=\linewidth]{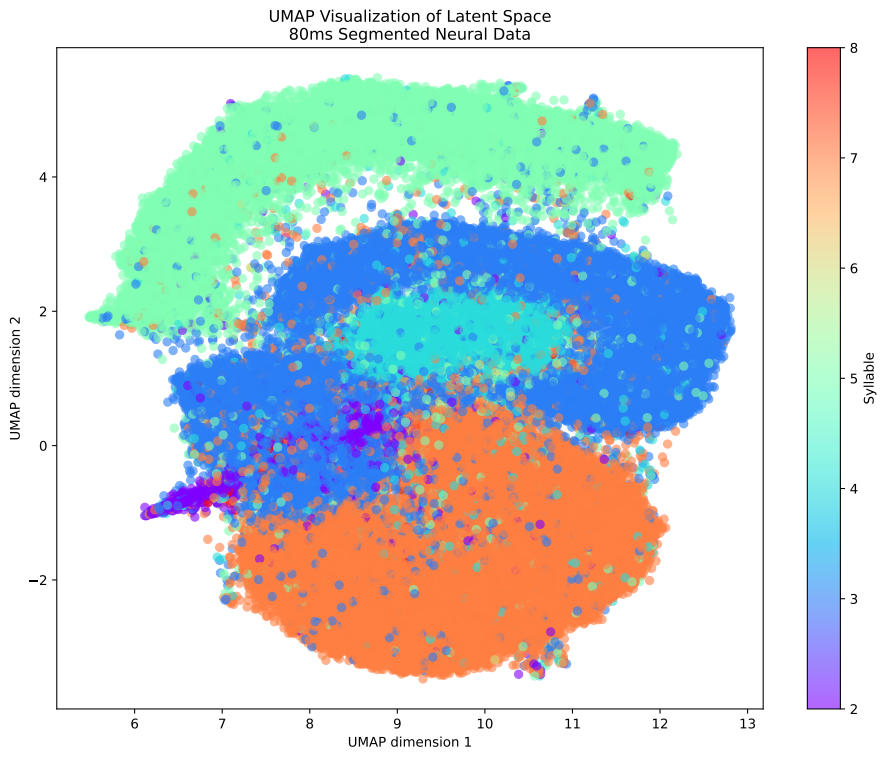}
       \caption{Neural - 80\,ms segments}
       \label{fig:umap_80ms_neural}
   \end{subfigure}

   \caption{UMAP visualization of latent representations from neuro2voc experiment}
   \label{fig:umap_neuro2voc}
\end{figure}

Figure~\ref{fig:umap_all_1} shows the UMAP projection of latent representations obtained from the generation experiment. Figure~\ref{fig:umap_neuro2voc} shows the result from the cross-modal transfer experiment. The neural representations only showed clustering effect after aligning to pre-trained vocal representations through contrastive learning. 

\begin{figure}[H]
   \centering
   \small
   \begin{subfigure}{0.32\linewidth}
       \includegraphics[width=\linewidth]{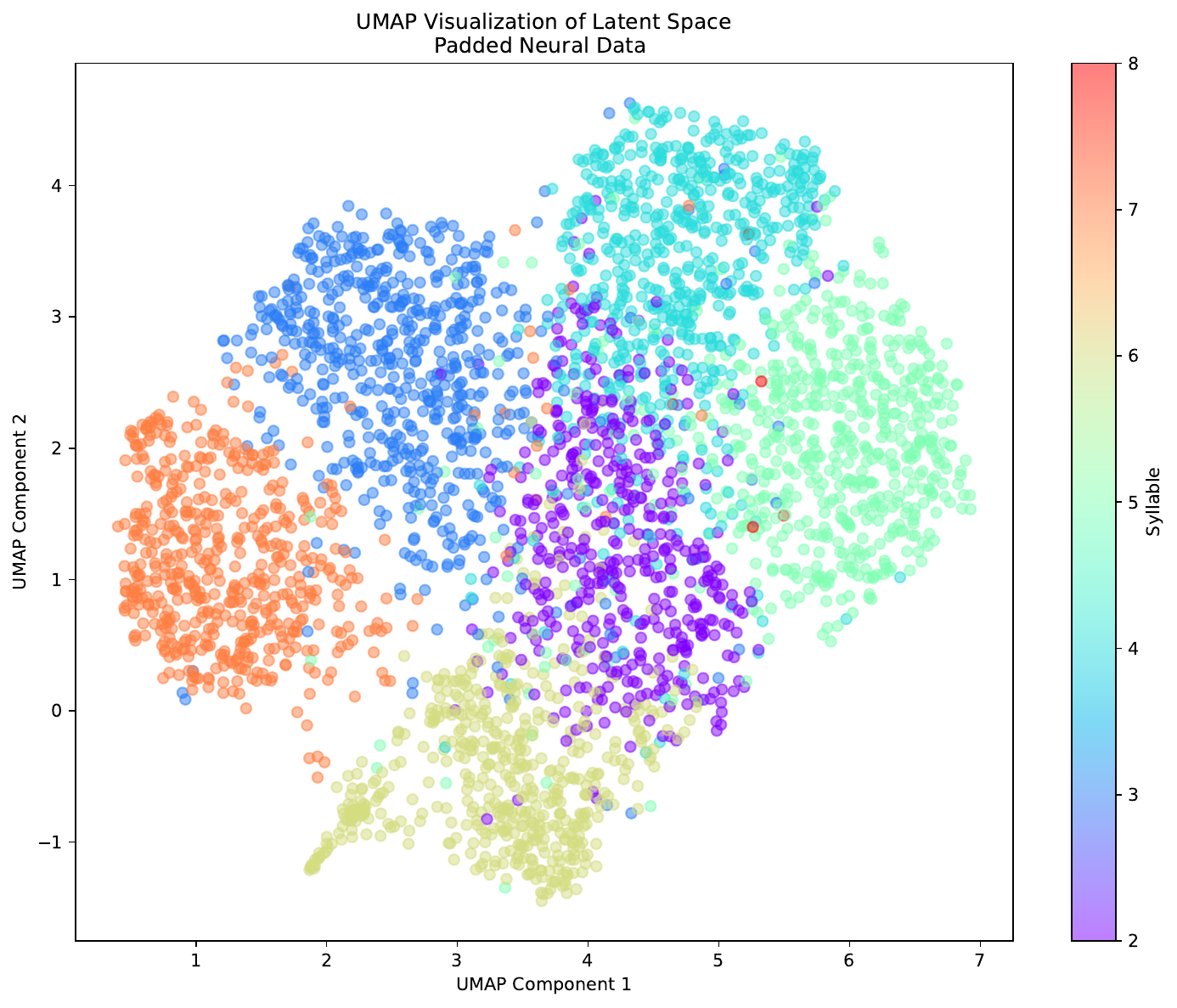}
       \caption{Neural - padded to 270\,ms}
       \label{fig:umap_padded_neural}
   \end{subfigure}
   \hfill
   \begin{subfigure}{0.32\linewidth}
       \includegraphics[width=\linewidth]{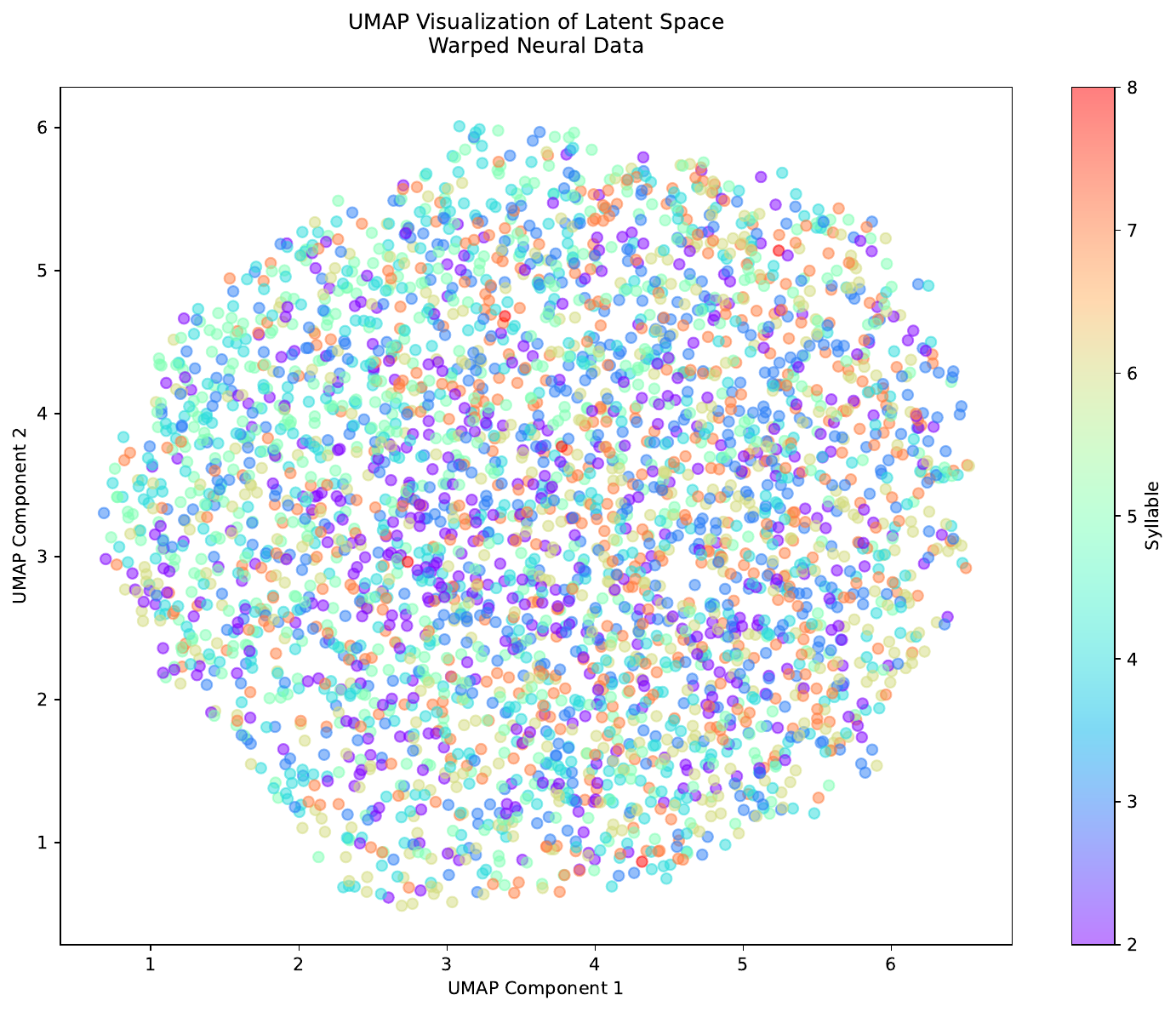}
       \caption{Neural - warped to 270\,ms}
       \label{fig:umap_warped_neural}
   \end{subfigure}
   \hfill
    \begin{subfigure}{0.32\linewidth}
        \phantom{\includegraphics[width=\linewidth]{images/UMAP_80ms_neural_joint.jpg}}
    \end{subfigure}

   \begin{subfigure}{0.32\linewidth}
        \includegraphics[width=\linewidth]{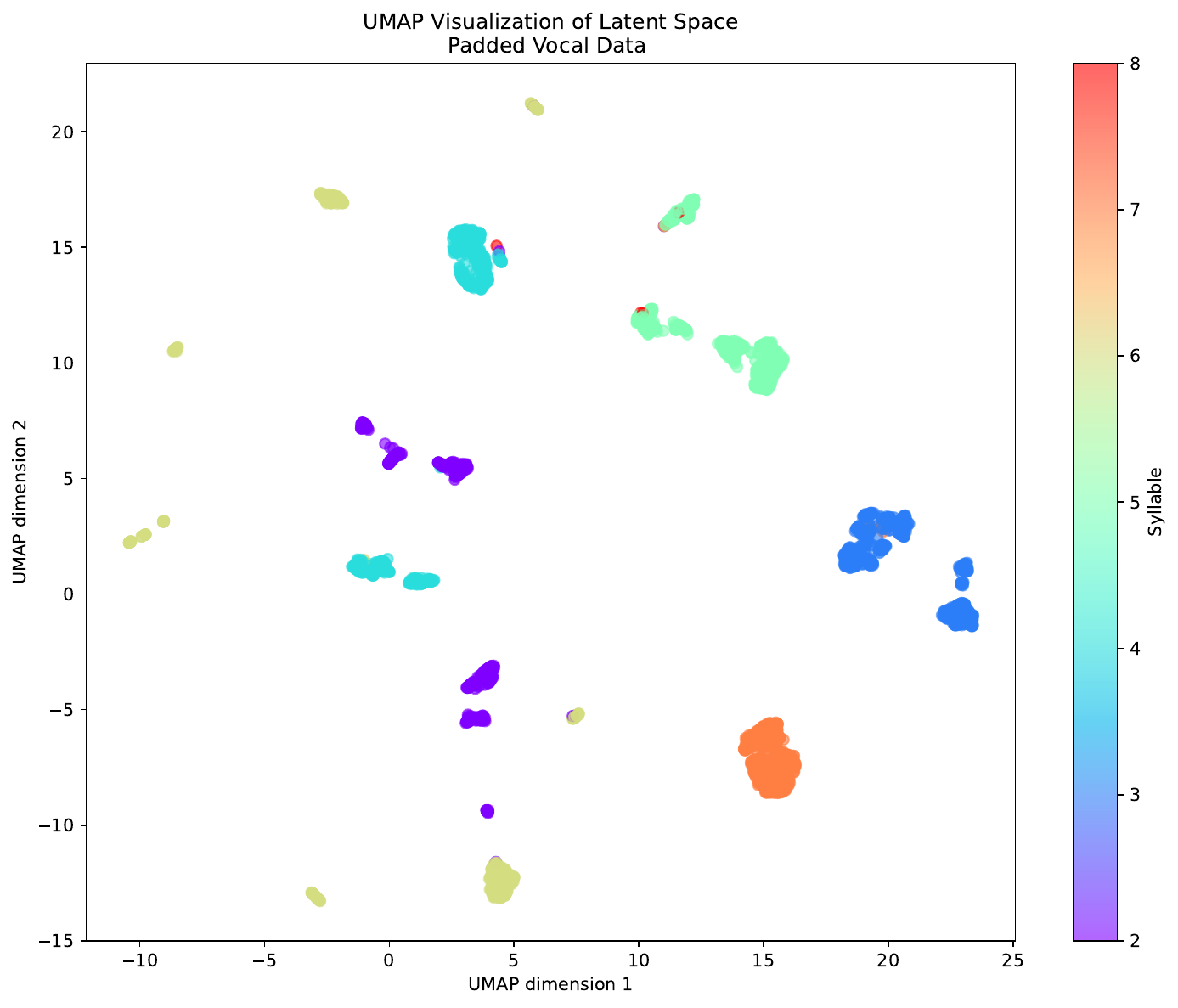}
       \caption{Vocal - padded to 224\,ms}
       \label{fig:umap_padded_vocal}
   \end{subfigure}
   \hfill
   \begin{subfigure}{0.32\linewidth}
       \includegraphics[width=\linewidth]{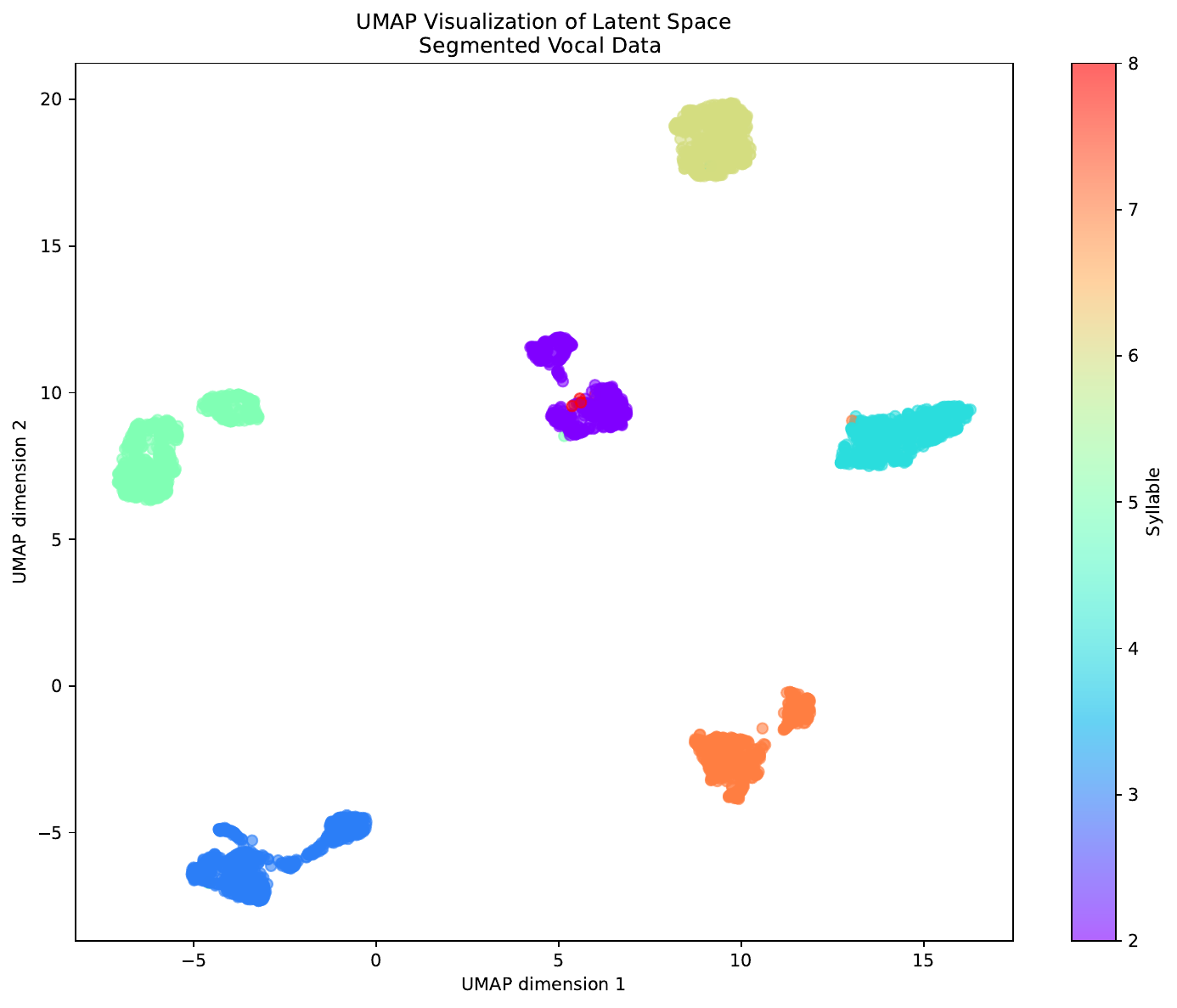}
       \caption{Vocal - warped to 224\,ms}
       \label{fig:umap_warped_vocal}
   \end{subfigure}
   \hfill
    \begin{subfigure}{0.32\linewidth}
        \phantom{\includegraphics[width=\linewidth]{images/UMAP_80ms_neural_joint.jpg}}
    \end{subfigure}

    \caption{UMAP visualization of latent representations from correlation experiment}
   \label{fig:umap_all_2}
\end{figure}

Figure~\ref{fig:umap_all_2} shows clear patterns of clustering in both vocal representations and neural representations, due to encoded temporal information during warping.

\subsection{Classification with Latent Representations}
The latent representations of both neural and vocal data were used to perform syllable classification tasks. From Table~\ref{tab:classification_accuracies}, the vocal data achieved high decoding accuracy, and thus conveyed the syllable information better. 

For the neural data, the padded and warped data in light gray have highest performance but encode temporal information. Trimmed neural data of length $40$\,ms, obtained at period $[-50\,\text{ms}, -10\,\text{ms}]$, achieved an accuracy of 49.9\,\% with SVM. This is better than the 27.7\,\% accuracy segmented neural data of the same length, which were obtained using sliding windows. The decoding efficiency of segmented neural data improved to 72.5\,\% after using contrastive learning to align with pre-trained vocal data.

\begin{table}[H]
\centering
\small
\caption{Syllable classification accuracies (\%)}
\begin{tabular}{lllrrccc}
\toprule
Data & Experiment & Method & \shortstack{Window\\(ms)} & \shortstack{Duration\\(ms)} & SVM & RF & XGBoost\\
\midrule
\multirow{8}{*}{\shortstack{Neural\\Data}} 
& Generation & Segmented & 4 & 40 & 27.7 & 25.8 & 26.8\\
& Generation & Trimmed & 10 & 40 & 49.9 & 45.5 & 48.1\\
& Generation & Trimmed & 10 & 80 & 53.5 & 45.2 & 50.1\\
& {\color{lightgray}Generation} & {\color{lightgray}Padded} &  {\color{lightgray}10} & {\color{lightgray}270} & {\color{lightgray}96.4} & {\color{lightgray}93.5} & {\color{lightgray}93.5}  \\
& {\color{lightgray}Correlation} & {\color{lightgray}Warped} &  {\color{lightgray}10} & {\color{lightgray}270} & {\color{lightgray}75.6} & {\color{lightgray}63.6} & {\color{lightgray}67.8}\\
& Contrastive & Segmented & 4 & 4 & 35.1 & 32.4 & 34.1\\
& \textBF{Contrastive} & \textBF{Segmented} & \textBF{4} & \textBF{40} & \textBF{72.5} & \textBF{65.2} & \textBF{69.3}\\
& \textBF{Contrastive} & \textBF{Segmented} & \textBF{4} & \textBF{80} & \textBF{91.0} & \textBF{84.4} & \textBF{89.6}\\
\midrule
\multirow{4}{*}{\shortstack{Vocal\\Data}}
& Generation & Segmented & 4 & 40 & 79.7 & 68.5 & 73.1\\
& Generation & Trimmed & 4 & 40 & 98.8 & 98.0 & 97.8\\
& {\color{lightgray}Correlation}& {\color{lightgray}Padded} & {\color{lightgray}4} &{\color{lightgray}224} & {\color{lightgray}96.4} &{\color{lightgray}93.5} & {\color{lightgray}93.5}\\
& {\color{lightgray}Correlation}& {\color{lightgray}Warped} & {\color{lightgray}4} &{\color{lightgray}224} & {\color{lightgray}99.4} &{\color{lightgray}99.3} & {\color{lightgray}98.8}\\
\bottomrule
\end{tabular}
\label{tab:classification_accuracies}
\end{table}


\section{Canonical Component Analysis}
\begin{figure}[H]
    \centering
    \includegraphics[width=0.7\textwidth]{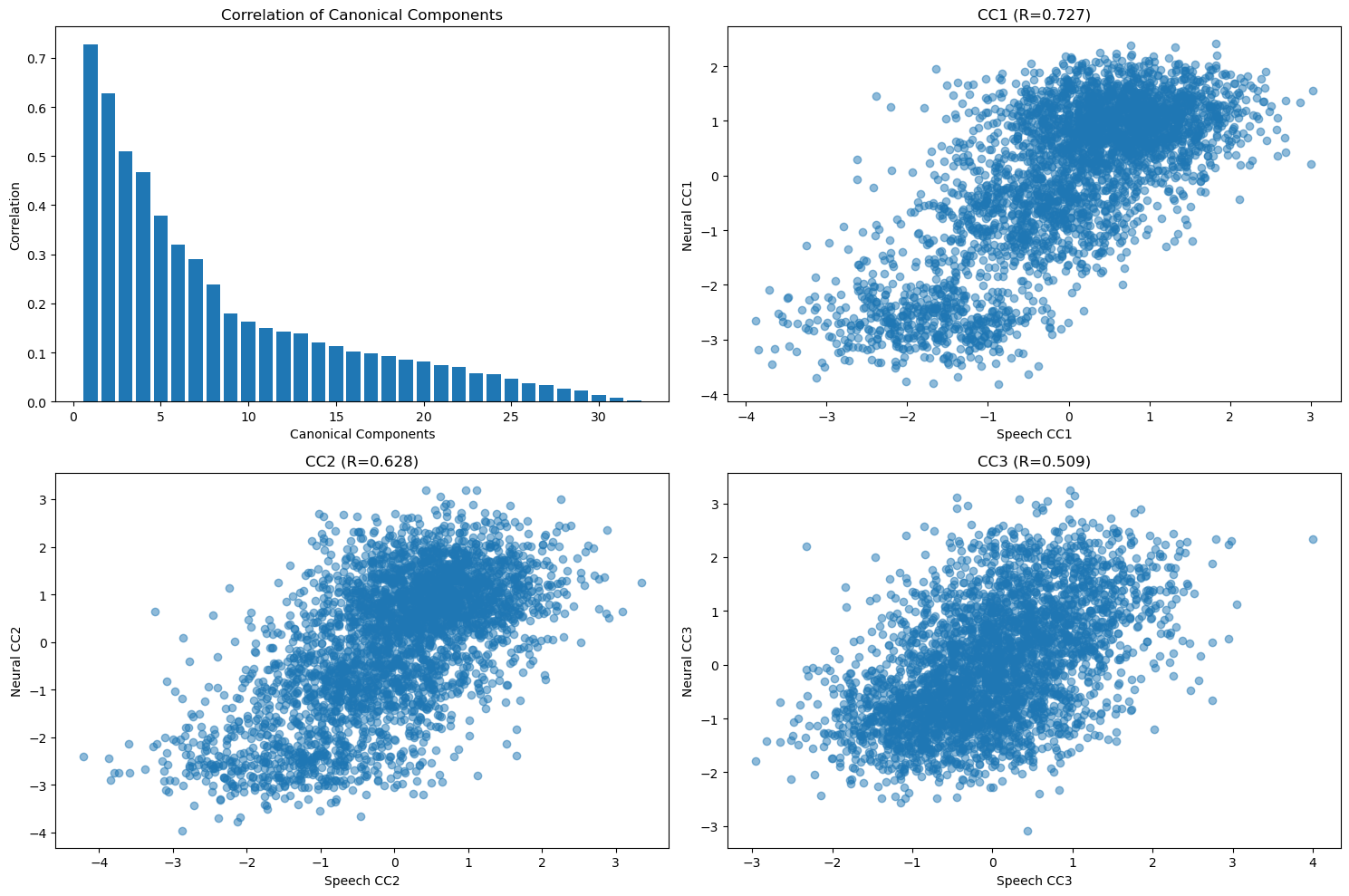}
    \caption{Result from canonical component analysis}
    \label{fig:CCA}
\end{figure}

Figure~\ref{fig:CCA} shows that the first two pairs of Canonical Components are strongly correlated ($R > 0.6$). This indicates that the variations among neural data is correlated with the variations among vocal data.

\newpage
\section{Discussion}

\subsection{Achievements of Latent Representations from VAE}
We showed different ways to process and reconstruct vocal and neural data, as well as the corresponding qualities of reconstruction. With the models and data from this project, the reconstruction quality was high for all vocal data and warped neural data, and the quality was acceptable for neural data in some modals. 

In the neuro2voc experiment, we combined contrastive learning with generative models, and successfully reconstructed vocal data using neural data. After contrastive learning methods, the decoding accuracy of segmented $40$\,ms data were enhanced from 27.7\,\% to 72.5\,\%. The highest decoding accuracy was $91.0$\,\% from neural data of length $80$\,ms, where the data segments were obtained with a sliding window.

\subsection{Limitations}
During training, the models showed different patterns for vocal and neural data. Training on vocal data exhibited clear learning progress in terms of loss reduction. In contrast, training on neural data models reached plateaux very early. This indicates two things that are essential for better reconstruction of neural data: (1) more training data, and (2) a more sophisticated VAE structure. 

\subsection{Future Directions}
A larger dataset of neural recordings is essential to enhance the model performance. We also observed that the 1D architecture performed better than the 2D model on neural data during training. The 2D VAE successfully captured relationships in spectrograms. However, it failed to detect connections among neuron populations. As EEGNet showed high decoding accuracy in Experiment~\ref{exp:3}, its structure could also be integrated into current VAE structure. These observations point to a clear direction for VAE on binned spike data.


\subfilebibliography

\chapter{Conclusion}%
\label{chap:conclusion}
\section{Achievements}
This project establishes several pilot frameworks in neural decoding and analysis. 
\subsection{Neural Activity Analysis and Classification}
We created a classification baseline using mean firing rates, achieving 77.3\,\% accuracy with SVM for syllable classification tasks. We combined XGBoost and SHAP analysis to reveal syllable-specific firing patterns and important neuron interactions. We also identified temporal relationships between LMAN burst and syllable production from the output of basic machine learning models, which aligns with neuroscience literature.

\subsection{Decoding Frameworks with New Architectures}
We developed a new framework to decode spike data, and demonstrated the potential of GPT-2 for neural decoding. We also investigated the decoding accuracy of simpler models using LSTM, CNN, Mamba-2, etc. EEGNet achieved the highest decoding accuracy (89.0\,\%). Mamba2 showed better performance when the data is sparse and long. We investigated the correlation of plateau point bin size of 1.5\,ms with the scale of AP.

\subsection{Generative Model with Contrastive Learning}
We implemented 15 models in CEBRA and achieved an accuracy as high as 75.7\,\% on syllable classification tasks. We confirmed the advantage that a contrastive learning framework offers. We also investigated the potential of VAE on all types of processed neural and vocal: segmented, trimmed, padded, warped, and in different durations. 

We used a combined contrastive-generative approach, achieving 91.0\,\% accuracy on syllable classification. The model offers the possibility to generate segmented vocal data from segmented neural data. In addition, CCA revealed strong correlations ($r=0.73$) between warped neural and vocal representations.

\section{Limitations}
This project had several limitations.

\subsection{Data Dependency and Generalization Challenges}
The decoding framework's reliance on annotated syllables creates a circular dependency, which limits the practical applications. The performance of the models on different birds was untested, and the transferability between different birds may be highly limited, as different birds have distinct syllables.
\subsection{Data Volume Constraints}
The data is insufficient for the current models to converge. A larger dataset would yield higher decoding accuracy of syllables or better reconstruction quality of neural data.
\subsection{Biological Plausibility}
The data came from the LMAN region, which primarily controls plasticity rather than production. Data from nuclei along the motor pathway would yield more accurate and interpretable results for decoding tasks.

\section{Future Direction}
The project establishes methodological foundations for future research, emphasizing the balance between computational performance, model interpretability, and biological plausibility through its achievements and limitations.

\subsection{Syllable-Based Approaches}
With data recorded from the motor pathway, the SHAP interaction values can be used to visualize how neurons integrate information and control the output syllables. 

Depending on the availability of data, a pre-trained model based on open source models like GPT-2 on multiple birds and syllables could be developed. With a pre-trained model, latent embeddings could serve as a target of interpretability analysis.

Mamba2 showed potential for decoding long and sparse spike trains. EEGNet and Mamba2 can be further investigated and developed in this direction to decode spike data in 20\,kHz or 30\,kHz with higher accuracies.

\subsection{Representation Learning}
Decoding neural activities with the CEBRA method using pretrained latent embeddings (like wav2vec) would improve interpretability on a finer scale. The CEBRA method could also be combined into the current joint VAE to better align neural representations to vocal representations in VAE. VAE framework itself could be further improved with statistical inference methods to better capture the patterns in the neural data.

\subsection{NeuroAI}
Recent advances in neuroscience and AI have given rise to a new field known as NeuroAI \citep{Benchetrit2023-pn, Conwell2023-ts, Francl2022-tm, Yamins2014-sn, Yamins2016-jf}. While current NeuroAI research focused primarily on visual perception, our project suggest promising direction of extending the work of NeuroAI to motor vocalization. Generative models can be built with song circuit inspired neural network architectures, while simultaneous recordings from the motor pathway and AFP could be used as inputs. Future work can perform ablation studies on layers representing LMAN, and evaluate the neural network by observing the variability in the reconstructed syllable.

\subfilebibliography

\renewcommand{\bibsection}{}

\phantomsection%
\addcontentsline{toc}{chapter}{References}%
\chapter*{References}
\begin{flushleft}
\bibliography{biblio/thesis, biblio/paperpile}
\end{flushleft}
\appendix


\chapter{Appendix}%

\label{appendix:Appendix}
\section{Neuron Depths}
\begin{table}[htbp]
\centering
\begin{tabular}{ll||ll||ll||ll}
\toprule
Neuron & Depth & Neuron & Depth & Neuron & Depth & Neuron & Depth \\
\midrule
N1  & 3380 & N20 & 2640 & N39 & 2460 & N58 & 2140 \\
N2  & 3220 & N21 & 2600 & N40 & 2440 & N59 & 2100 \\
N3  & 3200 & N22 & 2600 & N41 & 2440 & N60 & 1880 \\
N4  & 3180 & N23 & 2600 & N42 & 2420 & N61 & 1820 \\
N5  & 2980 & N24 & 2600 & N43 & 2400 & N62 & 1680 \\
N6  & 2960 & N25 & 2580 & N44 & 2380 & N63 & 1680 \\
N7  & 2880 & N26 & 2580 & N45 & 2380 & N64 & 1340 \\
N8  & 2860 & N27 & 2580 & N46 & 2360 & N65 & 1320 \\
N9  & 2820 & N28 & 2580 & N47 & 2360 & N66 & 1320 \\
N10 & 2740 & N29 & 2580 & N48 & 2340 & N67 & 1120 \\
N11 & 2720 & N30 & 2560 & N49 & 2320 & N68 & 1080 \\
N12 & 2720 & N31 & 2540 & N50 & 2300 & N69 & 1060 \\
N13 & 2700 & N32 & 2540 & N51 & 2280 & N70 & 1060 \\
N14 & 2680 & N33 & 2520 & N52 & 2280 & N71 & 840  \\
N15 & 2680 & N34 & 2520 & N53 & 2260 & N72 & 660  \\
N16 & 2660 & N35 & 2500 & N54 & 2240 & N73 & 480  \\
N17 & 2660 & N36 & 2480 & N55 & 2220 & N74 & 480  \\
N18 & 2640 & N37 & 2480 & N56 & 2180 & N75 & 200  \\
N19 & 2640 & N38 & 2460 & N57 & 2160 &     &      \\
\bottomrule
\end{tabular}
\caption{Depth ($\mu$m) to probe tip for all neurons used in experiment~\ref{exp:1}}
\label{tab:neuron_depths}
\end{table}

\section{Neural Information Integration}
\label{appendix:integration}
\begin{figure}[H]
    \centering
    \includegraphics[height=0.9\textheight]{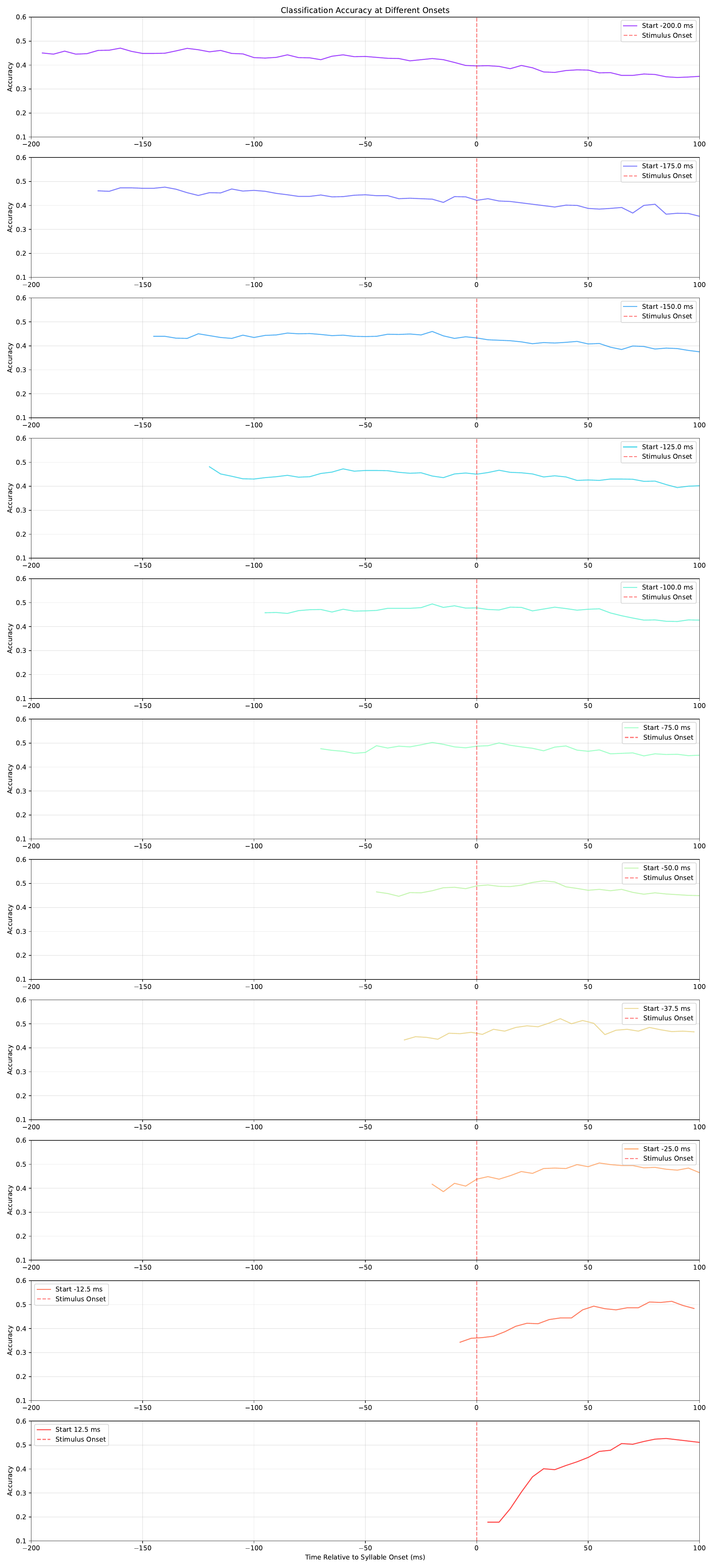}
    \caption{Classification accuracy at different onsets}
    \label{fig:neural_integration}
\end{figure}

\section{CEBRA Results}
\label{appendix:CEBRA}
\begin{figure}[htbp]
    \centering
    \includegraphics[width=0.9\textwidth]{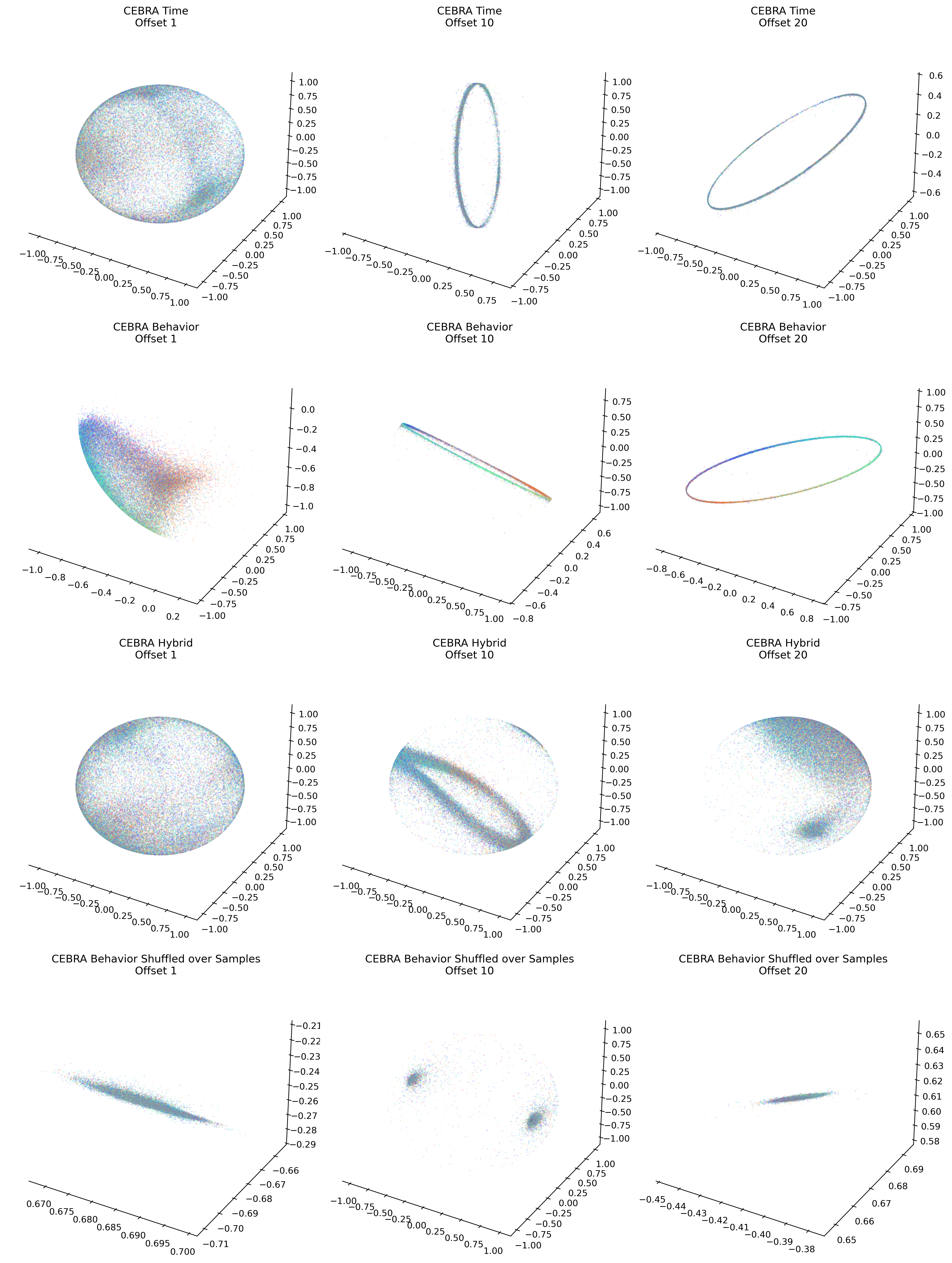}
    \caption{Extra CEBRA embedding visualization from different parameters}
    \label{fig:all_cebra}
\end{figure}

\begin{figure}[H]
    \centering
    \includegraphics[width=0.9\textwidth]{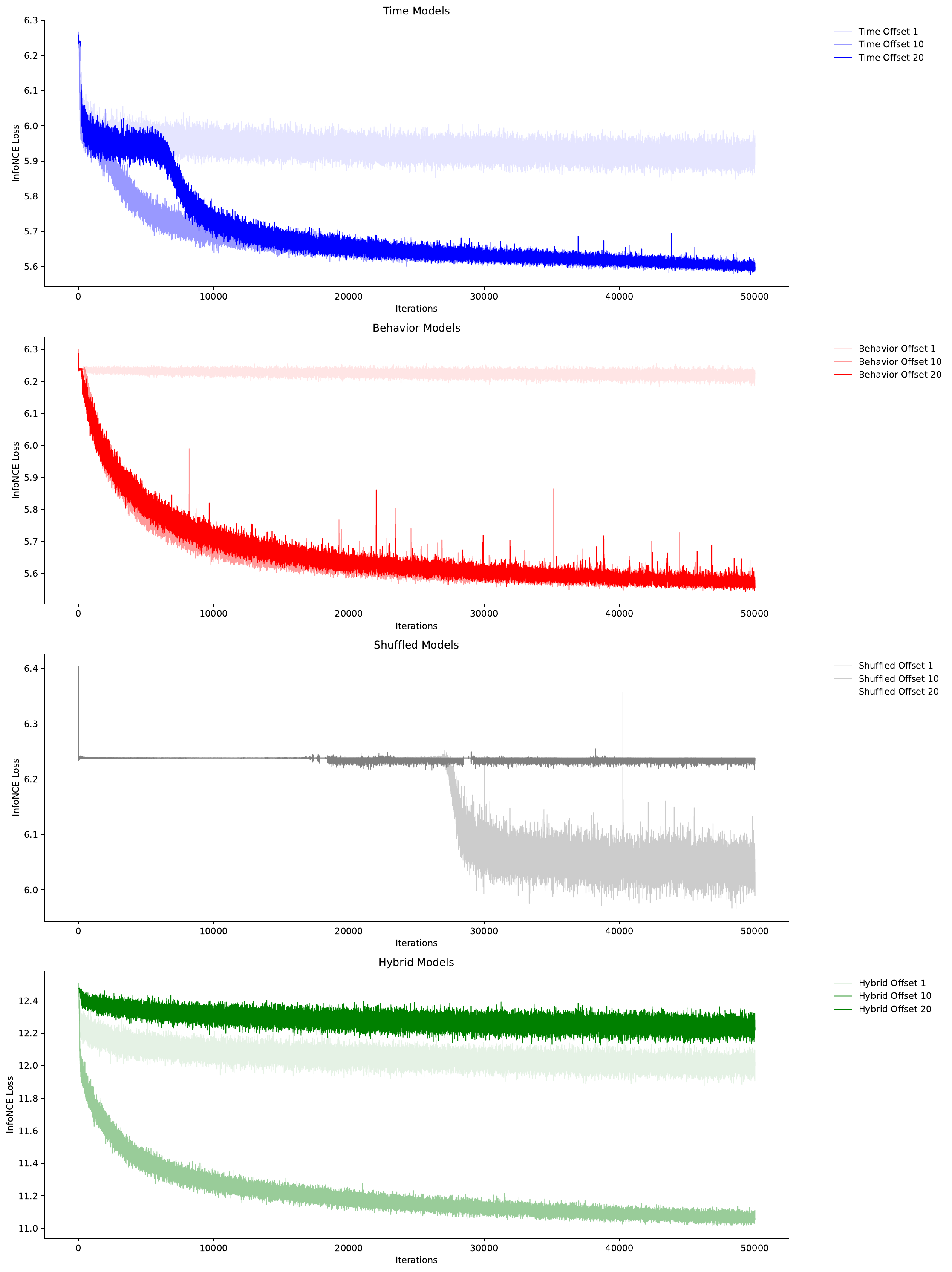}
    \caption{CEBRA training loss}
    \label{fig:cebra_loss}
\end{figure}

\begin{figure}[H]
    \centering
    \includegraphics[width=0.45\textwidth]{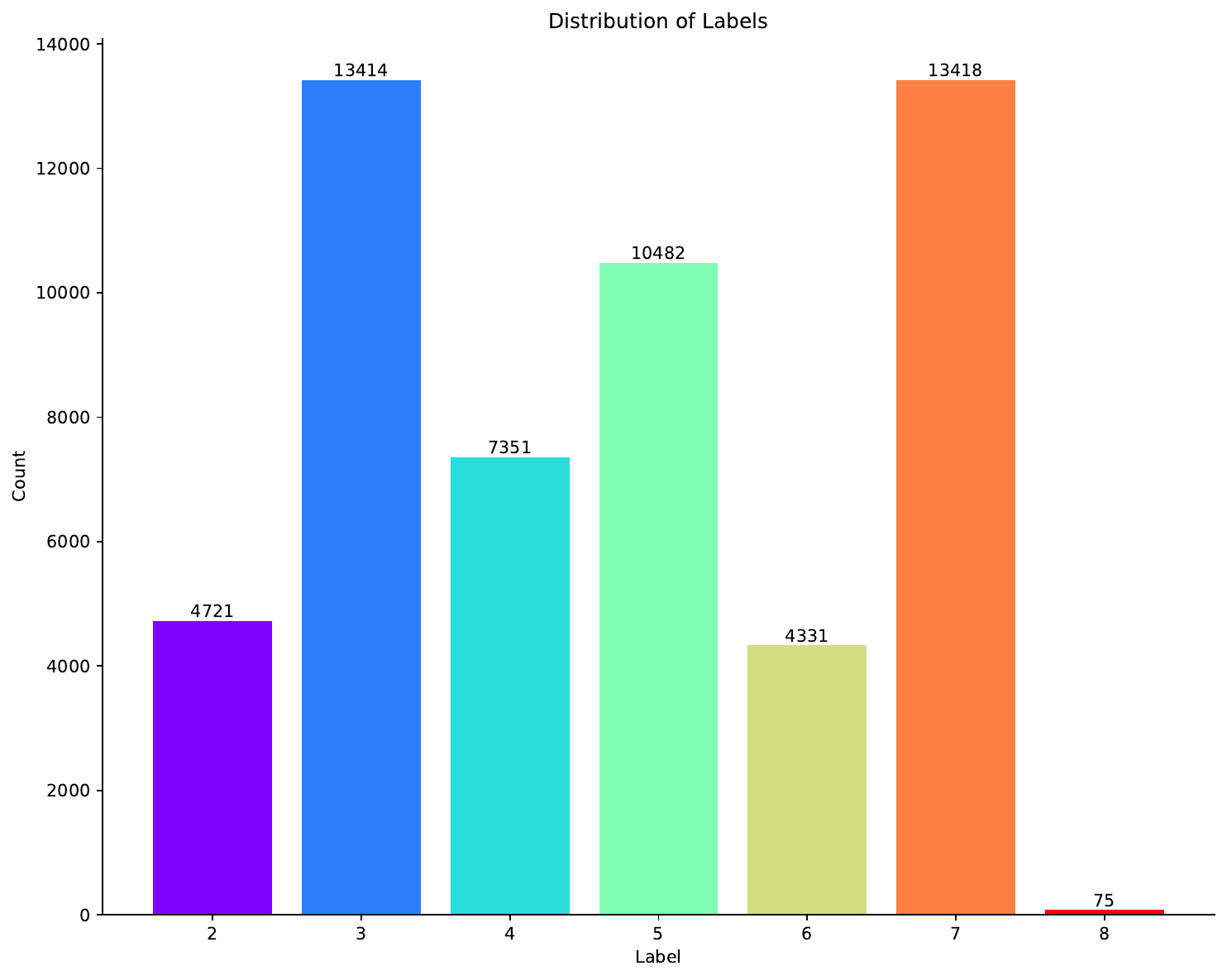}
    \caption{Data distribution in CEBRA}
    \label{fig:cebra_labels}
\end{figure}

\section{VAE Results}
\label{appendix:VAE}
\begin{figure}[H]
   \begin{subfigure}[b]{0.32\textwidth}
       \centering
       \includegraphics[width=\textwidth]{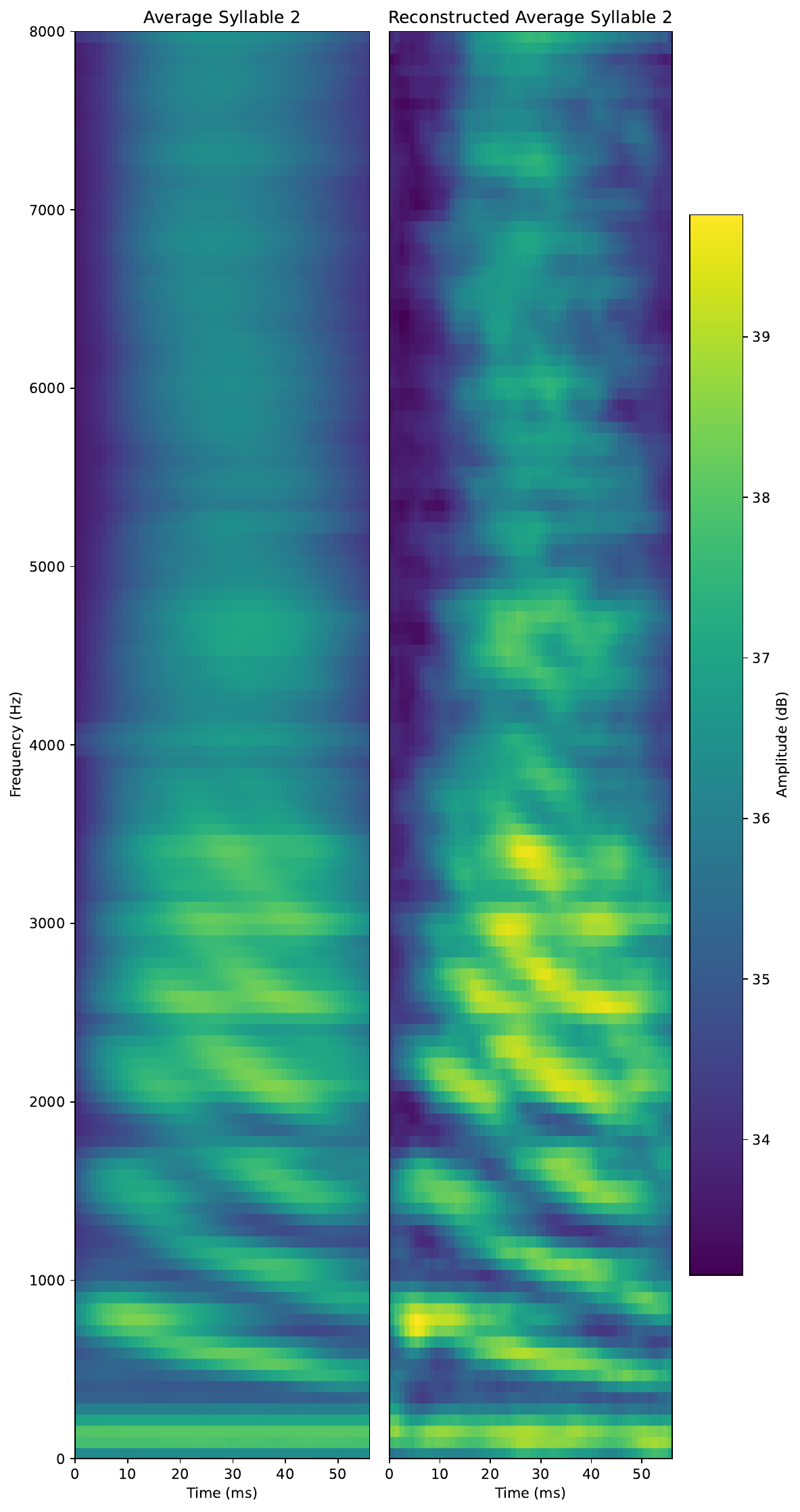}
       \caption{Syllable 2}
       \label{fig:syllable_2}
   \end{subfigure}
   \hfill
   \begin{subfigure}[b]{0.32\textwidth}
       \centering
       \includegraphics[width=\textwidth]{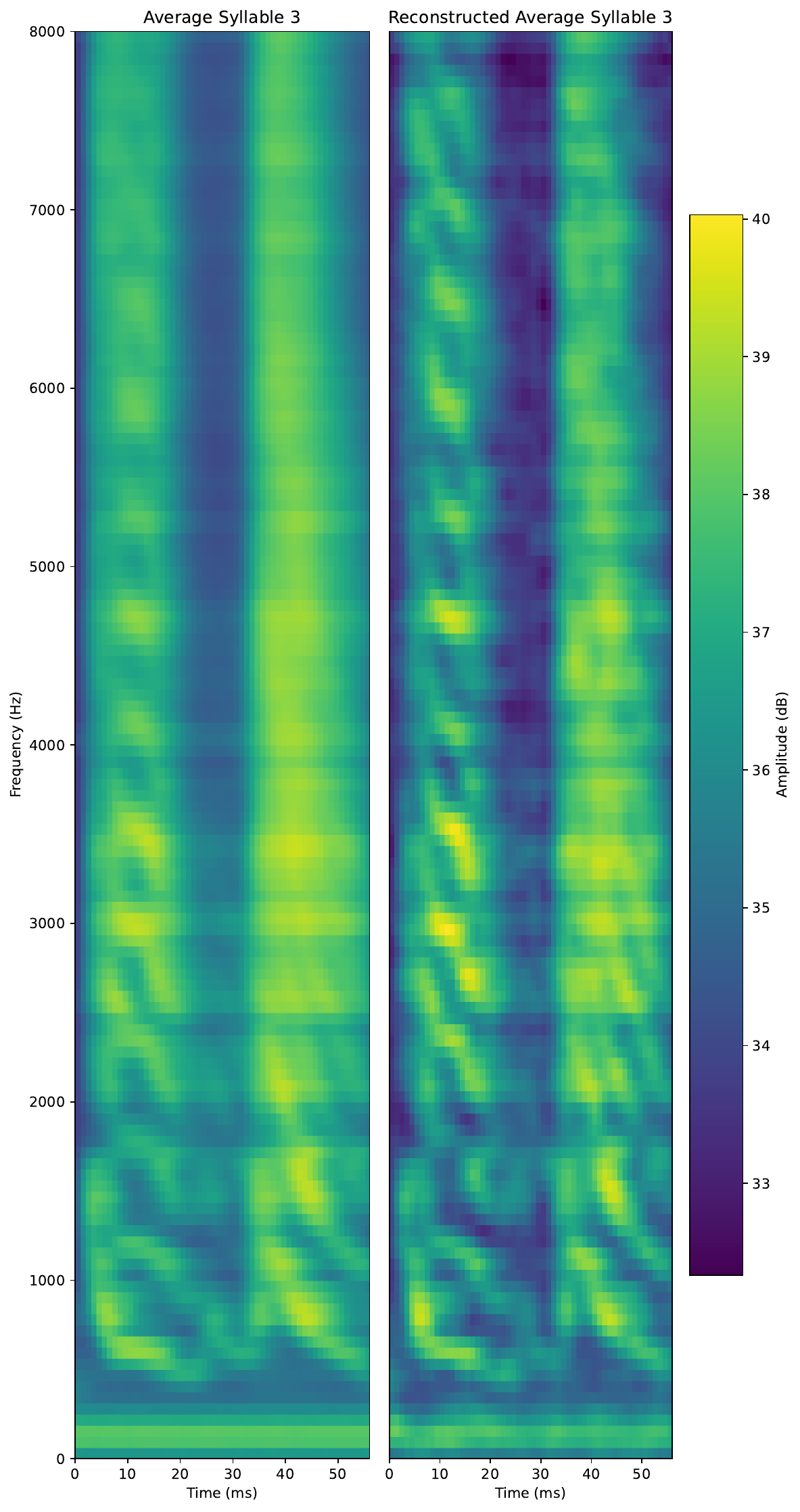}
       \caption{Syllable 3}
       \label{fig:syllable_3}
   \end{subfigure}
   \hfill
   \begin{subfigure}[b]{0.32\textwidth}
       \centering
       \includegraphics[width=\textwidth]{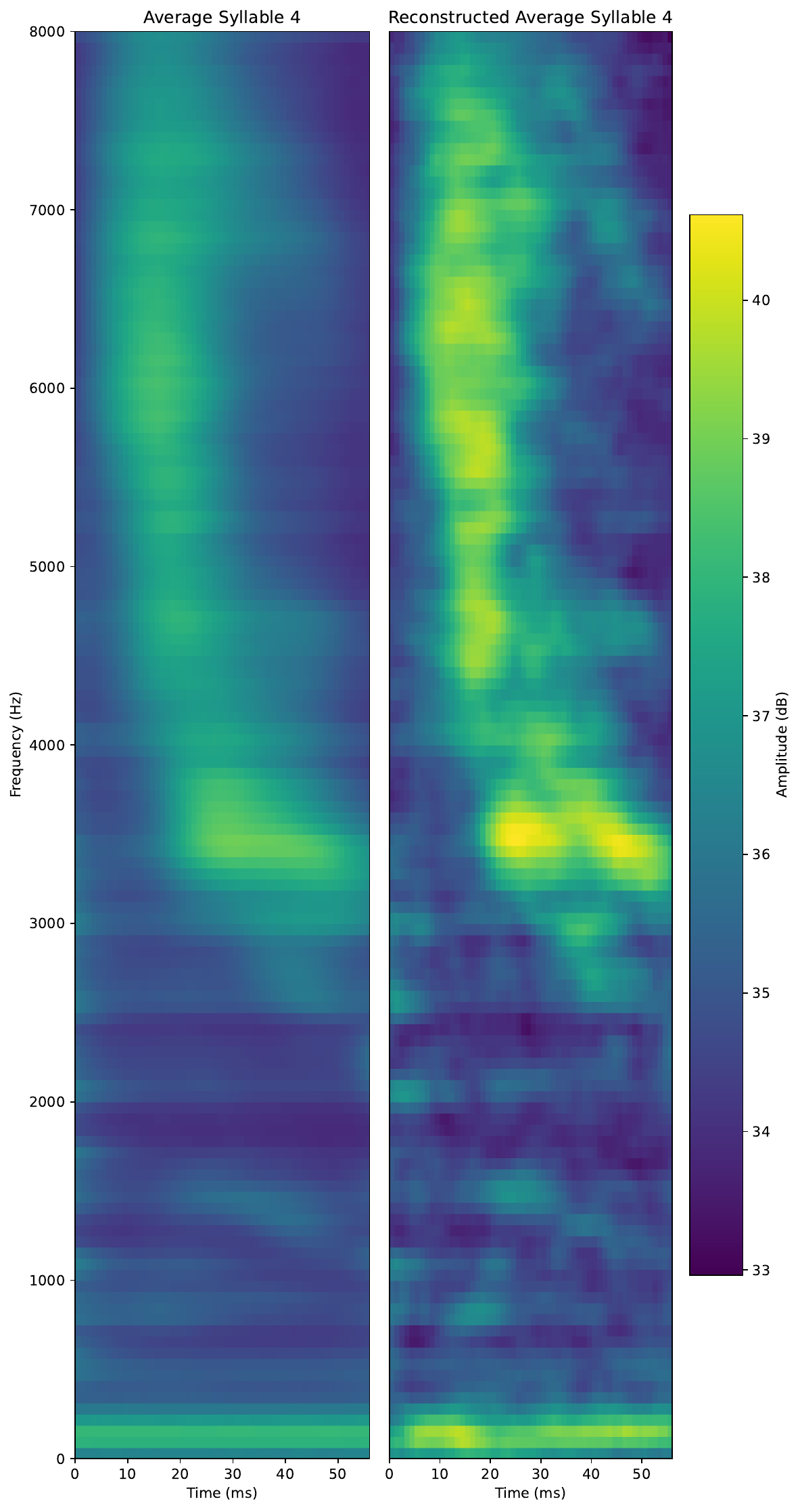}
       \caption{Syllable 4}
       \label{fig:syllable_4}
   \end{subfigure}
\end{figure}
\begin{figure}[H]
   \begin{subfigure}[b]{0.32\textwidth}
       \centering
       \includegraphics[width=\textwidth]{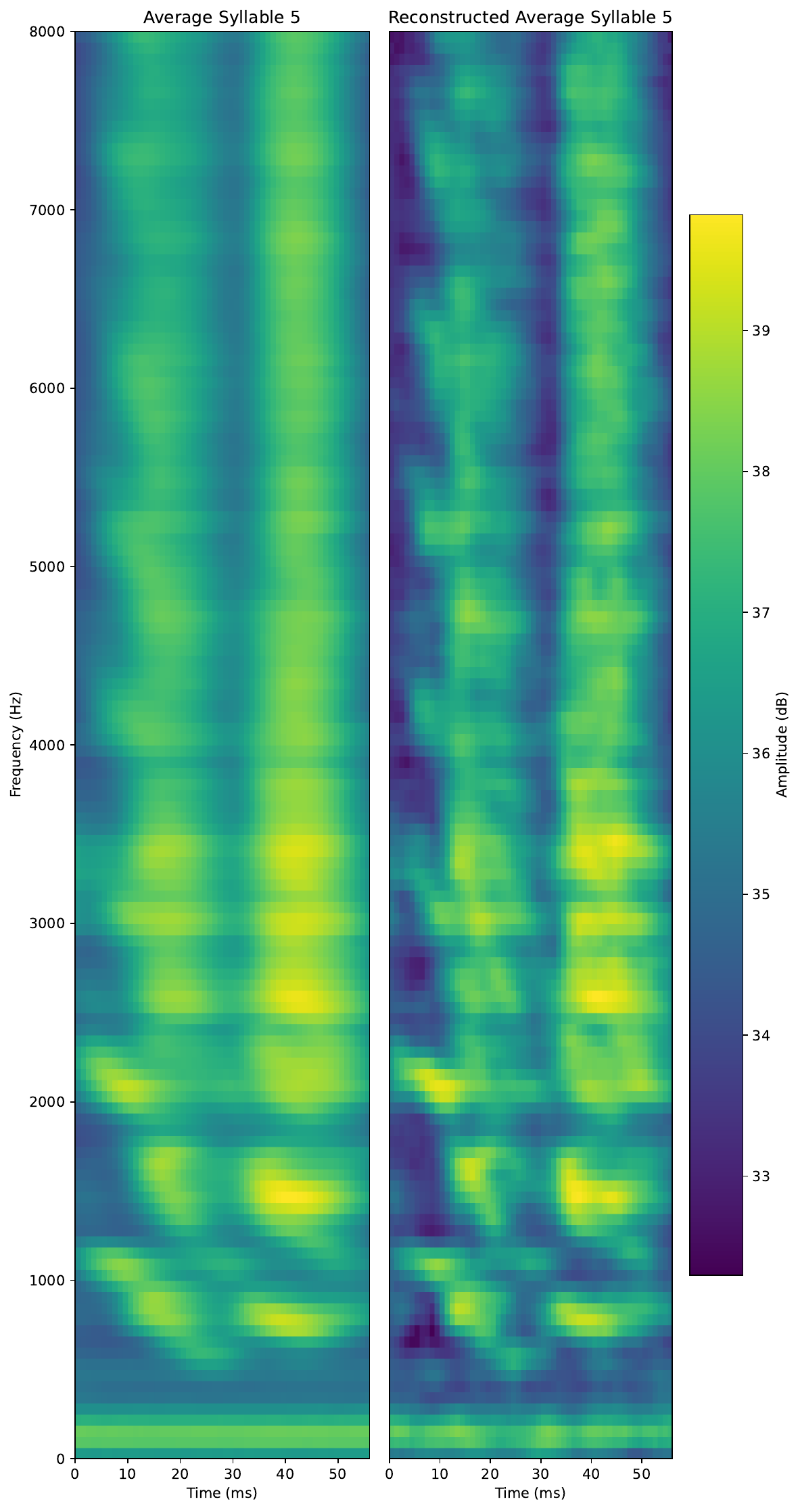}
       \caption{Syllable 5}
       \label{fig:syllable_5}
   \end{subfigure}
   \hfill
   \begin{subfigure}[b]{0.32\textwidth}
       \centering
       \includegraphics[width=\textwidth]{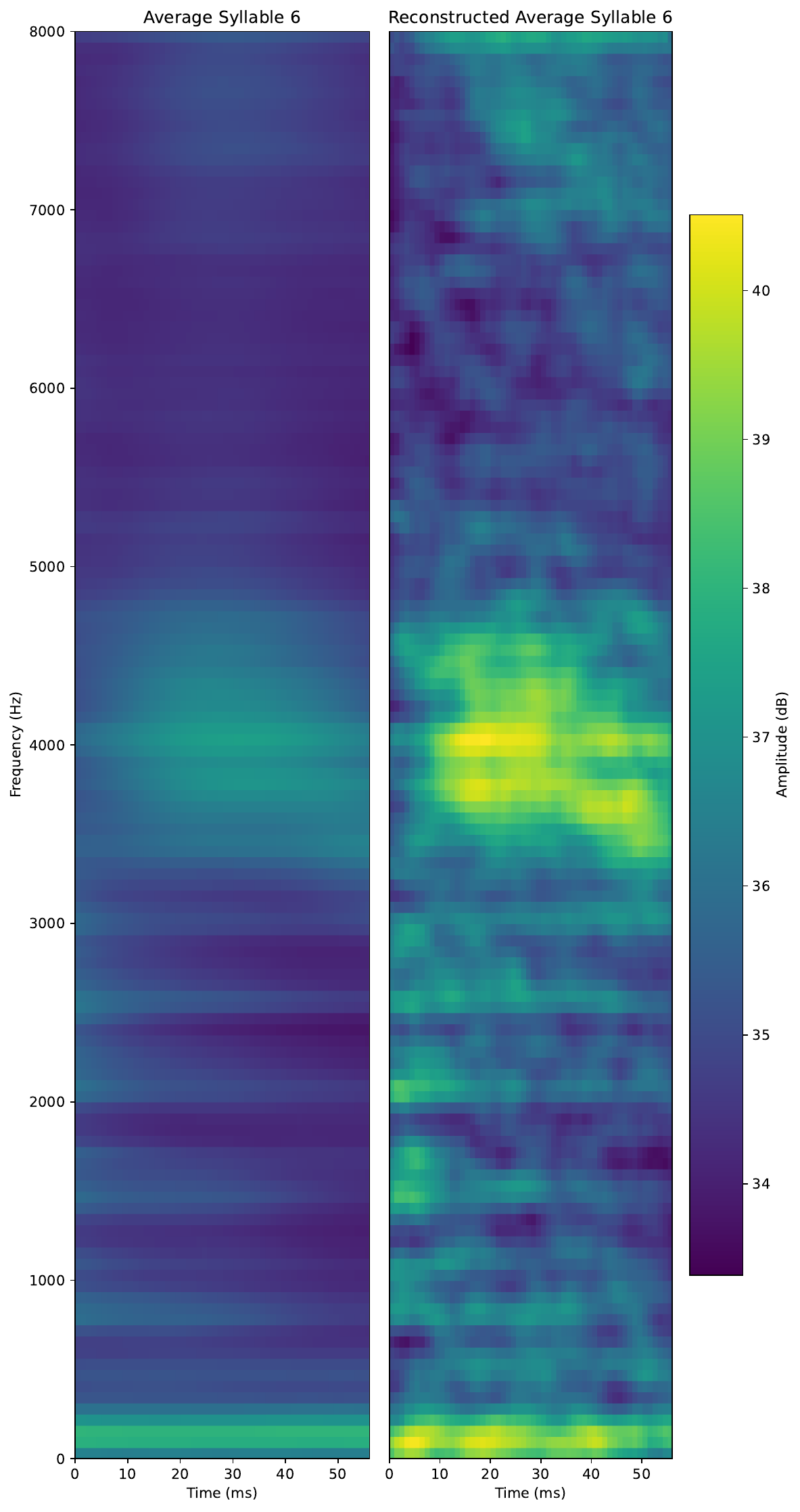}
       \caption{Syllable 6}
       \label{fig:syllable_6}
   \end{subfigure}
   \hfill
   \begin{subfigure}[b]{0.32\textwidth}
       \centering
       \includegraphics[width=\textwidth]{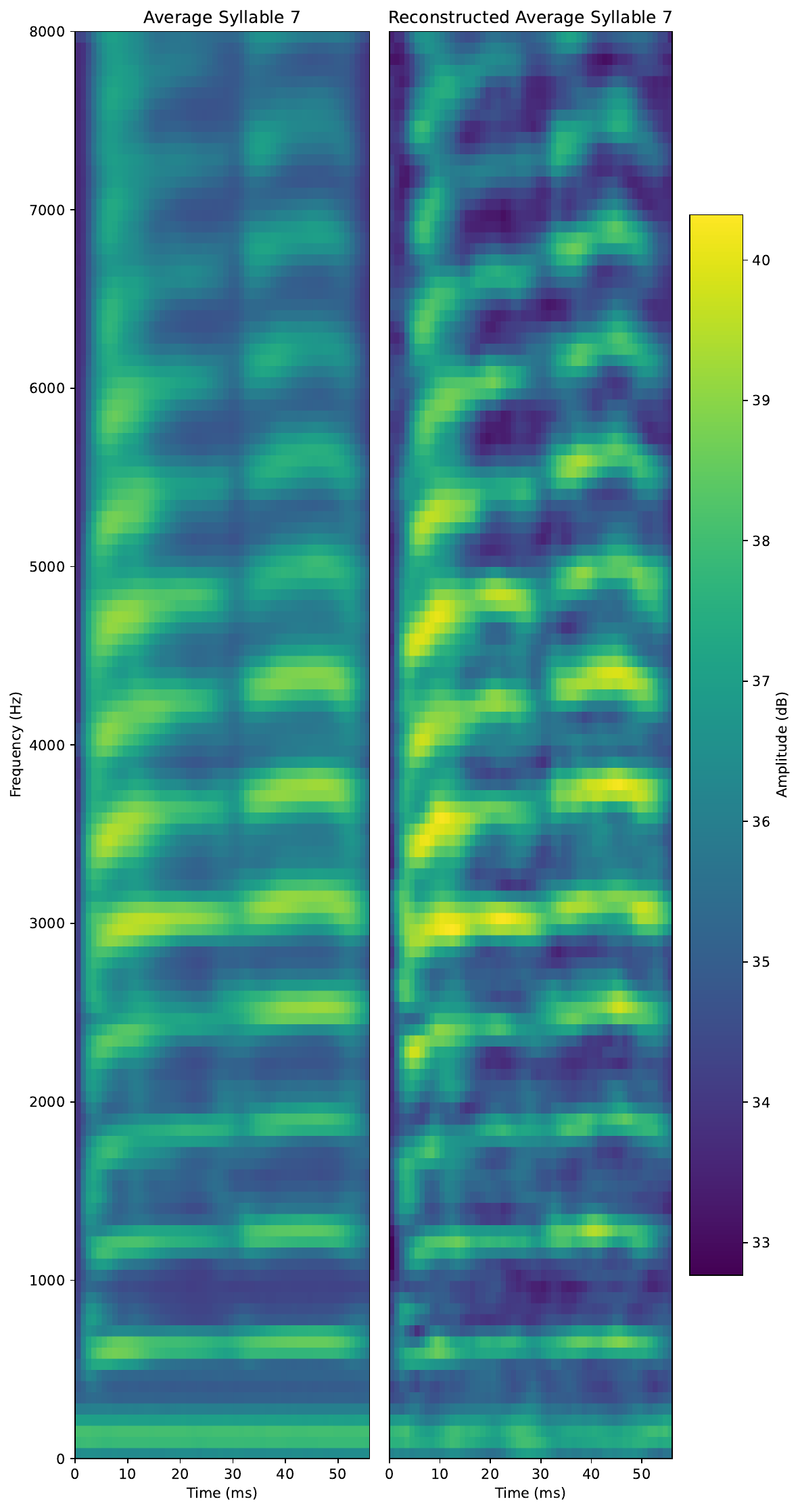}
       \caption{Syllable 7}
       \label{fig:syllable_7}
   \end{subfigure}

   \begin{subfigure}[b]{0.32\textwidth}
       \centering
       \includegraphics[width=\textwidth]{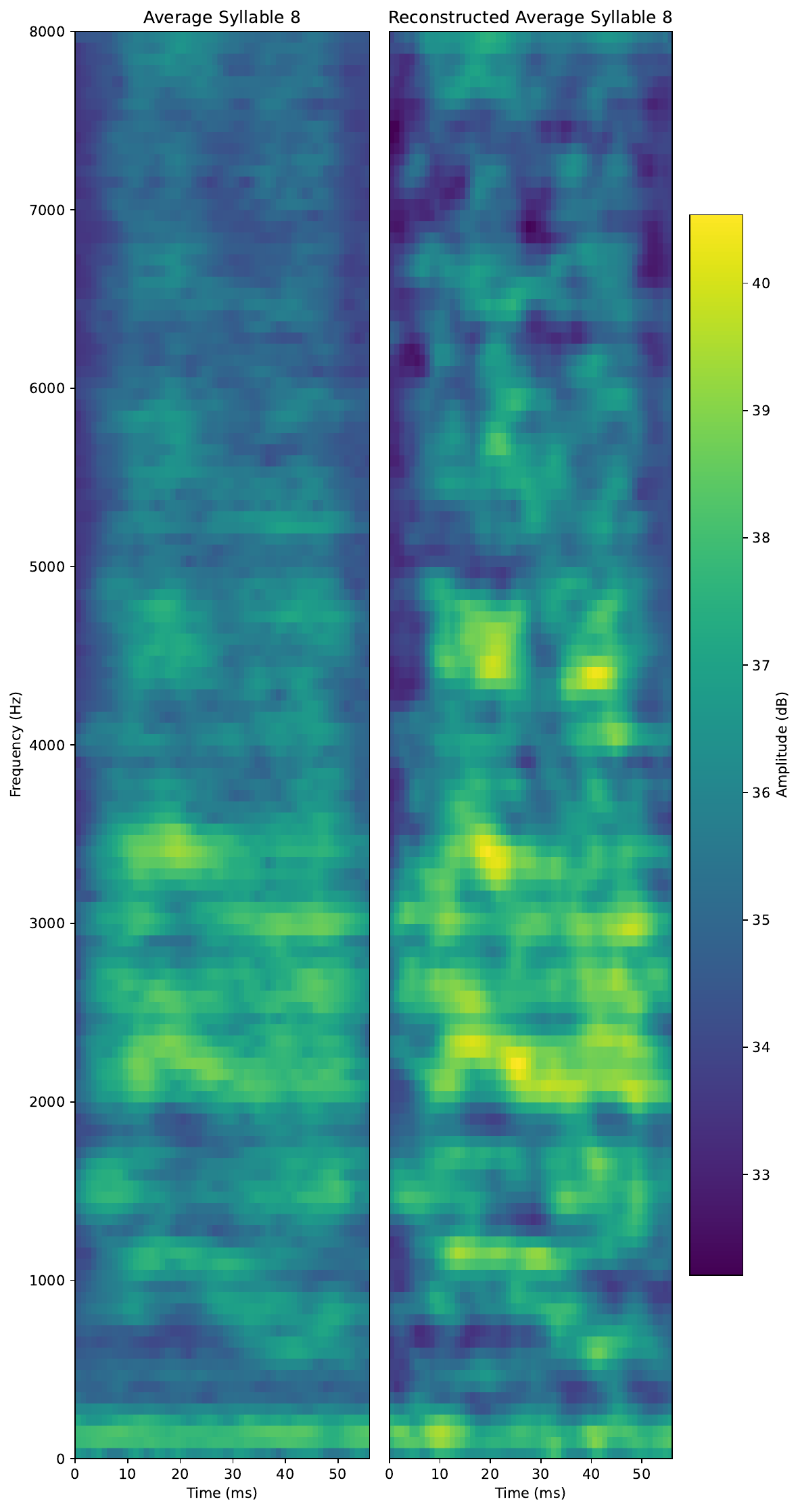}
       \caption{Syllable 8}
       \label{fig:syllable_8}
   \end{subfigure}
   
   \caption{Original and reconstruction syllables of a motif}
   \label{fig:all_syllables}
\end{figure}

\begin{figure}[H]
    \centering
    \includegraphics[width=\linewidth]{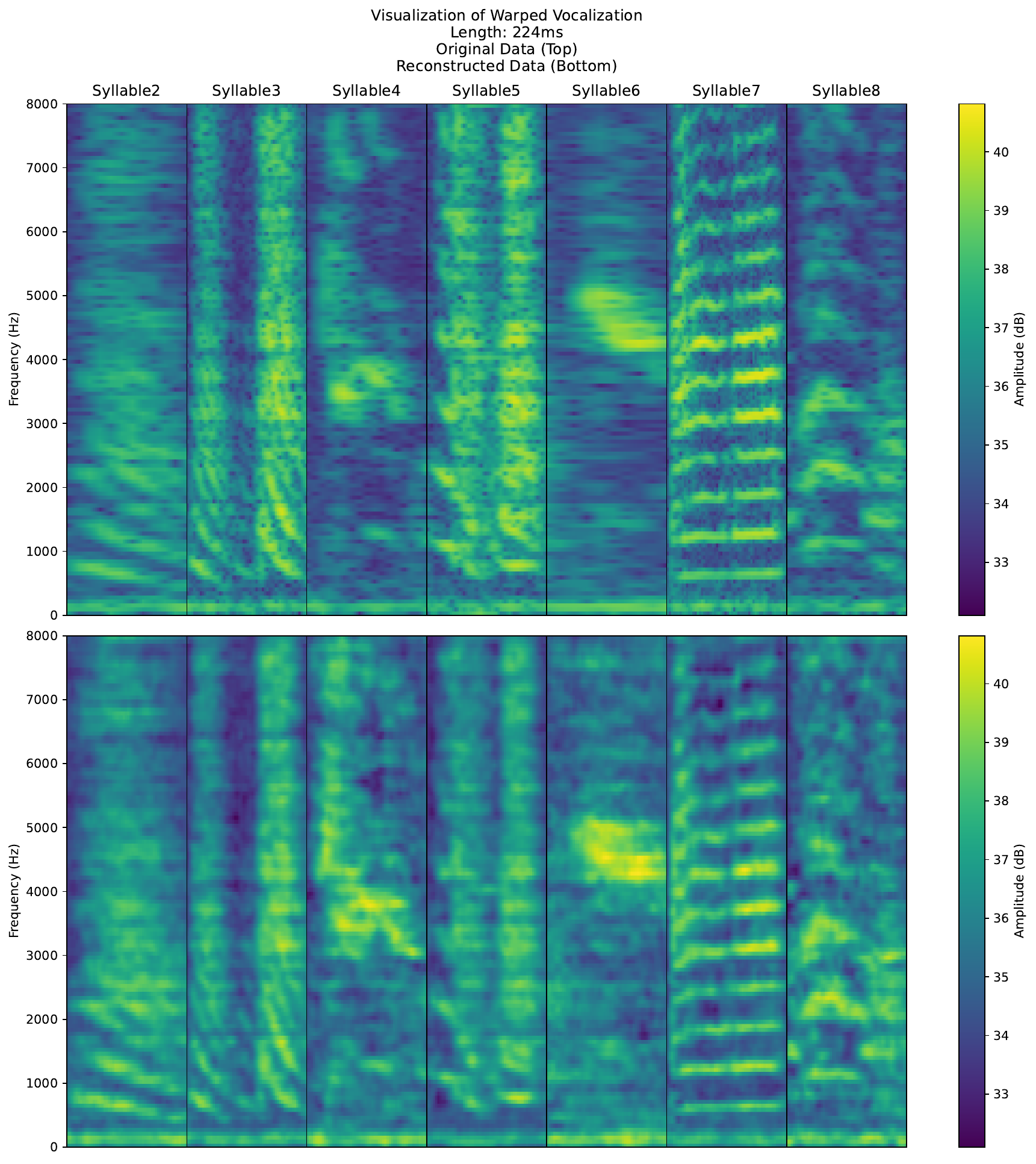}
    \caption{Reconstruction of warped vocal data}
    \label{fig:whole_motif}
\end{figure}

\begin{figure}[H]
    \centering
    \includegraphics[width=\linewidth]{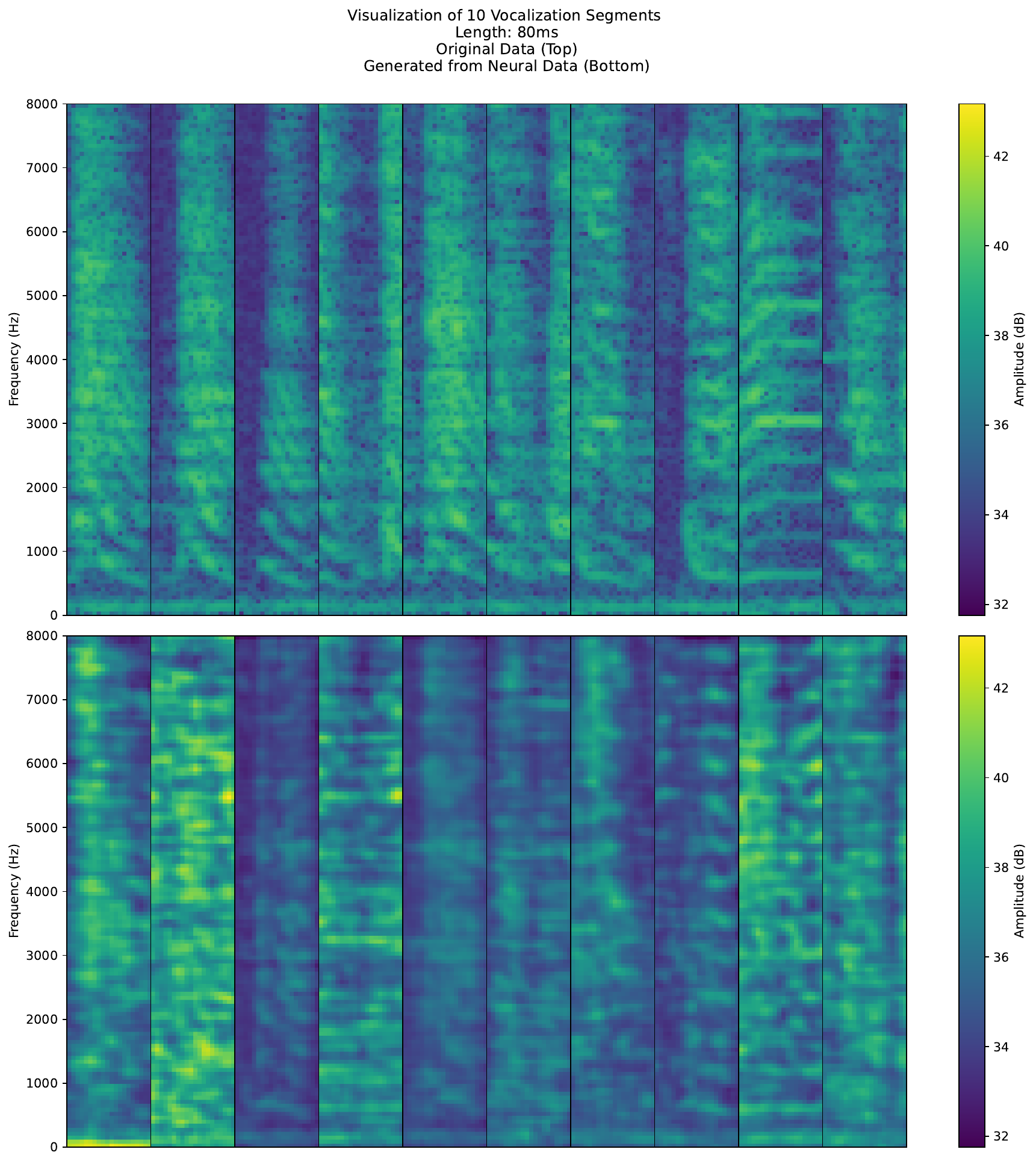}
    \caption{Generate 80\,ms vocalization from 80\,ms neural data}
    \label{fig:neuro2voc_80ms}
\end{figure}




\clearpage
\listoffigures

\clearpage
\listoftables

\clearpage

\chapter*{List of Acronyms}

\begin{tabular}{lp{12.5cm}}\\
AFP & anterior forebrain pathway\\
area X & area X of the paraolfactory lobe\\
CCA & Canonical Correlation Analysis\\
CNN & Convolutional Neural Network\\
CVE & Continuous Value Embedding\\
DLM & dorsolateral division of the medial thalamus\\
EEG & Electroencephalogram\\
HVC & Hyperstriatum Ventralis pars caudalis\\
KLD & Kullback-Leibler Divergence\\
LMAN & Lateral nucleus Magnocellularis of the Anterior Neostriatum\\
LSTM & Long Short-Term Memory\\
MEG & Magnetoencephalogram\\
MSE & Mean Squared Error\\
NLP & Natural Language Processing\\
RA & Nucleus Robustus of the Archistriatum\\
RBF & Radial Basis Function\\
RF & Random Forest\\
SVM & Support Vector Machine\\
SSM & State-Space Model\\
t-SNE & t-distributed Stochastic Neighbor Embedding\\
UMAP & Uniform Manifold Approximation and Projection\\
VAE & Variational Auto Encoder \\
\end{tabular}




\end{document}